\documentclass[aps,prl,reprint,showpacs,superscriptaddress,nofootinbib,twocolumn]{revtex4-2}
\usepackage{graphicx}
\usepackage{subfigure}
\usepackage{booktabs}
\usepackage{graphicx}
\usepackage{upgreek}
\usepackage{subfigure}
\usepackage{soul}
\usepackage{color}
\usepackage{amsmath}
\usepackage{makecell}
\usepackage[percent]{overpic}
\usepackage{hyperref}   % 
\usepackage{cleveref}   %
\begin{document}
\newcommand\mean[1]{\ensuremath{\langle #1\rangle}}
\newcommand\ket[1]{\ensuremath{|#1\rangle}}
\newcommand\bra[1]{\ensuremath{\langle#1|}}

\newcommand{\y}[1]{{\color{blue} #1}}
\newcommand{\yc}[1]{{\color{red} #1}}
\newcommand{\yr}[2]{{\color{blue}{\st{#1}} #2}}
\newcommand{\yd}[1]{{\color{blue} \st{#1}}}

\newcommand{\jc}[1]{{\color{green} #1}}

\title{\textbf{1-Mbps Twin-Field Quantum Key Distribution over 200 km Using Independent Dissipative Kerr Solitons}}

\author{Hao Dong*}
\affiliation{Hefei National Research Center for Physical Sciences at the Microscale and School of Physical Sciences,
University of Science and Technology of China, Hefei, Anhui, 230026, China}
\affiliation{Jinan Institute of Quantum Technology and Hefei National Laboratory Jinan Branch, Jinan, Shandong, 250101, China}

\author{Tian-Jiao Zhang*}
\affiliation{Hefei National Research Center for Physical Sciences at the Microscale and School of Physical Sciences,
University of Science and Technology of China, Hefei, Anhui, 230026, China}
\affiliation{Jinan Institute of Quantum Technology and Hefei National Laboratory Jinan Branch, Jinan, Shandong, 250101, China}

\author{Yan-Wei Chen*}
\affiliation{Hefei National Research Center for Physical Sciences at the Microscale and School of Physical Sciences,
University of Science and Technology of China, Hefei, Anhui, 230026, China}
\affiliation{Jinan Institute of Quantum Technology and Hefei National Laboratory Jinan Branch, Jinan, Shandong, 250101, China}

\author{Wei Sun}
\affiliation{International Quantum Academy and Shenzhen Futian SUSTech Institute for Quantum Technology and Engineering, Shenzhen, Guangdong, 518048, China}

\author{Cong Jiang}
\affiliation{Jinan Institute of Quantum Technology and Hefei National Laboratory Jinan Branch, Jinan, Shandong, 250101, China}
\affiliation{Hefei National Laboratory, University of Science and Technology of China, Hefei, Anhui, 230088, China}

\author{Sanli Huang}
\affiliation{International Quantum Academy and Shenzhen Futian SUSTech Institute for Quantum Technology and Engineering, Shenzhen, Guangdong, 518048, China}

\author{Shuyi Li}
\affiliation{International Quantum Academy and Shenzhen Futian SUSTech Institute for Quantum Technology and Engineering, Shenzhen, Guangdong, 518048, China}

\author{Di Ma}
\affiliation{Jinan Institute of Quantum Technology and Hefei National Laboratory Jinan Branch, Jinan, Shandong, 250101, China}

\author{Xiang-Bin Wang}
\affiliation{Jinan Institute of Quantum Technology and Hefei National Laboratory Jinan Branch, Jinan, Shandong, 250101, China}
\affiliation{Hefei National Laboratory, University of Science and Technology of China, Hefei, Anhui, 230088, China}
\affiliation{State Key Laboratory of Low Dimensional Quantum Physics, Department of Physics, Tsinghua University, Beijing 100084, China}

\author{Yang Liu}
\affiliation{Hefei National Research Center for Physical Sciences at the Microscale and School of Physical Sciences,
University of Science and Technology of China, Hefei, Anhui, 230026, China}
\affiliation{Jinan Institute of Quantum Technology and Hefei National Laboratory Jinan Branch, Jinan, Shandong, 250101, China}
\affiliation{Hefei National Laboratory, University of Science and Technology of China, Hefei, Anhui, 230088, China}
\author{Junqiu Liu}
\affiliation{International Quantum Academy and Shenzhen Futian SUSTech Institute for Quantum Technology and Engineering, Shenzhen, Guangdong, 518048, China}
\affiliation{Hefei National Laboratory, University of Science and Technology of China, Hefei, Anhui, 230088, China}

\author{Qiang Zhang}
\affiliation{Hefei National Research Center for Physical Sciences at the Microscale and School of Physical Sciences,
University of Science and Technology of China, Hefei, Anhui, 230026, China}
\affiliation{Jinan Institute of Quantum Technology and Hefei National Laboratory Jinan Branch, Jinan, Shandong, 250101, China}
\affiliation{Hefei National Laboratory, University of Science and Technology of China, Hefei, Anhui, 230088, China}

\author{Jian-Wei Pan}
\affiliation{Hefei National Research Center for Physical Sciences at the Microscale and School of Physical Sciences,
University of Science and Technology of China, Hefei, Anhui, 230026, China}
\affiliation{Hefei National Laboratory, University of Science and Technology of China, Hefei, Anhui, 230088, China}

\date{\today}

\begin{abstract}
Twin-field quantum key distribution (TF-QKD) dramatically enhances the secure key rate (SKR) over inter-city distances through its square-root scaling. Further improvements in aggregate SKR can be achieved by wavelength-division multiplexing (WDM) of parallel QKD channels. However, direct implementation in TF-QKD poses significant challenges, as each wavelength channel requires an independent ultra-stable seed laser, narrow-linewidth transmitters, and optical phase-locked loops (OPLLs), which are not easily scalable. Here, we circumvent these limitations by employing two independent, integrated dissipative Kerr soliton (DKS) microcombs at Alice and Bob as multi-wavelength sources. High-visibility single-photon interference across all wavelength channels is achieved by stabilizing the frequencies of every comb line—requiring only the stabilization of the pump wavelength and repetition rates of the two microcombs. Based on this architecture, we perform a full TF-QKD experiment using the sending-or-not-sending protocol, achieving a total SKR of 1.57~Mbps over 201.1~km of fiber using 16 DWDM channels. This result represents more than an order-of-magnitude enhancement compared with single-wavelength TF-QKD at the same distance. Given that a single DKS comb can support over 100 coherent lines across the C-band, this approach offers a scalable pathway toward high-rate quantum key distribution over inter-city distances.
\end{abstract}

\maketitle

\section{Introduction}\label{sec1}

Quantum key distribution (QKD) provides information-theoretic security, guaranteed by the laws of quantum physics~\cite{inproceedings,RevModPhys.92.025002}. In practice—particularly in high-bandwidth encrypted services—the performance metrics such as the SKR, transmission distance are equally critical for real-world deployments. Over the past decade, significant advances have enabled GHz-clock rates and high-count-rate single photon detection using multipixel superconducting nanowire single-photon detectors (SNSPDs), pushing SKR to 115.8~Mbps~\cite{li2023,Grünenfelder2023} over 10~km of optical fiber. However, for long-haul links, the performance of conventional protocols such as decoy-state BB84 is fundamentally limited by the repeaterless secret key capacity bound~\cite{takeoka2014,pirandola2017}, which indicates that the SKR scales linearly with channel transmittance $\eta$ (i.e., SKR $\propto \eta$). Consequently, the SKR drops to just a few Mbps~\cite{10.1063/5.0021468} at 100~km, and will further decrease beyond metropolitan distances.

TF-QKD~\cite{lucamarini2018} overcomes the rate–loss limitation by achieving a square-root scaling of SKR with channel transmittance (i.e., SKR$\propto \sqrt{\eta}$), a performance previously attainable only with quantum repeaters. What's more, TF-QKD is a measurement-device-independent (MDI) protocol that inherently eliminates all detection-side attacks. These properties make it uniquely suited for inter-city quantum trunk lines. This potential has been validated by rapid experimental progress: recent laboratory demonstrations have extended TF-QKD distance up to 1000 km~\cite{Minder2019,Liu2019TF300km,PhysRevX.9.021046,Chen2020TF509km,wang2022_830km}, while field trials have reached distances beyond 500 km~\cite{LiuHui2021,Chen2021_511km,Zhou2024_546km,Pittaluga2025} using the sending-or-not-sending (SNS) protocol~\cite{Wang_SNS2018}.
 
\begin{figure}[h!]
    \centering
    \includegraphics[width=0.5\textwidth]{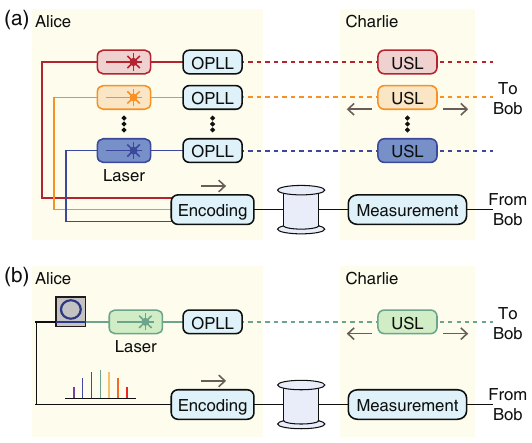}
    \caption{\textbf{Principle of WDM-based TF-QKD architectures.} \textbf{(a)} TF-QKD using an array of independent narrow-linewidth lasers for WDM-parallelization. Each laser is phase-locked via an OPLL to a remote USL. The quantum signals are combined and transmitted via a fiber channel for interference and detection at the central untrusted node. \textbf{(b)} TF-QKD using DKS microcombs as multi-wavelength sources. A single laser is phase-locked to a remote USL via a single OPLL. This pump laser drives a chip-integrated microresonator to generate a frequency comb. In both configurations, the sender setup at Bob's side is identical to that at Alice's. }
    \label{fig:setup_comparison}
\end{figure}

Despite the record-breaking distance, a higher SKR of TF-QKD is still demanded to meet the requirement of high-throughput backbone encryptions. A natural route to enhance aggregate SKR is WDM—a foundational technology in classical optical networks that enabled hundred-terabit-per-second capacities~\cite{Hu2018,Kemal:19}. However, high-visibility single-photon interference in TF-QKD requires exactly frequency-matching between the lasers at Alice and Bob. In typical implementations, this is achieved by distributing an ultra-stable seed laser from a central node and locking the narrow linewidth lasers via OPLLs~\cite{Minder2019,wang2022_830km,Liu2023TF1000km} to the seed. Unfortunately, such an architecture suffers from poor scalability across multiple wavelength channels. As shown in Fig.~\ref{fig:setup_comparison}(a), straightforward WDM extending TF-QKD would require $N$ independent USLs, $N$ OPLLs, and $N$ local lasers—one for each wavelength channel. This leads to a drastic increase in hardware complexity that scales linearly with the number of channels, presenting a fundamental bottleneck for the development of WDM-based TF-QKD systems.

Over the past decade, DKS microcomb has emerged as a transformative multi-wavelength platform. The high‑$Q$ microresonators, which are essential to DKS microcomb, can be fully integrated on-chip~\cite{Yi:15, Brasch:15, Joshi:16, Bao:19, LiuX:21, He:19, Gong:20, Guidry:21} and manufactured using established CMOS foundries~\cite{Liu:21, Ye:23, Girardi:25}. DKS microcomb features high coherence, broad bandwidth, and line spacings that naturally align with the dense wavelength-division multiplexing (DWDM) grid. These properties have enabled their use in system-level information and metrology applications. Notably, DKS microcomb has enabled massively parallel optical communications with high throughput and low cost~\cite{Marin-Palomo:17, Corcoran:20, Mazur:21}. 

The properties have also spurred interest as multi-wavelength sources for parallel QKD architectures~\cite{DKS_BB84_2019}. High-visibility Hong–Ou–Mandel interference~\cite{doi:10.1126/sciadv.adq8982} and MDI-QKD network based on optical frequency comb has been demonstrated~\cite{yan2025measurement}. A TF-QKD network architecture employing a microcomb at the server node and injection-locked lasers at the user nodes has also been explored~\cite{microcomb_TFQKDnet2026}. However, in this architecture, an external seed laser is injected into each user’s QKD encoder to align their wavelengths, raising security concerns related to Trojan-horse attacks~\cite{gisin2006trojan,jain2014trojan}. Practical countermeasures, such as watchdog detectors~\cite{wiesemann2025evaluation,juarez2026reference}, are nonetheless required in practical systems. Furthermore, the total number of laser sources at the user end is not reduced compared to that in a conventional TF-QKD network. 

Here, we employ two independent, integrated DKS microcombs as scalable multi-wavelength sources for TF-QKD. As illustrated in Fig.~\ref{fig:setup_comparison}(b), each microcomb source is pumped by a single laser, which is phase-locked to a single ultra-stable seed via a single OPLL at Alice and Bob. The generated comb lines enable parallel TF-QKD channels through WDM. Crucially, once the pump wavelength and the comb’s repetition rates ($f_{\text{rep}}$) are stabilized, all comb lines at Alice and Bob are automatically aligned in frequency—thereby eliminating the need for per-channel USLs, OPLLs, and narrow line-width lasers. This dramatically reduces hardware overhead and enables scalable parallel TF-QKD over a single fiber. To validate this approach, we implemented 16 parallel TF-QKD channels using selected comb lines transmitted simultaneously through 201.1~km of ultra-low-loss fiber. This proof-of-principle experiment achieves a total SKR of 1.57~Mbps using the SNS-TF-QKD protocol—an approximately 16-fold improvement over the system based on narrow-linewidth lasers as the light sources.

\section{Soliton generation and microcomb performance}

\begin{figure*}[htbp]
  \centering
  \includegraphics[width=\textwidth]{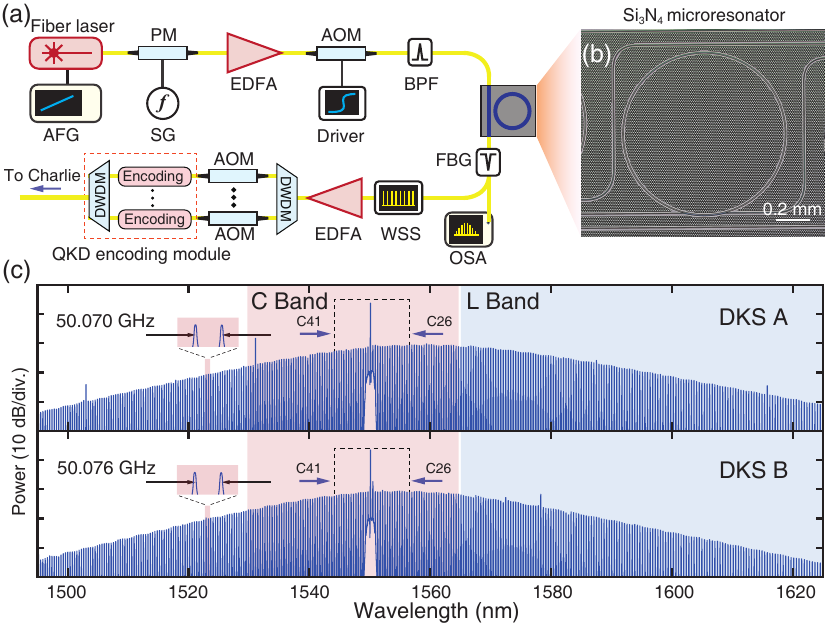}
  \caption{\textbf{Soliton microcomb generation.}
  \textbf{(a)} The experimental setup for DKS microcomb generation. A fiber laser serves as the pump for the Si$_3$N$_4$ microresonator, with its wavelength controlled by an arbitrary function generator (AFG). The pump light is amplified by an erbium-doped fiber amplifier (EDFA), and the amplified spontaneous emission (ASE) noise is suppressed using a bandpass filter (BPF). A phase modulator (PM) controled by the signal generator (SG) is inserted to extend the soliton steps and facilitate multi-soliton generation; an acousto-optic modulator (AOM) is used to lock the pump frequency and the DKS microcomb repetition rate. The resulting soliton spectrum is recorded by an optical spectrum analyzer (OSA). At the output of each DKS microcomb source, a wavelength-selective switch (WSS) isolates the designated QKD channels, and an additional AOM fine-tunes the optical frequency of each comb line. 
  \textbf{(b)} Photograph of one leveraged Si$_3$N$_4$ microresonators. 
  \textbf{(c)} The optical spectra of the generated single solitons. The line spacings are 50.070~GHz and 50.076~GHz, respectively. The wavelength channels used for TF-QKD are indicated by the dashed box, spanning C26 to C41 (1544.53~nm to 1556.55~nm).}
  \label{optical_setup}
\end{figure*}

The experimental setup for the DKS microcomb generation is shown in Fig.~\ref{optical_setup}(a). Independent DKS microcombs for Alice and Bob are generated in high-$Q$ silicon nitride (Si$_{3}$N$_{4}$) integrated microresonators, illustrated in Fig.~\ref{optical_setup}(b). The microresonators are fabricated using a foundry-compatible process~\cite{Ye:23}, featuring a waveguide thickness of 800~nm and a width of 2.2~$\upmu$m, and a designed inner radius of 455~$\upmu$m. Both devices exhibit intrinsic quality factors exceeding $Q_0>1 \times 10^7$. The free spectral range (FSR, $D_1/2\pi$) is approximately 50~GHz at 1550~nm wavelength. The second-order dispersion parameter ($D_2/2\pi$) is positive, corresponding to anomalous group-velocity dispersion (GVD)—a prerequisite for DKS generation. Further details on fabrication and characterization are provided in the Supplementary Information.

To generate soliton, the Si$_3$N$_4$ microresonator is pumped by a narrow-linewidth CW fiber laser. By scanning the microresonator resonance mode from effective blue detuning to red detuning, a multi-soliton state is typically excited. Subsequently, by reversing the scan direction, a single-soliton state can be accessed deterministically~\cite{Guo2017}. A phase modulator is introduced into the setup to create a blue sideband of the pump laser, enabling thermal compensation and facilitating robust multi-soliton excitation~\cite{Zheng2023}. Fig.~\ref{optical_setup}(c) shows the optical spectra of the DKS microcombs for Alice (DKS A) and Bob (DKS B). The measured mode spacings are 50.070~GHz and 50.076~GHz, respectively. Further details on DKS microcomb generation are provided in the Supplementary Information.

\begin{figure*}[htbp]
    \centering
    \includegraphics[width=\textwidth]{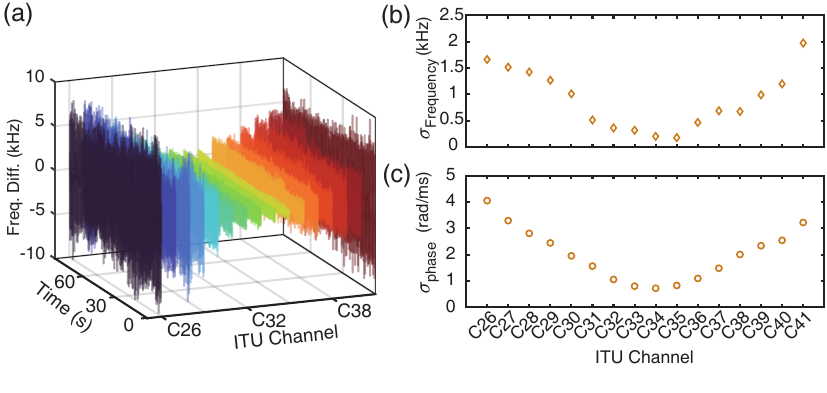}
    \caption{\textbf{(a)} The frequency difference between the corresponding comb lines (C26 to C41) from Alice and Bob. \textbf{(b)} The standard deviation of the relative frequency difference across the 16 microcomb line pairs. \textbf{(c)} The standard deviation of the phase drift rate between the corresponding comb lines.}
    \label{interfere}
\end{figure*}

Single-photon interference in TF-QKD requires precise wavelength matching between corresponding microcomb lines from DKS A and DKS B. To achieve this, the pump lasers of both soliton sources are locked to a remote USL at 1550.1~nm (193.4~THz); the resonance modes of their Si$_3$N$_4$ microresonators are temperature controlled to 34.52~$^\circ$C (Alice) and  30.32~$^\circ$C (Bob), to align with the pump frequency. Once established, the soliton state persists over a detuning range of approximately 500~MHz. This wide operating window provides sufficient flexibility to lock the pump laser to the remote USL via an OPLL, ensuring long-term stability of the DKS microcombs. In addition to pump-wavelength locking, we actively lock the repetition rates of the two DKS microcombs. A phase modulator is leveraged to generate sub-electro-optic (sub-EO) sidebands between adjacent microcomb lines. The beat frequency between adjacent sub-EO sidebands down-converts the DKS repetition rate to approximately 10~MHz. This signal enables a feedback control to stabilize the repetition rate. Details of both locking schemes are provided in the Supplementary Information.

At the output of each DKS microcomb source, a wavelength-selective switch (WSS) is used to filter out 16 comb lines spanning the ITU-T C-band channels C26 to C41 (1544.53~nm to 1556.55~nm) with 100-GHz spacing. These comb lines are then amplified collectively by a single erbium-doped fiber amplifier (EDFA) to achieve the required power levels for TF-QKD encoding. As the repetition rates of Alice’s and Bob’s microresonators are not perfectly matched, a frequency offset arises between corresponding comb lines. To compensate for this mismatch, an additional AOM is inserted in each wavelength channel to shift the frequency of the respective comb line, thereby achieving perfect frequency alignment. 

 The frequency offsets for all 16 comb-line pairs were measured using a frequency counter with a gate time of 1 ms, as shown in Fig.~\ref{interfere}(a). The corresponding standard deviation of the frequency offsets is shown in Fig.~\ref{interfere}(b), revealing an increasing trend for comb lines away from the pump wavelength. This behavior likely stems from the combined effects of the increasing linewidth~\cite{Lei2022} and the amplified feedback noise from the repetition-rate stabilization. The maximum standard deviation across all channels is measured to be less than 2~kHz, which remains within the phase compensation bandwidth of the TF-QKD system.

The performance of single-photon interference between corresponding microcomb lines is critical to the success of TF-QKD. We measured the interference for all 16 comb lines between Alice’s and Bob’s microcombs. The phase drift rate recorded is shown in Fig.~\ref{interfere}(c), which also exhibits a trend of increasing for comb lines away from the pump wavelength. We attribute this phenomenon mainly to the increasing frequency mismatch between corresponding comb lines. The maximum drift rate is below 4.1~rad/ms, which is comparable to that in hundred-kilometer fiber spools and can be effectively compensated in our TF-QKD system.

We performed full TF-QKD tests sequentially on each of the 16 wavelength channels. To emulate the worst-case crosstalk noise expected in realistic network scenarios, the non-target channels are set to the highest phase reference intensity. All 16 wavelength channels are then multiplexed and transmitted through a symmetric fiber to the detection node. At Charlie’s measurement station, the incoming light is first demultiplexed by a $1\times16$ DWDM module to separate the 16 wavelength channels, followed by a 50-GHz DWDM filter to suppress crosstalk noise. The signal in each channel is then directed into a polarizing beam splitter (PBS) followed by a 50:50 polarization-maintaining beam splitter (PMBS) for interference. The idler port of the PBS is used to monitor the polarization and delay drifts during the experiment for real-time stabilization. The interference outputs are detected by SNSPDs and recorded with a Time Tagger.

\section{Results}

\begin{figure*}[htbp]
     \centering
     \includegraphics[width=1\textwidth]{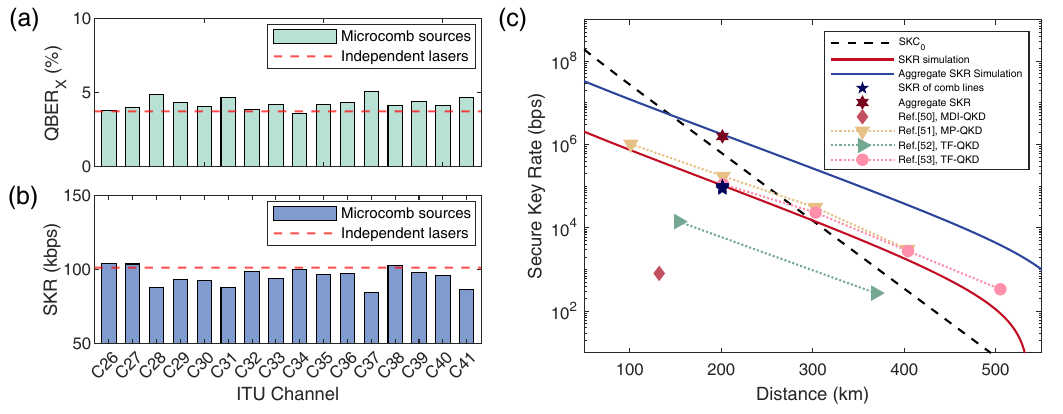}
     \caption{\textbf{Performance of the WDM-based TF-QKD system.} The bars in (a) and (b) represent the X-basis QBERs and SKRs of TF-QKD using 16 comb line pairs from two independent microcombs. The red dashed line in (a) and (b) represents the QBER and SKR of TF-QKD using two independent lasers as the source. (c) The blue stars and the red hexagon are the experimentally obtained SKRs of 16 comb-line pairs from independent microcombs and the aggregate SKR. The solid lines represent the theoretical simulation of SKR using a single wavelength (red) and 16 wavelength channels (blue). The black dashed line denotes the repeaterless secret key capacity bound ($\mathrm{SKC_0}$). The remaining symbols represent state-of-the-art SKRs from recent QKD experiments for comparison.}
     \label{fig:skr}
\end{figure*}

We implemented the three-state SNS-TF-QKD protocol, in which Alice and Bob each employ a vacuum, a decoy, and a signal source. In this scheme, vacuum pulses used for decoy-state analysis and key generation are not distinguished during state preparation and measurement. After error correction, the positions of all vacuum-related pulses are identified from the locations of error bits, enabling accurate estimation of vacuum counts. In the experiment, the signal state is sent with a probability of 27\% and a mean photon number of $\mu_y=0.48$, while the decoy state is sent with a probability of 3\% and a mean photon number of $\mu_x=0.05$. These parameters are identical across all 16 comb lines.

The system operated at a clock frequency of 1 GHz, with an effective signal rate of 800~MHz and a pulse width of 200~ps. Within each 100~ns time slot, the first 20~ns serves as the phase reference, while the remaining~80 ns is used for the quantum signal. For each wavelength channel, a total of $1.44\times10^{12}$ pulses is transmitted, corresponding to approximately 30 minutes of data acquisition. In calculating the aggregate SKR, the pulses across all wavelength channels were treated as a single ensemble, rather than simply summing the SKRs from individual channels. The fiber distances from Alice and Bob to Charlie are 99.5~km and 101.55~km, with losses of 16.25~dB and 16.49~dB, respectively. The total insertion loss at the receiver is approximately 2~dB. The detection efficiencies of the SNSPDs are 82.1\% and 82.9\%, with dark count rates of 77.5~Hz and 69.8~Hz, respectively. Throughout the experiment, the total power of the 15 non-target channels is set to –50.6~dBm. The resulting noise contributions are measured to be below 32.0~Hz and 35.0~Hz for the two SNSPD channels from all wavelength channels ~(see Supplementary Information for details).

In calculating the secure key rate, we adopt the incorporated decoy-state analysis combined with active odd-parity pairing (AOPP) during post-processing to maximize the secure key rate.  Alice and Bob can distil the secure keys according to the following formula~\cite{jiang_2020,jiang_2021,hu_2022}:
\begin{equation}
R=\frac{1}{N}\{n_1[1-H(e_1^{\mbox{ph}})]-\text{leak}_{EC}\}-R_{\mbox{tail}},
\label{Eq:KeyRate}
\end{equation}
where $R$ is the key rate of per sending-out pulse pair; $n_1$ is the number of untagged bits after AOPP and $e_1^{\mbox{ph}}$ is the corresponding phase-flip error rate; $\text{leak}_{EC}$ is the number of leaked information about the raw keys in the error correction process and generally $\text{leak}_{EC}=fn_t H(E_t)$, where $n_t$ is the number of survived bits after AOPP and $E_t$ is the corresponding bit-flip error rate in those survived bits, $f$ is the error correction inefficiency which we set to $f=1.16$; $H(x)=-x\log_2x-(1-x)\log_2(1-x)$ is the Shannon entropy. And $R_{\mbox{tail}}$ is for the security with finite-data size and the advanced decoy state analysis when calculating the SKR in the non-asymptotic case~(Detailed definition of $R_{\mbox{tail}}$ can be found in Supplementary Information).

For each of the 16 wavelength channels, an average of $4.533\times10^9$ valid detections per channel is recorded. As shown in Fig.~\ref{fig:skr}(a), the average QBER after AOPP is $5.78\times10^{-4}$ in the Z basis and 4.29\% in the X basis. The low interference QBERs indicate that the bandwidth of our phase compensation is sufficient to suppress the additional phase fluctuations induced in comb lines. At a total fiber distance of 201.1 km, the average SKR per comb line is 95.39 kbps, with detailed results shown in Fig.~\ref{fig:skr}(b). The total SKR across all 16 channels reaches 1.57 Mbps—approximately twice the fundamental repeaterless secret key capacity bound at this distance. As shown in Fig.~\ref{fig:skr}(c), the aggregates SKR achieved in this work shows a significant advantage at distances of 200 km and beyond, outperforming previous experiments based on decoy-state BB84~\cite{10.1063/5.0021468,Grünenfelder2023,li2023}, measurement-device-independent~\cite{PhysRevA.108.022605}, mode-pairing~\cite{PhysRevX.15.021066}, and TF-QKD~\cite{Pittaluga2021,Liu2023_1000km_finite-key_analysis} protocols.

For comparison, we implemented a TF-QKD using independent narrow-linewidth lasers at Alice and Bob, each locked to the remote USL via an OPLL, while keeping all other system components unchanged. Under this configuration, the system achieved a Z basis QBER after AOPP of $7.433\times10^{-4}$, an X basis QBER of 3.75\%, and an SKR of 102.16 kbps, as indicated by the red dashed lines in Fig.~\ref{fig:skr}(a) and Fig.~\ref{fig:skr}(b). The slightly lower QBER observed with discrete lasers likely stems from their narrower intrinsic linewidth and the absence of inter-channel crosstalk. Nevertheless, the performance difference is marginal, confirming that the microcomb source introduces only minor degradation in interference fidelity.

\section{Discussion and Conclusion}
In summary, we present a proof-of-principle demonstration of TF-QKD over a 201.1 km fiber link using DKS microcombs as scalable multi-wavelength sources. The system achieves a total SKR of 1.57 Mbps, representing nearly a 16-fold improvement compared to TF-QKD implementations based on narrow-linewidth lasers. The performance is enabled by full stabilization of both the pump wavelengths and repetition rates of two independent microcombs, allowing 16 DWDM channels to interfere with high fidelity through a single fiber. 
Compared with the previous microcomb-based TF-QKD network demonstration~\cite{microcomb_TFQKDnet2026}, our point-to-point TF-QKD system achieves an aggregate SKR that is about two orders of magnitude higher than the total SKR of the entire network, over the same transmission distance. Notably, we employ independent microcomb sources at Alice and Bob, instead of optical injection locking that requires external light to be injected into the encoding system. We implement phase randomization for decoy-state and random intensity and phase modulation during QKD encoding, which are essential for a practical TF-QKD system. 

The performance and practicality of microcomb-based TF-QKD can be further enhanced by fully harnessing the capabilities of DKS microcombs. A single DKS microcomb provides over two hundred coherent lines spanning the C+L telecom bands. Exploiting this broad spectral coverage—combined with an increased system clock rate and the use of low-loss hollow-core fiber—could enable aggregate SKRs approaching the gigabit-per-second level at this inter-city distance. Realizing this vision will require a highly integrated architecture with enhanced component performance. For instance, linewidth broadening at large mode numbers must be mitigated—e.g., via Kerr-induced synchronization~\cite{moille2025all} or self-injection locking~\cite{lei2024self}. 
Higher pump-to-soliton conversion efficiency and improved spectral flatness can be achieved using dual-microresonator designs~\cite{HelgasonO:2023, Girardi:25}, thereby reducing overall power consumption and enabling broadband parallelization. The pump lasers and microresonators can be integrated using hybrid or heterogeneous photonic platforms~\cite{Kondratiev:23, SternB:2018, Shen:20, Xiang:21, Sun:25}. High-speed modulators can likewise be heterogeneously integrated using thin-film lithium niobate~\cite{ChuraevM:2023} or lithium tantalate~\cite{NielsM:2026}, enabling scalable electro-optic modulation for QKD encoding. By combining these techniques, we envision a pathway toward a compact, energy-efficient, and high-SKR TF-QKD system for QKD trunk lines.

\section{Acknowledgments}
We thank Baoqi Shi and Jinbao Long for assisting the experiment. 
This work was supported by 
Quantum Science and Technology-National Science and Technology Major Project (2021ZD0300700, 
2023ZD0301500), 
National Key R\&D Program of China (Grant No. 2024YFA1409300), 
the National Natural Science Foundation of China (Grants No. T2125010, 
No. 12374470,  
No. 12404436, 
No. 62405202, 
No. U25D9005 
), 
the Chinese Academy of Sciences, 
Shenzhen-Hong Kong Cooperation Zone for Technology and Innovation (HZQB-KCZYB2020050), 
and Shenzhen Science and Technology Program (Grant No. RCJC20231211090042078).  
C.J., X.-B.W., and Q.Z. acknowledge support from the Taishan Scholar Program of Shandong Province. Q.Z. was supported by the New Corner Stone Science Foundation through the Xplorer Prize.

\section{Supplementary materials}

\subsection{The three-intensity SNS-TF-QKD protocol}

The three-intensity SNS protocol~\cite{physreva.98.062323} incorporating advanced decoy-state analysis combined with active odd-parity pairing (AOPP) during post-processing is adopted here. The source parameters are symmetric for Alice and Bob: there are three sources on each side which are the vacuum source $v$, the decoy source $x$, and the signal source $y$ with intensities $\mu_v=0, \mu_x, \mu_y$ and probabilities $p_0, p_x, p_y$ respectively. In each time window, Alice (Bob) randomly prepares and sends out a pulse from the three candidate sources to Charlie. Let $n_{lr}$ ($l,r=v,x,y$) be the number of clicking event from sources $lr$. Here, the clicking events from sources where Alice and Bob choose the sources $v$ or $y$ are used to extract the secure keys. By publicly announcing the position of clicking event where Alice or Bob chooses the $x$ source, Alice and Bob can know the values of $n_{vx}$ and $n_{vy}$ tother with $m_{xx}$ whose value can be used to estimate the phase error rate. After AOPP and error correction, Alice and Bob can know the values of $n_{vv},n_{vy},n_{yv}$ while kept the positions of sources $vy,yv$ privately. With those values, Alice and Bob can perform the decoy-state analysis and obtains $n_1$, the lower bound of the number of survived untagged bits after AOPP and  $e_1^{\mbox{ph}}$, the upper bound of the phase-flip error rate of those survived untagged bits after AOPP. And then Alice and Bob can distil the secure keys according to the following formula~\cite{jiang_2020,jiang_2021,hu_2022}:
\begin{equation}
R=\frac{1}{N}\{n_1[1-H(e_1^{\mbox{ph}})]-\text{leak}_{EC}\}-R_{\mbox{tail}},
\label{Eq:KeyRate}
\end{equation}
where $R$ is the key rate of per sending-out pulse pair; $\text{leak}_{EC}$ is the number of leaked information about the raw keys in the error correction process and generally $\text{leak}_{EC}=fn_t H(E_t)$ where $n_t$ is the number of survived bits after AOPP and $E_t$ is the corresponding bit-flip error rate in those survived bits, $f$ is the error correction inefficiency which we set to $f=1.16$; $H(x)=-x\log_2x-(1-x)\log_2(1-x)$ is the Shannon entropy. And $R_{\mbox{tail}}$ is
\begin{equation}\label{r2}
\begin{split}
R_{\mbox{tail}}=\frac{1}{N}[2\log_2{\frac{2}{\varepsilon_{cor}}}+4\log_2{\frac{1}{\sqrt{2}\varepsilon_{PA}\hat{\varepsilon}}}+2\log_2(n_{vy}+n_{yv})],
\end{split}
\end{equation}
where $\varepsilon_{cor}$ is the failure probability of error correction, $\varepsilon_{PA}$ is the failure probability of privacy amplification, $\hat{\varepsilon}$ is the coefficient while using the chain rules of smooth min- and max- entropy~\cite{vitanov2013chain}, and $2\log_2(n_{vy}+n_{yv})$ is the extra cost of the advanced decoy state analysis~\cite{hu_2022}. In this work, we set $\varepsilon_{cor}=\varepsilon_{PA}=\hat{\varepsilon}=\varepsilon_{PE}=10^{-10}$, where $\varepsilon_{PE}$ is the failure probability in the statistical fluctuation analysis.

\subsection{DKS microcomb light sources}

\subsubsection{Fabrication and characterization of the Si$_3$N$_4$ microresonator}

\begin{figure*}[h]
  \centering
  \includegraphics[width=\textwidth]{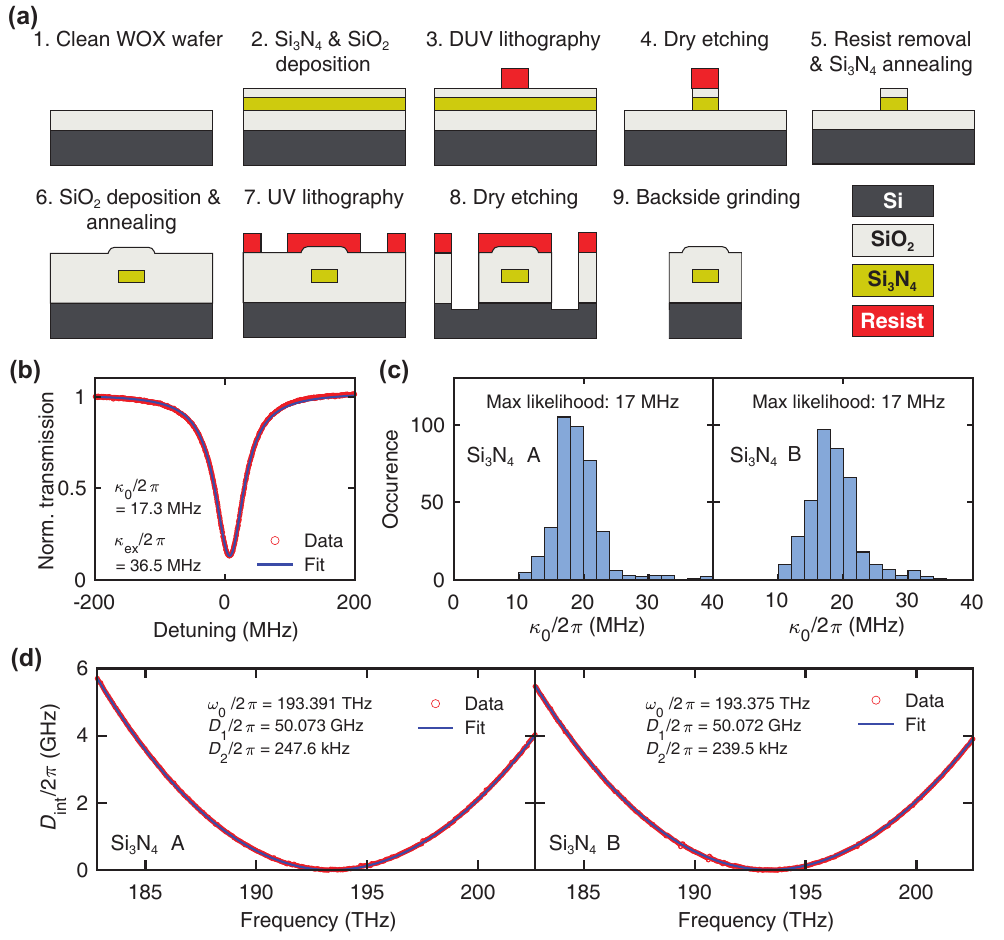}
  \caption{
  Fabrication and characterization of silicon nitride microresonators.
  \textbf{(a)}. 
  The DUV subtractive process flow of 6-inch-wafer Si$_3$N$_4$ foundry fabrication. 
  WOX, thermal wet oxide (SiO$_2$).
  \textbf{(b)} A typical transmission of the microresonators.
  The intrinsic/external linewidth is 17.3/36.5~MHz.
  \textbf{(c)} Statistical distribution of the intrinsic linewidths of the resonances of two Si$_3$N$_4$ microresonators.
  The maximum-likelihood intrinsic linewidths are $\kappa^\mathrm{A}_0/2\pi = 17~\mathrm{MHz}$, $\kappa^\mathrm{B}_0/2\pi = 17~\mathrm{MHz}$.
  \textbf{(d)} The integrated dispersions of the two Si$_3$N$_4$ microresonators.
  The FSRs ($D_1/2\pi$) are 50.073~GHz and 50.072~GHz, which are quite close to each other.
  }
  \label{Fig:fab}
\end{figure*}

The fabrication process flow for Si$_3$N$_4$ integrated waveguides and microresonators is shown in Fig.~\ref{Fig:fab}(a).
The process is based on 6-inch (150-mm-diameter) wafers and uses an optimized deep-ultraviolet (DUV) subtractive process~\cite{Ye:23,Sun2025}. The process starts with the deposition of a 800-nm-thick Si$_3$N$_4$ film on a clean thermal wet SiO$_2$ substrate by low-pressure chemical vapor deposition (LPCVD). A SiO$_2$ layer is then deposited on Si$_3$N$_4$ as an etch hardmask, again using LPCVD. After spin-coating a DUV photoresist, KrF (248~nm) stepper lithography defines the waveguide pattern in the photoresist. Subsequent dry etching with C$_4$F$_8$, CHF$_3$, and O$_2$ transfers the pattern from the photoresist to the SiO$_2$ hardmask and then into the Si$_3$N$_4$ layer to form waveguides and microresonators. The dry etch process is optimized to yield smooth, vertical sidewalls. High-quality photolithography and dry etching are critical for minimizing optical scattering loss in the waveguides. The photoresist is then removed, followed by thermal annealing in a nitrogen atmosphere to drive out hydrogen.
A SiO$_2$ cladding layer is then deposited on top of the wafer and thermally annealed again.
Smooth chip facets are created by contact UV photolithography and additional deep dry etching, which is critical for hybrid integration and packaging. 
The chip size is also defined in this step.
Finally, the wafer is separated into individual chips by backside grinding.

The fabricated Si$_3$N$_4$ microresonators are characterized using a homemade vector spectrum analyzer~\cite{Luo2024,Shi2025,shi2026} under the ambient laboratory temperature conditions (approximately 22~°C). The resonance modes $\omega_\mu$ of the microresonator can be expressed as
\begin{align}
  \omega_{\mu}
    &= \omega_{0} + D_{1}\mu + \frac{1}{2} D_{2}\mu^{2} + \frac{1}{6} D_{3}\mu^3 + \cdots \\
    &= \omega_{0} + D_{1}\mu + D_\mathrm{int},
  \label{eq:modes}
\end{align}
where $\mu$ is the mode number, $D_1/2\pi$ is the FSR, $D_2/2\pi$ is the group velocity dispersion (GVD), $D_\mathrm{int}/2\pi$ is the integrated dispersion, and $D_n/2\pi$ ($n>2$) is the higher-order dispersion.
Fig.~\ref{Fig:fab}(b) shows a representative transmission spectrum of a Si$_3$N$_4$ microresonator resonance.
The intrinsic and external linewidths are fitted as $\kappa_0/2\pi = 17.3~\mathrm{MHz}$ and $\kappa_\mathrm{ex}/2\pi = 36.5~\mathrm{MHz}$, respectively.
Fig.~\ref{Fig:fab}(c) shows the statistical distribution of intrinsic linewidths over the measured resonance modes $\omega_\mu$.
Over 380 modes from 1480 nm to 1640 nm are characterized for each of the two Si$_3$N$_4$ microresonators (A and B). 
The maximum-likelihood intrinsic linewidths are $\kappa^\mathrm{A}_0/2\pi = 17~\mathrm{MHz}$, $\kappa^\mathrm{B}_0/2\pi = 17~\mathrm{MHz}$, corresponding to an intrinsic $Q$ of $1.1\times 10^7$.
Fig.~\ref{Fig:fab}(d) shows the integrated dispersions $D_\mathrm{int}$ of microresonators A and B.
The $0\mathrm{th}$-mode resonance frequencies are slightly different, i.e., $\omega_0^\mathrm{A}/2\pi = 193.391$~THz and $\omega_0^\mathrm{B}/2\pi = 193.375$~THz.
The mode resonance can be tuned by changing the chip temperature, which is discussed below in section~\ref{temperature_influence}. The FSRs are $D_1^\mathrm{A}/2\pi = 50.073$~GHz and $D_1^\mathrm{B}/2\pi = 50.072$~GHz, which are close to each other.
The GVDs are $D_2^\mathrm{A}/2\pi = 247.6$~kHz and $D_2^\mathrm{B}/2\pi =239.5$~kHz, corresponding to anomalous dispersion required for soliton generation. 

\subsubsection{Soliton generation}

\begin{figure*}[h]  
  \centering
  \includegraphics[width=0.8\textwidth]{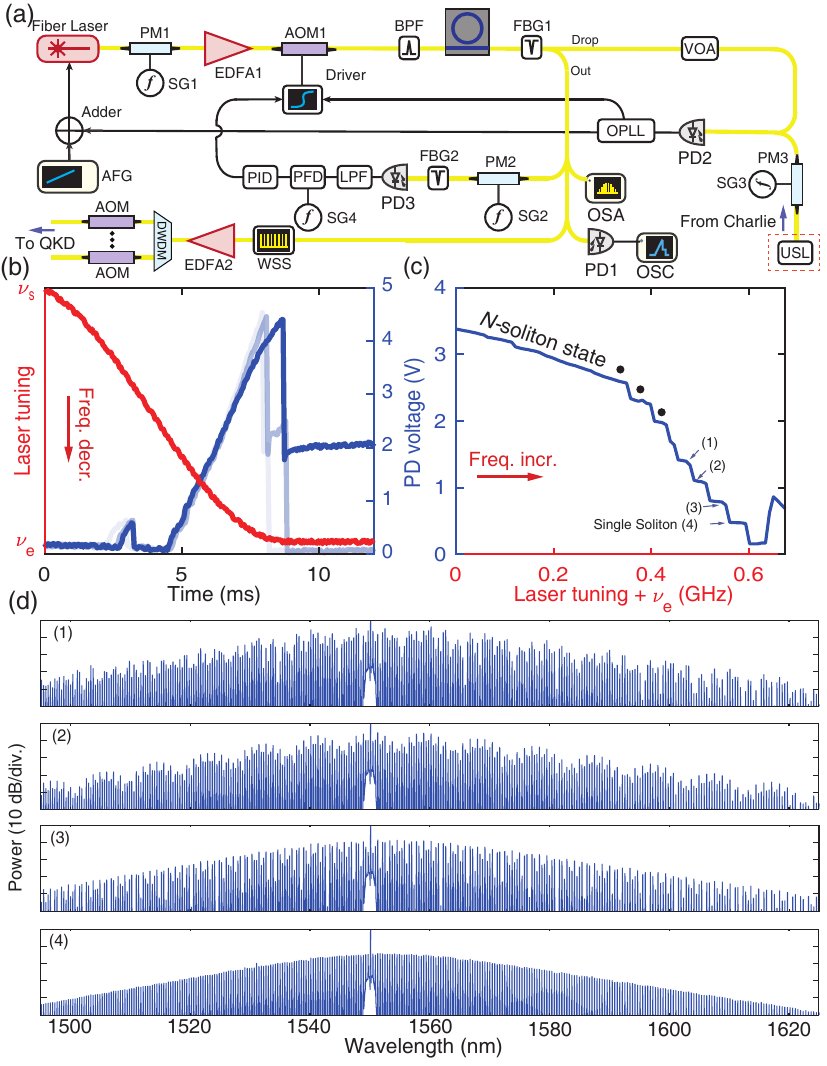}
  \caption{Soliton microcomb generation.
  \textbf{(a)} The experimental setup for soliton generation, pump wavelength locking and microcomb repetition rate locking. PM: phase modulator, SG: signal generator, AOM: acousto-optic modulator, BPF: bandpass filter, FBG: fiber Bragg grating, VOA: variable optical attenuator, OPLL: optical phase-locked loop, PD: photodetector, AFG: arbitrary function generator, PID: proportional-integral-derivative, PFD: phase-frequency detector, LPF: low-pass filter, OSA: optical spectrum analyzer, DWDM: dense wavelength division multiplexing, EDFA: erbium-doped fiber amplifier, WSS: wavelength-selective switch, OSC: oscilloscope, USL: ultra-stable laser.
  \textbf{(b)} Time sequence for exciting the soliton state in the Si$_3$N$_4$ microresonator.
 The red curve represents the laser frequency. Three typical PD voltage curves are shown in blue, where the darkest curve corresponds to soliton excitation. Freq.\ decr., frequency decreasing.
  \textbf{(c)} Single soliton addressing. 
  The single-soliton state can be accessed by increasing the pump frequency after exciting the soliton state.
  Freq.\ incr., frequency increasing.
   \textbf{(d)} The typical soliton state spectra in Fig.~\ref{Fig:soliton_setup}(c).
  }
  \label{Fig:soliton_setup}
\end{figure*}

The experimental setup for soliton generation is shown in Fig.~\ref{Fig:soliton_setup}(a). 
The pump laser comes from a fiber laser. To synchronize the pump wavelength of two DKS microcombs, temperature control was employed. Specifically, by setting the temperatures of the Alice and Bob microresonators to $T\mathrm{_A} = 34.52~^\circ \text{C}$ and $T\mathrm{_B} = 30.32~^\circ \text{C}$, the two resonance frequencies of the pumped modes are aligned to the pump frequency $\nu_\mathrm{pump}^\mathrm{A}$=193.4~THz and $\nu_\mathrm{pump}^\mathrm{B}$=193.4~THz, which can be locked to a remote USL at 1550.1~nm (193.4~THz). 

A phase modulator ($\mathrm{PM1}$) is introduced to create a blue sideband for soliton step extension~\cite{Zheng23}. The modulation frequency of $\mathrm{PM1}$ at Alice is set to 780~MHz, and that at Bob to 720~MHz. The modulated pump laser is amplified by an erbium-doped fiber amplifier (EDFA), the amplified spontaneous emission (ASE) noise is filtered using a bandpass filter (BPF). The soliton state is monitored after the out port of fiber Bragg grating (FBG) by an optical spectrum analyzer (OSA) and an oscilloscope (OSC) after a photodetector (PD). The process of exciting soliton state is shown in  Fig.~\ref{Fig:soliton_setup}(b). 
The pump frequency is scanned from $\nu_\mathrm{s}$ to $\nu_\mathrm{e}$ ($\nu_\mathrm{s}>\nu_\mathrm{e}$), where soliton steps appear on the relative red-detuned side. 

By adjusting the offset voltage of the arbitrary function generator (AFG) that controls the pump laser, we can tune the pump laser frequency. In the initial stage, the AFG offset is decreased, driving the pump frequency from blue-detuned side to red-detuned side (frequency decrease). During this process, the start frequency \(\nu_\mathrm{s}\) and the stop frequency \(\nu_\mathrm{e}\) of the pump laser are moving together to maintain a suitable scanning range and center frequency. The system can excite and enter a stable multiple-soliton state by repeatedly triggering the AFG's burst mode once the resonance is crossed. 
While the soliton state is excited, the soliton number is usually large.

Once the multi-soliton state was generated, we tuned the pump laser from the red-detuned to the blue-detuned side of the resonance by adjusting the offset voltage of the AFG. At this moment, the number of solitons can be gradually reduced by scanning \(\nu_\mathrm{e}\). As shown in Fig.~\ref{Fig:soliton_setup}(c), by monitoring the OSC and OSA, a stepwise decrease in the soliton number can be clearly observed. Fig.~\ref{Fig:soliton_setup}(d) displays the corresponding spectral evolution from the region marked in Fig.~\ref{Fig:soliton_setup}(c). The initial state exhibits a multi-soliton spectrum with multiple sidebands; as the frequency moved to the blue-detuned side, the spectrum progressively smoothens, sideband amplitudes decrease, and eventually a characteristic single-soliton spectrum emerges.

\begin{figure*}[h]
     \centering
     \includegraphics[width=0.7\textwidth]{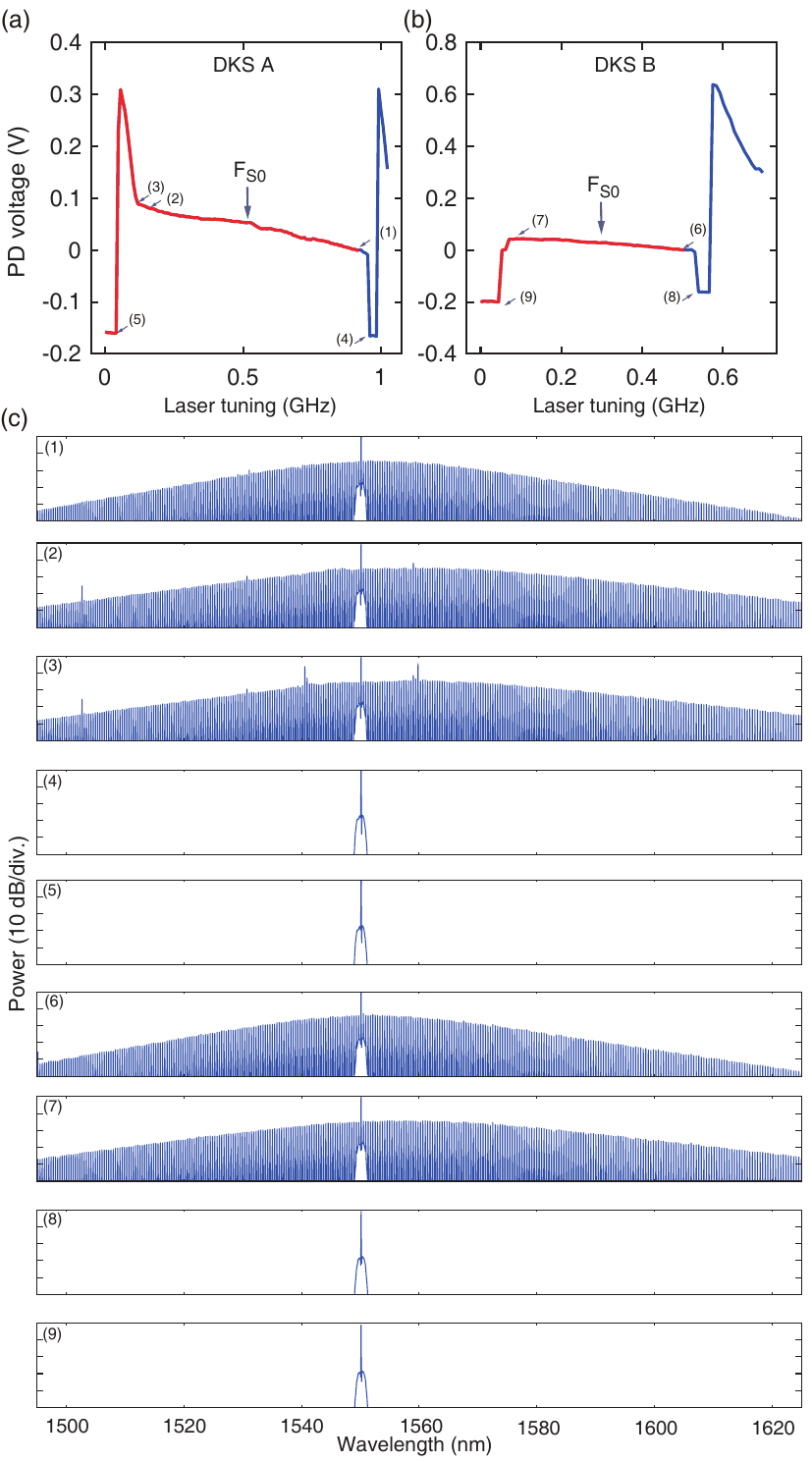}
     \caption{\textbf{(a)} and \textbf{(b)} The frequency tuning range for single-soliton state. The center frequency of the soliton step \({F_\mathrm{S0}}\) was shown. \textbf{(c)} The typical spectra in Fig.~\ref{solitonstep}(a) and (b).}
    \label{solitonstep}
\end{figure*}

After obtaining a stable single-soliton state, we further characterized the frequency tuning range over which this single-soliton state can be maintained. The pump laser frequency was continuously increased or decreased by precisely controlling the offset voltage of the AFG, until the soliton state can no longer be maintained. Throughout this process, we simultaneously monitored the OSC and OSA to determine the existence boundaries of the single-soliton state. The resulting tuning boundaries are shown as the red curve (red-detuning side, corresponding to decreasing frequency) and the blue curve (blue-detuning side, corresponding to increasing frequency) in the Fig.~\ref{solitonstep}(a) and (b).
 
For DKS A, the tuning range toward the red-detuned side is approximately 800 MHz, and the tuning range toward the blue-detuned side is approximately 20 MHz. Therefore, the total tuning range for DKS A is approximately 820 MHz. For DKS B, the tuning range toward the red-detuned side is approximately 440 MHz, and the tuning range toward the blue-detuned side is approximately 40 MHz, corresponding to a total tuning range of approximately 480 MHz.

In addition, the optical spectra corresponding to several characteristic frequency points marked in Fig.~\ref{solitonstep}(a) and (b) are summarized in Fig.~\ref{solitonstep}(c). These spectra further verify the stability and spectral integrity of the single-soliton state at different tuning positions.
\clearpage  

\subsubsection{Influence of pump laser frequency and pump power}

\begin{figure*}[h]
    \centering
    \includegraphics[width=\textwidth]{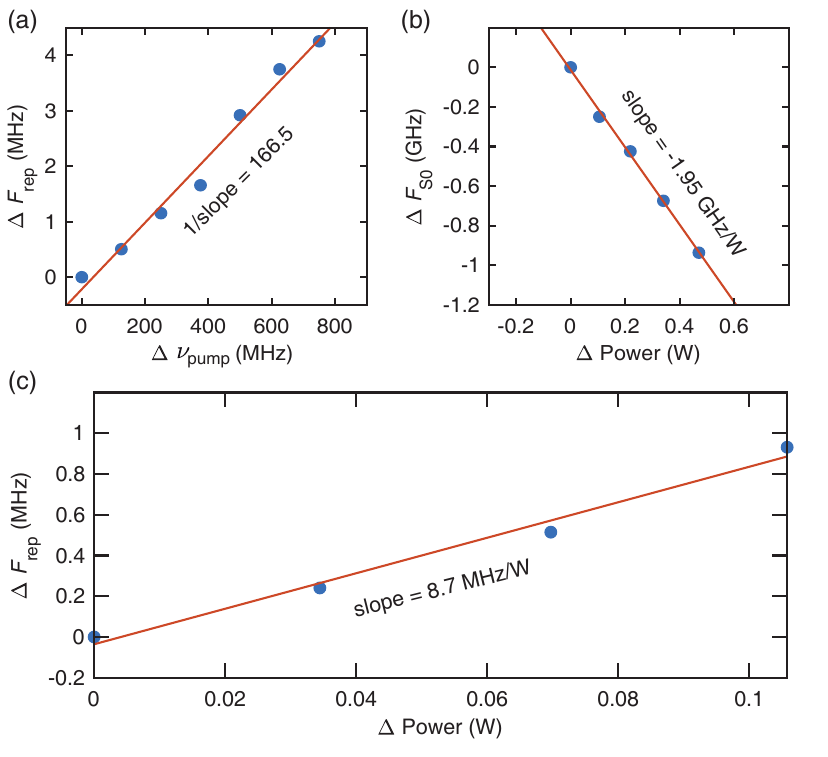}
    \caption{\textbf{(a)} The relationship between pump frequency and the repetition rate of the DKS microcomb in the single-soliton state. \textbf{(b)} The center frequency of the soliton step \({F_\mathrm{S0}}\) changed by the pump power. \textbf{(c)} The repetition rate of the DKS microcomb changed by the pump power.}
    \label{Frep_F0}
\end{figure*} 

Under identical experimental conditions for all other parameters, the dependence of the DKS microcomb repetition rate on the pump frequency $\nu_\mathrm{pump}$ was experimentally characterized, as shown in Fig.~\ref{Frep_F0}(a). The experimental results indicate that within the soliton step range, for every 166~MHz adjustment in the pump frequency, the repetition rate changes correspondingly by 1~MHz, demonstrating a positive correlation. This indicates that upon entering the single-soliton state, the repetition rate can be adjusted by tuning the pump laser frequency.

During soliton excitation, we investigated the effect of input pump power on the center frequency of the soliton step (i.e., \({F_\mathrm{S0}}\) in Fig.~\ref{solitonstep}(a) and Fig.~\ref{solitonstep}(b)), result is shown in Fig.~\ref{Frep_F0}(b). The initial power of the pump is 1.479~W (31.7~dBm). The results indicate a negative correlation between \({F_\mathrm{S0}}\) and the pump power. We performed a linear fit on the results, and the linear fitting coefficient is -1.95~GHz/W.

After entering the single-soliton state we tested the dependence of the repetition frequency on the pump power. The results are shown in Fig.~\ref{Frep_F0}(c). We performed a linear fit on the results and the linear fitting coefficient is 8.7~MHz/W. The test results demonstrate that we can tune the pump power to lock the repetition rate of the DKS microcomb, as discussed in Section~\ref{pump_and_repetition_locking}.

The above results are the test results of DKS A under this experimental condition, and DKS B exhibits similar behavior.

\subsubsection{DKS comb lines locking scheme}
\label{pump_and_repetition_locking}

In this work, we employ two independent DKS microcombs as multi-wavelength light sources for TF-QKD implementations. We select a total of 16 comb lines near the pump light as parallel quantum channels. The optical frequency $\nu_\mathrm{n}$ of the $n$-th comb-line is determined by the pump laser frequency $\nu_\mathrm{pump}$ and the comb’s repetition rate $F_\mathrm{rep}$, with the relationship expressed as:
\[
\nu_n = \nu_\mathrm{pump} + n \cdot F_\mathrm{rep}
\]
where $n$ is the comb-line order relative to the pump (which can be a positive or negative integer). 

The implementation of TF-QKD relies on single-photon interference between two light sources. Using DKS microcombs as the light sources for TF‑QKD requires that each pair of comb lines achieve the performance of two lasers locked to an ultrastable optical reference. However, microcombs initially generated through microresonators suffer from two key limitations: frequency offsets between the pump lasers and mismatches in their repetition rates. To address these challenges, the pump laser frequency and repetition rate of the DKS microcomb must be precisely locked.

\begin{figure*}[h]
    \centering
    \includegraphics[width=\textwidth]{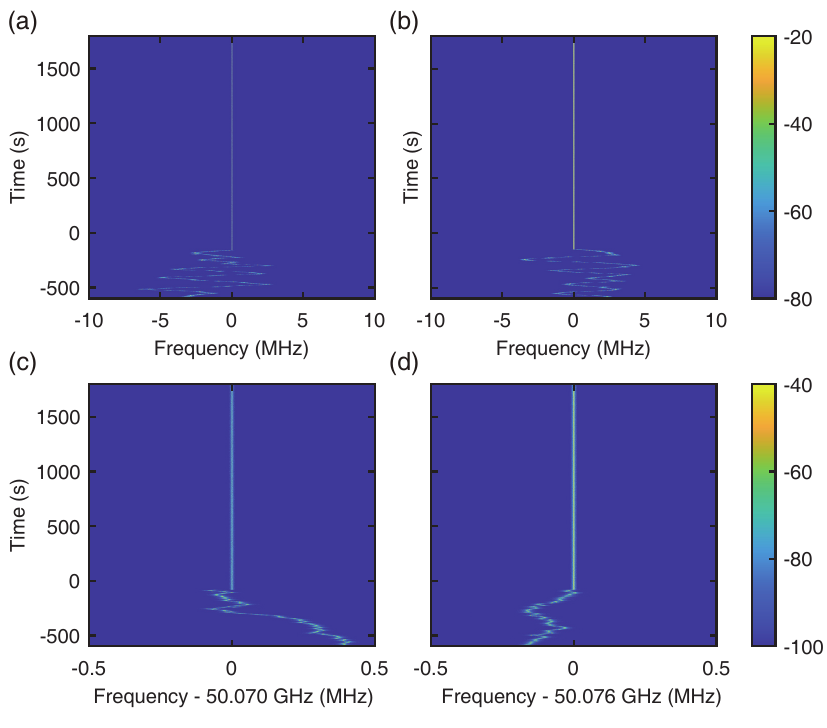}
    \caption{\textbf{(a)} and \textbf{(b)} show the 30 minutes states of pump frequency locking from unlock to lock for DKS A and DKS B. \textbf{(c)} and \textbf{(d)} show the 30 minutes states for repetition rate locking from unlock to lock for DKS A and DKS B.}
    \label{unlock_lock}
\end{figure*}

The experimental setup for pump wavelength locking and DKS microcomb repetition rate locking is shown in Fig.~\ref{Fig:soliton_setup}(a). For pump wavelength locking, the output light from the drop port of FBG1 is attenuated, then interfered with the ultrastable laser from Charlie and coupled into a PD, generating their beat signal. Subsequently, the beat signal is fed into the OPLL feedback control. One of the feedback signal from the locking electronics is used to adjust the driving frequency of the AOM, while the another one feedback signal is used to control the piezoelectric ceramic (PZT) inside the pump laser to compensate for slow drift of the laser frequency. The pump wavelength locking performance of Alice and Bob was monitored using an optical spectrum analyzer, which recorded the spectral evolution over 30 minutes from the unlocked to the locked state, as shown in Fig.~\ref{unlock_lock}(a) and (b). Without locking, the frequency of the pump lasers drifts within $\pm$5~MHz; with locking, it is stabilized to the kHz level. 

After the established pump frequency locking and stable temperature conditions, the repetition rate locking can be established. A portion of the light from the output port of FBG1 is used for the repetition-rate locking of the microcomb. This light first passes through $\mathrm{PM2}$ to generate sidebands, where the $\mathrm{PM2}$ frequency of A is set to $f_{\mathrm{PM2}}^\mathrm{A}=25.040$ GHz and that of B to $f_{\mathrm{PM2}}^\mathrm{B}=25.043$ GHz. Subsequently, $\mathrm{FBG2}$ is used to filter out the sidebands from two adjacent comb lines. The +1st-order sideband and the -1st-order sideband of the adjacent comb lines generate a beat signal with a frequency difference of $f_{\Delta} =2f_\mathrm{PM}-F_\mathrm{rep} = 10 \text{MHz}$. After detection by PD3 and filtering by the low-pass filter (LPF), this signal is fed into a phase-frequency detector (PFD) together with the 10~MHz reference signal generated by the $\mathrm{SG4}$, producing an error signal that serves as the input to the PID controller. The output signal of the PID controller is used to adjust the amplitude of the AOM driving signal, which modifies the pump power to lock the repetition rate. The repetition rate locking results of Alice and Bob's microcombs are monitored using a spectrum analyzer. The results without and with repetition rate locking are shown in Fig.~\ref{unlock_lock}(c) and Fig.~\ref{unlock_lock}(d). 

\subsubsection{Temperature influence on the Si$_3$N$_4$ microcomb}\label{temperature_influence}

Due to the thermo-optic effect\cite{Arbabi:13}, temperature variation induces a shift in the resonance frequency of the microresonator. The relationship between the refractive index and temperature is given by:
\[
n(T) = n_0 + \frac{dn}{dT} \Delta T,
\]
where $n_0$ is the refractive index at temperature $T_0$, and $\frac{dn}{dT}$ is the thermo-optic coefficient, approximately $2.45 \times 10^{-5} \, \text{K}^{-1}$. It is noted that the thermal expansion effect of the Si$_3$N$_4$ microresonator is neglected, as its influence is much smaller than that of the thermo-optic effect.

\begin{figure*}[h]
    \centering
    \includegraphics[width=\textwidth]{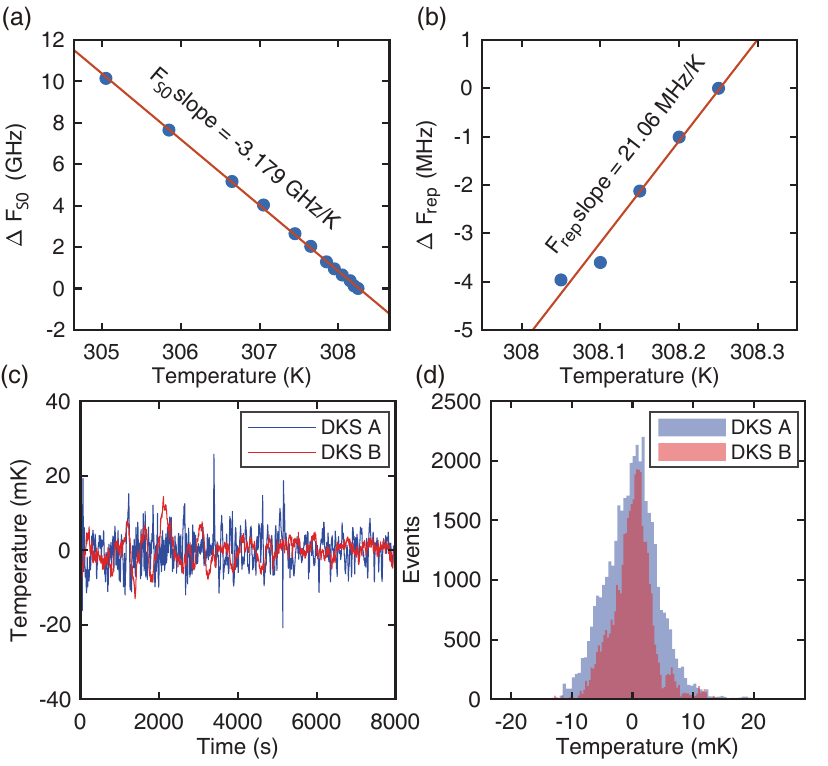}
    \caption{\textbf{(a)} Variation of the soliton step center frequency \({F_\mathrm{S0}}\) with temperature. \textbf{(b)} Variation of the repetition rate $F_\mathrm{rep}$ with temperature. \textbf{(c)} The long-term temperature stability for DKS A and DKS B under temperature control. \textbf{(d)} Temperature distribution of DKS A and DKS B under temperature control.}
    \label{temperature}
\end{figure*}

The temperature tuning characteristics of the resonance frequency were experimentally measured. We investigated the soliton formation by controlling the temperature of the microresonator under different conditions. The variation of the \({F_\mathrm{S0}}\) with temperature is shown in Fig.~\ref{temperature}(a). Through linear fitting, we obtain a tuning coefficient of -3.179~GHz/K. Furthermore, after entering the soliton, we tested the variation of the repetition rate $F_\mathrm{rep}$ of microcomb with temperature under the pump laser frequency, and the result is shown in Fig.~\ref{temperature}(b). A coefficient of 21.06~MHz/K was obtained through liner fitting.

To prevent temperature variations from affecting the duration of the soliton state, we implemented feedback control of the ambient temperature of the microresonator after soliton microcomb generation. Fig.~\ref{temperature}(c) shows the ambient temperature of Alice’s and Bob’s microresonators over time, and Fig.~\ref{temperature}(d) presents the corresponding temperature distributions, with standard deviations of 4.46~mK for DKS A and 3.45~mK for DKS B, respectively.

\subsection{Experimental implementation of the parallel-architecture TF-QKD}

\subsubsection{Experimental Setup for TF-QKD}

\begin{figure*}[h]
  \centering
  \includegraphics[width=1\linewidth]{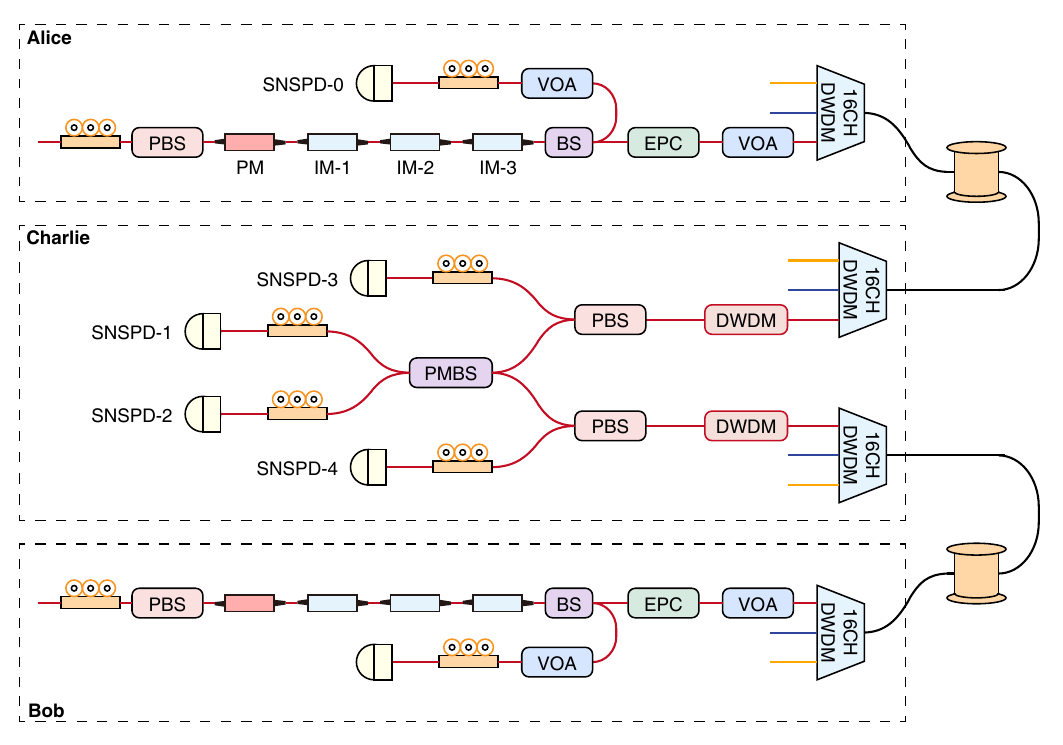}
  \caption{\textbf{The experimental setup.} Alice and Bob employ DKS microcombs as the light source. One comb line of the microcomb is attenuated to the single-photon level after intensity and phase encoding, and then transmitted to Charlie together with 15 continuous-wave comb lines. At Charlie’s side, after 16-channel DWDM demultiplexing and 50~GHz-bandwidth DWDM filtering, the signals interfere on a PMBS and are detected by SNSPDs. PBS: polarization beam splitter, PM: phase modulator, IM: intensity modulator, BS: beam splitter, VOA: variable optical attenuator, EPC: electronic polarization controller, DWDM: dense wavelength division multiplexing, PMBS: polarization maintaining beam splitter, SNSPD: superconducting nanowire single-photon detector.}
  \label{Fig:encoding_setup}
\end{figure*}

The configuration of the system’s encoding and detection modules is depicted in Fig.~\ref{Fig:encoding_setup}. Due to constraints on available signal generators and essential encoding components, encoding and detection are performed sequentially on a single comb line at a time, while the remaining 15 comb lines maintain continuous-wave (CW) transmission. The wavelength of the encoded comb line is $\lambda _{i} \left ( i\in\left \{ 1,2,\dots ,16 \right \}  \right ) $. The PMs and IMs used in the experiment are all polarization-maintaining components; therefore, before modulating the $\lambda _{i}$, it is first passed through a PC and a PBS to stabilize its polarization state, ensuring alignment with the principal axis of the polarization-maintaining fiber (PMF). 

The PM performs phase randomization on quantum signals, while also applying fixed phase modulation to the phase reference light pulses to estimate the phase fluctuation introduced by the fiber link. During each 100~ns period, Alice successively modulates the $\left \{ 0,0,\pi ,\pi  \right \} $ phases, and Bob successively modulates the $\left \{ 0,\frac{\pi }{2} ,\frac{\pi }{2} ,0  \right \} $ phases in the initial 20~ns interval, with each phase value held for 5~ns, leading to a phase difference of $\delta _{AB}=\left \{ 0,-\frac{\pi }{2},\frac{\pi }{2},\pi    \right \}$ between Alice and Bob. The sequence is recorded in data post-processing. During the remaining 80~ns, Alice and Bob apply random phases to the quantum signals. The phase values are randomly chosen from 16 equally spaced discrete phase values $\left \{ 0, \frac{\pi }{16},\dots ,\frac{15\pi }{16}   \right \} $. IM-1 chops the continuous-wave light into optical pulses with a clock rate of 1 GHz and a pulse width of 200~ps. IM-2 performs three-intensity decoy-state modulation. IM-3 conducts intensity modulation on the strong phase reference light and quantum light with a period of 100~ns.

The output light from IM-3 is split into two paths by a 90:10 PMBS. 10\% of the light is attenuated by an attenuator and detected by SNSPD-0. Based on the detection events of SNSPD-0, Alice and Bob generate statistical histograms for the vacuum, decoy, and signal states, and control the bias of IM-2 in real time according to their intensities, such that the intensities of the three states remain stable around the target values. Similarly, the intensity ratio between the phase reference and the quantum signal can be stabilized by controlling the bias of IM-3 based on the 100-ns statistical histogram of the detection events of SNSPD-0. 90\% of the output light from PMBS first passes through an electrically controlled polarization controller (EPC) and is then attenuated to the single-photon level. Subsequently, the $\lambda _{i}$ is combined with the remaining 15 comb lines via a DWDM multiplexer and sent to the detection side through a symmetric low-loss channel. In this step, the intensity of each unmodulated comb line is set to the peak intensity of the $\lambda _{i}$'s strong reference to simulate the worst-case scenario.

At the detection side, the multi-wavelength light from each transmitter is first demultiplexed by a 16-channel DWDM. To reduce crosstalk noise between different wavelengths, each channel is further filtered by a 50 GHz DWDM filter centered at the corresponding wavelength. The $\lambda _{i}$ passes through a PBS, whose reflection port is connected to an SNSPD (SNSPD-3 or SNSPD-4 in Fig.~\ref{Fig:encoding_setup}) for polarization feedback. Based on the count rate of SNSPD-3 (SNSPD-4), we adjust the EPC at the sender in real time to minimize the output from the reflection port of the PBS. The light transmitted through the PBS is sent to a 50:50 PMBS for interference. The interference signals are detected by SNSPDs (SNSPD-1 and SNSPD-2) and recorded by a Time Tagger. 

\subsubsection{Noise Measurement}

The quantum signal light ($\lambda _{i}$) is combined with 15 uncoded comb lines via a DWDM before being transmitted to the detection side. The intensity of the phase reference of $\lambda _{i}$ is -76.3~dBm. Taking into account the pulse width and the duty cycle of the phase reference, we set the intensity of each comb line to -62.3~dBm, in order to evaluate the worst-case scenario. The rosstalk noises may arise from the Raman scattering during transmission and finite isolation between DWDM elements, etc. To test the crosstalk noise, we first switch off the light for each wavelength channel $\lambda _{i}$ while keeping the intensities of the remaining comb lines to -62.3 dBm; then we measure the noise in the selected wavelength channel using the SNSPDs as in the main experiment. The measured noises, subtracting the SNSPD dark counts, are presented in table~\ref{noise_fiber}. The noise contributions are below 32.0 cps and 35.0 cps for the two SNSPD channels for all wavelength channels.

\begin{table*}[h]
\caption{Noise Measurement in Different ITU Channels.}
\centering
\begin{tabular*}{\textwidth}{@{\extracolsep\fill}l|cccccccc}
\toprule
ITU channel & C26 & C27 & C28 & C29 & C30 & C31 & C32 & C33 \\
\hline
Counts of $\text{D}_\text{1}$ (cps) & 30.7 & 28.7 & 21.7 & 18.3 & 13.8 & 12.9 & 24.2 & 14.8\\
Counts of $\text{D}_\text{2}$ (cps) & 31.4 & 22.3 & 17.8 & 16.7 & 12.6 & 9.3 & 20.6 & 16.2\\
\botrule
\end{tabular*}
\label{noise}
\vspace{0.3cm}

\begin{tabular*}{\textwidth}{@{\extracolsep\fill}l|cccccccc}
\toprule
ITU channel & C34 & C35 & C36 & C37 & C38 & C39 & C40 & C41 \\
\hline
Counts of $\text{D}_\text{1}$ (cps) & 17.7 & 22.0 & 30.8 & 31.8 & 16.9 & 22.3 & 29.7 & 30.4\\
Counts of $\text{D}_\text{2}$ (cps) & 18.2 & 24.4 & 31.6 & 30.7 & 17.5 & 19.3 & 27.5 & 34.5\\
\botrule
\end{tabular*}
\label{noise_fiber}
\end{table*}

\subsubsection{Detailed Experimental Parameters and Results}

At the measurement side, the light from Alice and Bob first passes through a $1\times16$ channel DWDM for demultiplexing. We then further filter each channel using a 50~GHz-bandwidth DWDM filter. The losses for each channel of the 1×16 DWDM and the losses for each corresponding 50~GHz DWDM are listed in the table.~\ref{Loss_of_DWDM}.

\begin{table*}[h]
\caption{The loss of 16-channel DWDM and 50~GHz DWDM at detection side.}
\label{Loss_of_DWDM}
\centering
\begin{tabular*}{\textwidth}{@{\extracolsep\fill}l|cccccccc}
\toprule
ITU channel & C26 & C27 & C28 & C29 & C30 & C31 & C32 & C33 \\
\hline
16ch-DWDM-A (dB) & 1.38 & 1.40 & 1.42 & 1.44 & 1.40 & 1.31 & 1.43 & 1.44 \\
16ch-DWDM-B (dB) & 1.36 & 1.31 & 1.40 & 1.44 & 1.35 & 1.44 & 1.32 & 1.37 \\
\hline
DWDM-A (dB) & 0.29 & 0.28 & 0.23 & 0.28 & 0.27 & 0.22 & 0.21 & 0.25 \\
DWDM-B (dB) & 0.28 & 0.27 & 0.25 & 0.21 & 0.21 & 0.23 & 0.28 & 0.24 \\
\botrule
\end{tabular*}

\vspace{0.3cm}

\begin{tabular*}{\textwidth}{@{\extracolsep\fill}l|cccccccc}
\toprule
ITU channel & C34 & C35 & C36 & C37 & C38 & C39 & C40 & C41 \\
\hline
16ch-DWDM-A (dB) & 1.40 & 1.41 & 1.35 & 1.36 & 1.40 & 1.33 & 1.31 & 1.30 \\
16ch-DWDM-B (dB) & 1.36 & 1.41 & 1.43 & 1.30 & 1.42 & 1.37 & 1.36 & 1.40 \\
\hline
DWDM-A (dB) & 0.30 & 0.23 & 0.26 & 0.28 & 0.22 & 0.23 & 0.28 & 0.27 \\
DWDM-B (dB) & 0.28 & 0.22 & 0.29 & 0.23 & 0.22 & 0.26 & 0.25 & 0.25 \\
\botrule
\end{tabular*}
\end{table*}

The experimental results are summarized in~\cref{result:C26toC29,result:C30toC33,result:C34toC37,result:C38toC41,result:independent_lasers_and_total_lines}. In the table, we denote $\text{N}_{\text{total}}$ as the total number of signal pulses, $n_{t}$ (After AOPP) as the remaining pairs after active odd parity pairing (AOPP), $n_{1}$ (Before AOPP)/$n_{1}$ (After AOPP) as the number of the untagged bits before/after AOPP, $e_{1}^{ph}$ (Before AOPP)/$e_{1}^{ph}$ (After AOPP) as the phase-flip error rate before/after AOPP, and QBER E (Before AOPP)/E (After AOPP) as the bit-flip error rate before/after the bit error rejection by AOPP. With all the parameters in the table, the final SKR per pulse and in one second is calculated as R (per pulse) and R (bps).

In calculation, the chosen phase difference is selected as $\text{D}_{\text{s}}$ (in degrees). $E_{X}$ represents the error rates when Alice and Bob send decoy states $\mathrm{\mu}_{\text{x}}$ with a phase difference range of $\text{D}_{\text{s}}$. In the following rows, we list the numbers of pulses Alice and Bob sent in different decoy states, labelled as “Sent-AB”, where “A” (“B”) is “0”, “1”, or “2”, indicating the intensity Alice (Bob) has chosen within “vacuum”, “$\mathrm{\mu}_{\text{x}}$”, or “$\mathrm{\mu}_{\text{y}}$”. With the same rule, the numbers of detections are listed as “Detected-AB”. The total detections reported by Charlie is denoted as “Detected-ch”, where “ch” can be “Det1” or “Det2” indicating the responsive detector of the recorded counts. The events falls in $\text{D}_{\text{s}}$ angle range is denoted as “Detected-11-$\text{D}_{\text{s}}$”, the numbers of correct detections in this $\text{D}_{\text{s}}$ range is denoted as “Correct-11-$\text{D}_{\text{s}}$”.

\begin{table*}[ht]
\caption{Experimental results for ITU Channel from C26 to C29.}
\label{result:C26toC29}
\centering
\begin{tabular*}{\textwidth}{@{\extracolsep\fill}l|cccc} 
    \toprule
    ITU Channel & C26 & C27 & C28 & C29\\
    Central Wavelength (nm) & 1553.33 & 1552.52 & 1551.72 & 1550.92\\
    \hline
    $\text{N}_{\text{total}}$&1440000000000&1440000000000&1440000000000&1440000000000\\
    R (per pulse)&$1.298\times10^{-4}$&$1.301\times10^{-4}$&$1.104\times10^{-4}$&$1.165\times10^{-4}$\\
    R (bps)&103855.2&104073.6&88344.8&93212\\
    \hline
    $n_{1}$ (Before AOPP)&$1.87558\times10^{9}$&$1.90687\times10^{9}$&$1.90399\times10^{9}$&$1.8658\times10^{9}$\\
    $n_{1}$ (After AOPP)&$3.16774\times10^{8}$&$3.23227\times10^{8}$&$3.20151\times10^{8}$&$3.13113\times10^{8}$\\
    $e_{1}^{ph}$ (Before AOPP)&3.96\%&4.12\%&5.49\%&4.81\%\\
    $e_{1}^{ph}$ (After AOPP)&7.64\%&7.94\%&10.42\%&9.20\%\\
    E (Before AOPP)&26.73\%&26.74\%&26.74\%&26.74\%\\
    E (After AOPP)&$4.67\times10^{-4}$&$4.91\times10^{-4}$&$5.01\times10^{-4}$&$4.99\times10^{-4}$\\
    $n_{t}$ (After AOPP)&901469305&931307516&937020515&919209099\\
    $E_{X}$&3.4\%&3.5\%&4.6\%&4.0\%\\
    $\text{D}_{\text{s}}$&10&10&10&10\\
    \hline
    Sent-00&699840000000&699840000000&699840000000&699840000000\\
    Sent-01&30780000000&30780000000&30780000000&30780000000\\
    Sent-10&31680000000&31680000000&31680000000&31680000000\\
    Sent-02&277380000000&277380000000&277380000000&277380000000\\
    Sent-20&276480000000&276480000000&276480000000&276480000000\\
    Sent-12&10440000000&10440000000&10440000000&10440000000\\
    Sent-21&11340000000&11340000000&11340000000&11340000000\\
    Sent-11&1080000000&1080000000&1080000000&1080000000\\
    Sent-22&100980000000&100980000000&100980000000&100980000000\\
    \hline
    Detected-Det1&2226708049&2279600389&2292323777&2243123929\\
    Detected-Det2&2262529812&2314867560&2329350952&2292610210\\
    \hline
    Detected-00&1011643&1088529&1119088&1092618\\
    Detected-01&18686158&18704623&18873078&18547478\\
    Detected-10&18271024&18922504&18740971&18319431\\
    Detected-02&1619988536&1631838421&1641560492&1611248807\\
    Detected-20&1541970766&1603874903&1613662540&1583156572\\
    Detected-12&64955664&65483789&65816412&64602810\\
    Detected-21&70700488&73253841&73718022&72205070\\
    Detected-11&1400426&1421828&1428217&1405811\\
    Detected-22&1152253156&1179879511&1186755909&1165155542\\
    \hline
    Detected-11-$\text{D}_{\text{s}}$&164673&167615&173620&169313\\
    Correct-11-$\text{D}_{\text{s}}$&159087&161709&165704&162547\\
    \botrule
\end{tabular*}
\end{table*}

\begin{table*}[ht]
\caption{Experimental results for ITU Channel from C30 to C33.}
\label{result:C30toC33}
\centering
\begin{tabular*}{\textwidth}{@{\extracolsep\fill}l|cccc} % 居中且填充
    \toprule
    ITU Channel & C30 & C31 & C32 & C33\\
    Central Wavelength (nm) & 1553.33 & 1552.52 & 1551.72 & 1550.92\\
    \hline
    $\text{N}_{\text{total}}$&1440000000000&1440000000000&1440000000000&1440000000000\\
    R (per pulse)&$1.157\times10^{-4}$&$1.099\times10^{-4}$&$1.237\times10^{-4}$&$1.174\times10^{-4}$\\
    R (bps)&92581.6&87909.6&98982.4&93918.4\\
    \hline
    $n_{1}$ (Before AOPP)&$1.82159\times10^{9}$&$1.84847\times10^{9}$&$1.86769\times10^{9}$&$1.83442\times10^{9}$\\
    $n_{1}$ (After AOPP)&$2.99331\times10^{8}$&$3.08974\times10^{8}$&$3.14961\times10^{8}$&$3.10394\times10^{8}$\\
    $e_{1}^{ph}$ (Before AOPP)&4.44\%&5.21\%&4.19\%&4.48\%\\
    $e_{1}^{ph}$ (After AOPP)&8.53\%&9.92\%&8.06\%&8.58\%\\
    E (Before AOPP)&26.74\%&26.74\%&26.76\%&26.75\%\\
    E (After AOPP)&$5.18\times10^{-4}$&$5.05\times10^{-4}$&$7.43\times10^{-4}$&$8.37\times10^{-4}$\\
    $n_{t}$ (After AOPP)&913170246&911051164&924199198&901764638\\
    $E_{X}$&3.7\%&4.3\%&3.5\%&3.8\%\\
    $\text{D}_{\text{s}}$&10&10&10&10\\
    \hline
    Sent-00&699840000000&699840000000&699840000000&699840000000\\
    Sent-01&30780000000&30780000000&30780000000&30780000000\\
    Sent-10&31680000000&31680000000&31680000000&31680000000\\
    Sent-02&277380000000&277380000000&277380000000&277380000000\\
    Sent-20&276480000000&276480000000&276480000000&276480000000\\
    Sent-12&10440000000&10440000000&10440000000&10440000000\\
    Sent-21&11340000000&11340000000&11340000000&11340000000\\
    Sent-11&1080000000&1080000000&1080000000&1080000000\\
    Sent-22&100980000000&100980000000&100980000000&100980000000\\
    \hline
    Detected-Det1&2219614642&2228483535&2252719369&2173404631\\
    Detected-Det2&2290726212&2273569080&2288174099&2264004655\\
    \hline
    Detected-00&1128384&1099204&1626327&1795725\\
    Detected-01&18395097&18488314&18481459&18103732\\
    Detected-10&17707241&18046215&18469807&18178814\\
    Detected-02&1605670251&1603698087&1600302016&1569845726\\
    Detected-20&1571338842&1567178082&1596795867&1554919521\\
    Detected-12&64160487&64249979&64207651&63120894\\
    Detected-21&71857853&71584062&72998690&70877766\\
    Detected-11&1364230&1384143&1391036&1379875\\
    Detected-22&1158718469&1156324529&1166620615&1139187233\\
    \hline
    Detected-11-$\text{D}_{\text{s}}$&163623&167686&167093&163966\\
    Correct-11-$\text{D}_{\text{s}}$&157543&160415&161169&157730\\
    \botrule
\end{tabular*}
\end{table*}

\begin{table*}[ht]
\caption{Experimental results for ITU Channel from C34 to C37.}
\label{result:C34toC37}
\centering
\begin{tabular*}{\textwidth}{@{\extracolsep\fill}l|cccc} % 居中且填充
    \toprule
    ITU Channel & C34 & C35 & C36 & C37\\
    Central Wavelength (nm) & 1550.12 & 1549.32 & 1548.52 & 1547.72\\
    \hline
    $\text{N}_{\text{total}}$&1440000000000&1440000000000&1440000000000&1440000000000\\
    R (per pulse)&$1.250\times10^{-4}$&$1.208\times10^{-4}$&$1.217\times10^{-4}$&$1.060\times10^{-4}$\\
    R (bps)&100024&96620&97328&84816.8\\
    \hline
    $n_{1}$ (Before AOPP)&$1.84487\times10^{9}$&$1.86088\times10^{9}$&$1.85333\times10^{9}$&$1.79774\times10^{9}$\\
    $n_{1}$ (After AOPP)&$3.05863\times10^{8}$&$3.14576\times10^{8}$&$3.12161\times10^{8}$&$3.01162\times10^{8}$\\
    $e_{1}^{ph}$ (Before AOPP)&3.82\%&4.40\%&4.41\%&5.30\%\\
    $e_{1}^{ph}$ (After AOPP)&7.38\%&8.45\%&8.47\%&10.08\%\\
    E (Before AOPP)&26.75\%&26.75\%&26.74\%&26.76\%\\
    E (After AOPP)&$7.63\times10^{-4}$&$7.26\times10^{-4}$&$4.83\times10^{-4}$&$5.19\times10^{-4}$\\
    $n_{t}$ (After AOPP)&922567802&925459687&911092013&876327093\\
    $E_{X}$&3.2\%&3.8\%&3.8\%&4.5\%\\
    $\text{D}_{\text{s}}$&10&10&10&10\\
    \hline
    Sent-00&699840000000&699840000000&699840000000&699840000000\\
    Sent-01&30780000000&30780000000&30780000000&30780000000\\
    Sent-10&31680000000&31680000000&31680000000&31680000000\\
    Sent-02&277380000000&277380000000&277380000000&277380000000\\
    Sent-20&276480000000&276480000000&276480000000&276480000000\\
    Sent-12&10440000000&10440000000&10440000000&10440000000\\
    Sent-21&11340000000&11340000000&11340000000&11340000000\\
    Sent-11&1080000000&1080000000&1080000000&1080000000\\
    Sent-22&100980000000&100980000000&100980000000&100980000000\\
    \hline
    Detected-Det1&2249567717&2244287390&2232817379&2164392019\\
    Detected-Det2&2293755376&2286520030&2261670137&2189386542\\
    \hline
    Detected-00&1672745&1584278&1050778&1088210\\
    Detected-01&18531630&18279064&18150905&17879527\\
    Detected-10&18046079&18547989&18460332&17622587\\
    Detected-02&1609967619&1587695750&1596438738&1560694549\\
    Detected-20&1589934453&1602725869&1569035476&1504721260\\
    Detected-12&64440148&63794357&64174604&62379728\\
    Detected-21&72608355&72971877&71614679&68944477\\
    Detected-11&1385460&1396058&1393524&1349413\\
    Detected-22&1166736604&1163812178&1154168480&1119098810\\
    \hline
    Detected-11-$\text{D}_{\text{s}}$&165300&164527&162732&158962\\
    Correct-11-$\text{D}_{\text{s}}$&159980&158325&156588&151772\\
    \botrule
\end{tabular*}
\end{table*}

\begin{table*}[ht]
\caption{Experimental results for ITU Channel from C38 to C41.}
\label{result:C38toC41}
\centering
\begin{tabular*}{\textwidth}{@{\extracolsep\fill}l|cccc} % 居中且填充
    \toprule
    ITU Channel & C38 & C39 & C40 & C41\\
    Central Wavelength (nm) & 1546.92 & 1546.12 & 1545.32 & 1544.53\\
    \hline
    $\text{N}_{\text{total}}$&1440000000000&1440000000000&1440000000000&1440000000000\\
    R (per pulse)&$1.289\times10^{-4}$&$1.232\times10^{-4}$&$1.200\times10^{-4}$&$1.084\times10^{-4}$\\
    R (bps)&103139.2&98595.2&96009.6&86710.4\\
    \hline
    $n_{1}$ (Before AOPP)&$1.93202\times10^{9}$&$1.94199\times10^{9}$&$1.88978\times10^{9}$&$1.83526\times10^{9}$\\
    $n_{1}$ (After AOPP)&$3.31223\times10^{8}$&$3.31971\times10^{8}$&$3.14204\times10^{8}$&$3.04142\times10^{8}$\\
    $e_{1}^{ph}$ (Before AOPP)&4.40\%&4.85\%&4.49\%&5.18\%\\
    $e_{1}^{ph}$ (After AOPP)&8.44\%&9.26\%&8.63\%&9.85\%\\
    E (Before AOPP)&26.74\%&26.74\%&26.73\%&26.75\%\\
    E (After AOPP)&$5.30\times10^{-4}$&$4.99\times10^{-4}$&$6.42\times10^{-4}$&$5.21\times10^{-4}$\\
    $n_{t}$ (After AOPP)&960631975&934786308&911215291&915119000\\
    $E_{X}$&3.7\%&4.1\%&3.8\%&4.4\%\\
    $\text{D}_{\text{s}}$&10&10&10&10\\
    \hline
    Sent-00&699840000000&699840000000&699840000000&699840000000\\
    Sent-01&30780000000&30780000000&30780000000&30780000000\\
    Sent-10&31680000000&31680000000&31680000000&31680000000\\
    Sent-02&277380000000&277380000000&277380000000&277380000000\\
    Sent-20&276480000000&276480000000&276480000000&276480000000\\
    Sent-12&10440000000&10440000000&10440000000&10440000000\\
    Sent-21&11340000000&11340000000&11340000000&11340000000\\
    Sent-11&1080000000&1080000000&1080000000&1080000000\\
    Sent-22&100980000000&100980000000&100980000000&100980000000\\
    \hline
    Detected-Det1&2311205680&2301077853&2241722793&2246216651\\
    Detected-Det2&2360971579&2325065913&2320891007&2271159804\\
    \hline
    Detected-00&1195477&1113939&1415560&1133358\\
    Detected-01&18653809&19041938&19154367&18199552\\
    Detected-10&19509166&19206652&18160247&18146377\\
    Detected-02&1618886785&1651450195&1661775095&1606000284\\
    Detected-20&1671582307&1606350043&1552225398&1575729215\\
    Detected-12&65243330&66365049&66463275&64369935\\
    Detected-21&76035611&73358114&71187131&71827714\\
    Detected-11&1448689&1451958&1416408&1381366\\
    Detected-22&1199622085&1187805878&1170816319&1160588654\\
    \hline
    Detected-11-$\text{D}_{\text{s}}$&171261&173767&168929&164589\\
    Correct-11-$\text{D}_{\text{s}}$&164854&166656&162514&157414\\
    \botrule
\end{tabular*}
\end{table*}

\begin{table*}[ht]
\caption{Experimental results for independent lasers and 16 comb lines.}
\label{result:independent_lasers_and_total_lines}
\centering
\begin{tabular*}{\textwidth}{@{\extracolsep\fill}l|cc} % 居中且填充
    \toprule
    Condition & Independent lasers & 16 comb lines\\
    \hline
    $\text{N}_{\text{total}}$&1440000000000&23040000000000\\
    R (per pulse)&$1.275\times10^{-4}$&$1.228\times10^{-4}$\\
    R (bps)&101965.6&1572416\\
    \hline
    $n_{1}$ (Before AOPP)&$1.87436\times10^{9}$&$2.99275\times10^{10}$\\
    $n_{1}$ (After AOPP)&$3.1103\times10^{8}$&$5.03781\times10^{9}$\\
    $e_{1}^{ph}$ (Before AOPP)&3.78\%&4.35\%\\
    $e_{1}^{ph}$ (After AOPP)&7.32\%&8.34\%\\
    E (Before AOPP)&26.74\%&26.74\%\\
    E (After AOPP)&$7.43\times10^{-4}$&$5.77\times10^{-4}$\\
    $n_{t}$ (After AOPP)&948369513&14696390850\\
    $E_{X}$&3.2\%&3.9\%\\
    $\text{D}_{\text{s}}$&10&10\\
    \hline
    Sent-00&699840000000&11197440000000\\
    Sent-01&30780000000&492480000000\\
    Sent-10&31680000000&506880000000\\
    Sent-02&277380000000&4438080000000\\
    Sent-20&276480000000&4423680000000\\
    Sent-12&10440000000&167040000000\\
    Sent-21&11340000000&181440000000\\
    Sent-11&1080000000&17280000000\\
    Sent-22&100980000000&1615680000000\\
    \hline
    Detected-Det1&2314958552&35907265803\\
    Detected-Det2&2326790776&36625252968\\
    \hline
    Detected-00&1670039&20215863\\
    Detected-01&18721390&296170731\\
    Detected-10&18476849&294355436\\
    Detected-02&1628471804&25777061351\\
    Detected-20&1641226927&25305201114\\
    Detected-12&65286372&1033828112\\
    Detected-21&74708616&1155743750\\
    Detected-11&1414260&22398442\\
    Detected-22&1191773071&18627543972\\
    \hline
    Detected-11-$\text{D}_{\text{s}}$&167341&2671586\\
    Correct-11-$\text{D}_{\text{s}}$&161955&2566799\\
    \botrule
\end{tabular*}
\end{table*}

\clearpage
\bibliography{ref}

\begin{thebibliography}{71}%
\makeatletter
\providecommand \@ifxundefined [1]{%
 \@ifx{#1\undefined}
}%
\providecommand \@ifnum [1]{%
 \ifnum #1\expandafter \@firstoftwo
 \else \expandafter \@secondoftwo
 \fi
}%
\providecommand \@ifx [1]{%
 \ifx #1\expandafter \@firstoftwo
 \else \expandafter \@secondoftwo
 \fi
}%
\providecommand \natexlab [1]{#1}%
\providecommand \enquote  [1]{``#1''}%
\providecommand \bibnamefont  [1]{#1}%
\providecommand \bibfnamefont [1]{#1}%
\providecommand \citenamefont [1]{#1}%
\providecommand \href@noop [0]{\@secondoftwo}%
\providecommand \href [0]{\begingroup \@sanitize@url \@href}%
\providecommand \@href[1]{\@@startlink{#1}\@@href}%
\providecommand \@@href[1]{\endgroup#1\@@endlink}%
\providecommand \@sanitize@url [0]{\catcode `\\12\catcode `\$12\catcode
  `\&12\catcode `\#12\catcode `\^12\catcode `\_12\catcode `\%12\relax}%
\providecommand \@@startlink[1]{}%
\providecommand \@@endlink[0]{}%
\providecommand \url  [0]{\begingroup\@sanitize@url \@url }%
\providecommand \@url [1]{\endgroup\@href {#1}{\urlprefix }}%
\providecommand \urlprefix  [0]{URL }%
\providecommand \Eprint [0]{\href }%
\providecommand \doibase [0]{https://doi.org/}%
\providecommand \selectlanguage [0]{\@gobble}%
\providecommand \bibinfo  [0]{\@secondoftwo}%
\providecommand \bibfield  [0]{\@secondoftwo}%
\providecommand \translation [1]{[#1]}%
\providecommand \BibitemOpen [0]{}%
\providecommand \bibitemStop [0]{}%
\providecommand \bibitemNoStop [0]{.\EOS\space}%
\providecommand \EOS [0]{\spacefactor3000\relax}%
\providecommand \BibitemShut  [1]{\csname bibitem#1\endcsname}%
\let\auto@bib@innerbib\@empty
%</preamble>
\bibitem [{\citenamefont {Bennett}\ and\ \citenamefont
  {Brassard}(1984)}]{inproceedings}%
  \BibitemOpen
  \bibfield  {author} {\bibinfo {author} {\bibfnamefont {C.}~\bibnamefont
  {Bennett}}\ and\ \bibinfo {author} {\bibfnamefont {G.}~\bibnamefont
  {Brassard}},\ }\bibfield  {title} {\bibinfo {title} {Withdrawn: Quantum
  cryptography: Public key distribution and coin tossing}\ }(\bibinfo {year}
  {1984})\ pp.\ \bibinfo {pages} {175--179}\BibitemShut {NoStop}%
\bibitem [{\citenamefont {Xu}\ \emph {et~al.}(2020)\citenamefont {Xu},
  \citenamefont {Ma}, \citenamefont {Zhang}, \citenamefont {Lo},\ and\
  \citenamefont {Pan}}]{RevModPhys.92.025002}%
  \BibitemOpen
  \bibfield  {author} {\bibinfo {author} {\bibfnamefont {F.}~\bibnamefont
  {Xu}}, \bibinfo {author} {\bibfnamefont {X.}~\bibnamefont {Ma}}, \bibinfo
  {author} {\bibfnamefont {Q.}~\bibnamefont {Zhang}}, \bibinfo {author}
  {\bibfnamefont {H.-K.}\ \bibnamefont {Lo}},\ and\ \bibinfo {author}
  {\bibfnamefont {J.-W.}\ \bibnamefont {Pan}},\ }\bibfield  {title} {\bibinfo
  {title} {Secure quantum key distribution with realistic devices},\ }\href
  {https://doi.org/10.1103/RevModPhys.92.025002} {\bibfield  {journal}
  {\bibinfo  {journal} {Rev. Mod. Phys.}\ }\textbf {\bibinfo {volume} {92}},\
  \bibinfo {pages} {025002} (\bibinfo {year} {2020})}\BibitemShut {NoStop}%
\bibitem [{\citenamefont {Li}\ \emph {et~al.}(2023)\citenamefont {Li},
  \citenamefont {Zhang}, \citenamefont {Tan}, \citenamefont {Lu}, \citenamefont
  {Liao}, \citenamefont {Huang}, \citenamefont {Li}, \citenamefont {Wang},
  \citenamefont {Mao}, \citenamefont {Yan}, \citenamefont {Li}, \citenamefont
  {Liu}, \citenamefont {Zhang}, \citenamefont {Peng}, \citenamefont {You},
  \citenamefont {Xu},\ and\ \citenamefont {Pan}}]{li2023}%
  \BibitemOpen
  \bibfield  {author} {\bibinfo {author} {\bibfnamefont {W.}~\bibnamefont
  {Li}}, \bibinfo {author} {\bibfnamefont {L.}~\bibnamefont {Zhang}}, \bibinfo
  {author} {\bibfnamefont {H.}~\bibnamefont {Tan}}, \bibinfo {author}
  {\bibfnamefont {Y.}~\bibnamefont {Lu}}, \bibinfo {author} {\bibfnamefont
  {S.-K.}\ \bibnamefont {Liao}}, \bibinfo {author} {\bibfnamefont
  {J.}~\bibnamefont {Huang}}, \bibinfo {author} {\bibfnamefont
  {H.}~\bibnamefont {Li}}, \bibinfo {author} {\bibfnamefont {Z.}~\bibnamefont
  {Wang}}, \bibinfo {author} {\bibfnamefont {H.-K.}\ \bibnamefont {Mao}},
  \bibinfo {author} {\bibfnamefont {B.}~\bibnamefont {Yan}}, \bibinfo {author}
  {\bibfnamefont {Q.}~\bibnamefont {Li}}, \bibinfo {author} {\bibfnamefont
  {Y.}~\bibnamefont {Liu}}, \bibinfo {author} {\bibfnamefont {Q.}~\bibnamefont
  {Zhang}}, \bibinfo {author} {\bibfnamefont {C.-Z.}\ \bibnamefont {Peng}},
  \bibinfo {author} {\bibfnamefont {L.}~\bibnamefont {You}}, \bibinfo {author}
  {\bibfnamefont {F.}~\bibnamefont {Xu}},\ and\ \bibinfo {author}
  {\bibfnamefont {J.-W.}\ \bibnamefont {Pan}},\ }\bibfield  {title} {\bibinfo
  {title} {High-rate quantum key distribution exceeding
  110{\thinspace}mb{\thinspace}s--1},\ }\href
  {https://doi.org/10.1038/s41566-023-01166-4} {\bibfield  {journal} {\bibinfo
  {journal} {Nature Photonics}\ }\textbf {\bibinfo {volume} {17}},\ \bibinfo
  {pages} {416} (\bibinfo {year} {2023})}\BibitemShut {NoStop}%
\bibitem [{\citenamefont {Gr{\"u}nenfelder}\ \emph {et~al.}(2023)\citenamefont
  {Gr{\"u}nenfelder}, \citenamefont {Boaron}, \citenamefont {Resta},
  \citenamefont {Perrenoud}, \citenamefont {Rusca}, \citenamefont {Barreiro},
  \citenamefont {Houlmann}, \citenamefont {Sax}, \citenamefont {Stasi},
  \citenamefont {El-Khoury}, \citenamefont {H{\"a}nggi}, \citenamefont
  {Bosshard}, \citenamefont {Bussi{\`e}res},\ and\ \citenamefont
  {Zbinden}}]{Grünenfelder2023}%
  \BibitemOpen
  \bibfield  {author} {\bibinfo {author} {\bibfnamefont {F.}~\bibnamefont
  {Gr{\"u}nenfelder}}, \bibinfo {author} {\bibfnamefont {A.}~\bibnamefont
  {Boaron}}, \bibinfo {author} {\bibfnamefont {G.~V.}\ \bibnamefont {Resta}},
  \bibinfo {author} {\bibfnamefont {M.}~\bibnamefont {Perrenoud}}, \bibinfo
  {author} {\bibfnamefont {D.}~\bibnamefont {Rusca}}, \bibinfo {author}
  {\bibfnamefont {C.}~\bibnamefont {Barreiro}}, \bibinfo {author}
  {\bibfnamefont {R.}~\bibnamefont {Houlmann}}, \bibinfo {author}
  {\bibfnamefont {R.}~\bibnamefont {Sax}}, \bibinfo {author} {\bibfnamefont
  {L.}~\bibnamefont {Stasi}}, \bibinfo {author} {\bibfnamefont
  {S.}~\bibnamefont {El-Khoury}}, \bibinfo {author} {\bibfnamefont
  {E.}~\bibnamefont {H{\"a}nggi}}, \bibinfo {author} {\bibfnamefont
  {N.}~\bibnamefont {Bosshard}}, \bibinfo {author} {\bibfnamefont
  {F.}~\bibnamefont {Bussi{\`e}res}},\ and\ \bibinfo {author} {\bibfnamefont
  {H.}~\bibnamefont {Zbinden}},\ }\bibfield  {title} {\bibinfo {title} {Fast
  single-photon detectors and real-time key distillation enable high
  secret-key-rate quantum key distribution systems},\ }\href
  {https://doi.org/10.1038/s41566-023-01168-2} {\bibfield  {journal} {\bibinfo
  {journal} {Nature Photonics}\ }\textbf {\bibinfo {volume} {17}},\ \bibinfo
  {pages} {422} (\bibinfo {year} {2023})}\BibitemShut {NoStop}%
\bibitem [{\citenamefont {Takeoka}\ \emph {et~al.}(2014)\citenamefont
  {Takeoka}, \citenamefont {Guha},\ and\ \citenamefont {Wilde}}]{takeoka2014}%
  \BibitemOpen
  \bibfield  {author} {\bibinfo {author} {\bibfnamefont {M.}~\bibnamefont
  {Takeoka}}, \bibinfo {author} {\bibfnamefont {S.}~\bibnamefont {Guha}},\ and\
  \bibinfo {author} {\bibfnamefont {M.~M.}\ \bibnamefont {Wilde}},\ }\bibfield
  {title} {\bibinfo {title} {Fundamental rate-loss tradeoff for optical quantum
  key distribution},\ }\href {https://doi.org/10.1038/ncomms6235} {\bibfield
  {journal} {\bibinfo  {journal} {Nature Communications}\ }\textbf {\bibinfo
  {volume} {5}},\ \bibinfo {pages} {5235} (\bibinfo {year} {2014})}\BibitemShut
  {NoStop}%
\bibitem [{\citenamefont {Pirandola}\ \emph {et~al.}(2017)\citenamefont
  {Pirandola}, \citenamefont {Laurenza}, \citenamefont {Ottaviani},\ and\
  \citenamefont {Banchi}}]{pirandola2017}%
  \BibitemOpen
  \bibfield  {author} {\bibinfo {author} {\bibfnamefont {S.}~\bibnamefont
  {Pirandola}}, \bibinfo {author} {\bibfnamefont {R.}~\bibnamefont {Laurenza}},
  \bibinfo {author} {\bibfnamefont {C.}~\bibnamefont {Ottaviani}},\ and\
  \bibinfo {author} {\bibfnamefont {L.}~\bibnamefont {Banchi}},\ }\bibfield
  {title} {\bibinfo {title} {Fundamental limits of repeaterless quantum
  communications},\ }\href {https://doi.org/10.1038/ncomms15043} {\bibfield
  {journal} {\bibinfo  {journal} {Nature Communications}\ }\textbf {\bibinfo
  {volume} {8}},\ \bibinfo {pages} {15043} (\bibinfo {year}
  {2017})}\BibitemShut {NoStop}%
\bibitem [{\citenamefont {Gr{\"u}nenfelder}\ \emph {et~al.}(2020)\citenamefont
  {Gr{\"u}nenfelder}, \citenamefont {Boaron}, \citenamefont {Rusca},
  \citenamefont {Martin},\ and\ \citenamefont {Zbinden}}]{10.1063/5.0021468}%
  \BibitemOpen
  \bibfield  {author} {\bibinfo {author} {\bibfnamefont {F.}~\bibnamefont
  {Gr{\"u}nenfelder}}, \bibinfo {author} {\bibfnamefont {A.}~\bibnamefont
  {Boaron}}, \bibinfo {author} {\bibfnamefont {D.}~\bibnamefont {Rusca}},
  \bibinfo {author} {\bibfnamefont {A.}~\bibnamefont {Martin}},\ and\ \bibinfo
  {author} {\bibfnamefont {H.}~\bibnamefont {Zbinden}},\ }\bibfield  {title}
  {\bibinfo {title} {Performance and security of 5ghz repetition rate
  polarization-based quantum key distribution},\ }\href
  {https://doi.org/10.1063/5.0021468} {\bibfield  {journal} {\bibinfo
  {journal} {Applied Physics Letters}\ }\textbf {\bibinfo {volume} {117}},\
  \bibinfo {pages} {144003} (\bibinfo {year} {2020})}\BibitemShut {NoStop}%
\bibitem [{\citenamefont {Lucamarini}\ \emph {et~al.}(2018)\citenamefont
  {Lucamarini}, \citenamefont {Yuan}, \citenamefont {Dynes},\ and\
  \citenamefont {Shields}}]{lucamarini2018}%
  \BibitemOpen
  \bibfield  {author} {\bibinfo {author} {\bibfnamefont {M.}~\bibnamefont
  {Lucamarini}}, \bibinfo {author} {\bibfnamefont {Z.~L.}\ \bibnamefont
  {Yuan}}, \bibinfo {author} {\bibfnamefont {J.~F.}\ \bibnamefont {Dynes}},\
  and\ \bibinfo {author} {\bibfnamefont {A.~J.}\ \bibnamefont {Shields}},\
  }\bibfield  {title} {\bibinfo {title} {Overcoming the rate--distance limit of
  quantum key distribution without quantum repeaters},\ }\href
  {https://doi.org/10.1038/s41586-018-0066-6} {\bibfield  {journal} {\bibinfo
  {journal} {Nature}\ }\textbf {\bibinfo {volume} {557}},\ \bibinfo {pages}
  {400} (\bibinfo {year} {2018})}\BibitemShut {NoStop}%
\bibitem [{\citenamefont {Minder}\ \emph {et~al.}(2019)\citenamefont {Minder},
  \citenamefont {Pittaluga}, \citenamefont {Roberts}, \citenamefont
  {Lucamarini}, \citenamefont {Dynes}, \citenamefont {Yuan},\ and\
  \citenamefont {Shields}}]{Minder2019}%
  \BibitemOpen
  \bibfield  {author} {\bibinfo {author} {\bibfnamefont {M.}~\bibnamefont
  {Minder}}, \bibinfo {author} {\bibfnamefont {M.}~\bibnamefont {Pittaluga}},
  \bibinfo {author} {\bibfnamefont {G.~L.}\ \bibnamefont {Roberts}}, \bibinfo
  {author} {\bibfnamefont {M.}~\bibnamefont {Lucamarini}}, \bibinfo {author}
  {\bibfnamefont {J.~F.}\ \bibnamefont {Dynes}}, \bibinfo {author}
  {\bibfnamefont {Z.~L.}\ \bibnamefont {Yuan}},\ and\ \bibinfo {author}
  {\bibfnamefont {A.~J.}\ \bibnamefont {Shields}},\ }\bibfield  {title}
  {\bibinfo {title} {Experimental quantum key distribution beyond the
  repeaterless secret key capacity},\ }\href
  {https://doi.org/10.1038/s41566-019-0377-7} {\bibfield  {journal} {\bibinfo
  {journal} {Nature Photonics}\ }\textbf {\bibinfo {volume} {13}},\ \bibinfo
  {pages} {334} (\bibinfo {year} {2019})}\BibitemShut {NoStop}%
\bibitem [{\citenamefont {Liu}\ \emph {et~al.}(2019)\citenamefont {Liu},
  \citenamefont {Yu}, \citenamefont {Zhang}, \citenamefont {Guan},
  \citenamefont {Chen}, \citenamefont {Zhang}, \citenamefont {Hu},
  \citenamefont {Li}, \citenamefont {Jiang}, \citenamefont {Lin}, \citenamefont
  {Chen}, \citenamefont {You}, \citenamefont {Wang}, \citenamefont {Wang},
  \citenamefont {Zhang},\ and\ \citenamefont {Pan}}]{Liu2019TF300km}%
  \BibitemOpen
  \bibfield  {author} {\bibinfo {author} {\bibfnamefont {Y.}~\bibnamefont
  {Liu}}, \bibinfo {author} {\bibfnamefont {Z.-W.}\ \bibnamefont {Yu}},
  \bibinfo {author} {\bibfnamefont {W.}~\bibnamefont {Zhang}}, \bibinfo
  {author} {\bibfnamefont {J.-Y.}\ \bibnamefont {Guan}}, \bibinfo {author}
  {\bibfnamefont {J.-P.}\ \bibnamefont {Chen}}, \bibinfo {author}
  {\bibfnamefont {C.}~\bibnamefont {Zhang}}, \bibinfo {author} {\bibfnamefont
  {X.-L.}\ \bibnamefont {Hu}}, \bibinfo {author} {\bibfnamefont
  {H.}~\bibnamefont {Li}}, \bibinfo {author} {\bibfnamefont {C.}~\bibnamefont
  {Jiang}}, \bibinfo {author} {\bibfnamefont {J.}~\bibnamefont {Lin}}, \bibinfo
  {author} {\bibfnamefont {T.-Y.}\ \bibnamefont {Chen}}, \bibinfo {author}
  {\bibfnamefont {L.}~\bibnamefont {You}}, \bibinfo {author} {\bibfnamefont
  {Z.}~\bibnamefont {Wang}}, \bibinfo {author} {\bibfnamefont {X.-B.}\
  \bibnamefont {Wang}}, \bibinfo {author} {\bibfnamefont {Q.}~\bibnamefont
  {Zhang}},\ and\ \bibinfo {author} {\bibfnamefont {J.-W.}\ \bibnamefont
  {Pan}},\ }\bibfield  {title} {\bibinfo {title} {Experimental twin-field
  quantum key distribution through sending or not sending},\ }\href
  {https://doi.org/10.1103/PhysRevLett.123.100505} {\bibfield  {journal}
  {\bibinfo  {journal} {Phys. Rev. Lett.}\ }\textbf {\bibinfo {volume} {123}},\
  \bibinfo {pages} {100505} (\bibinfo {year} {2019})}\BibitemShut {NoStop}%
\bibitem [{\citenamefont {Wang}\ \emph {et~al.}(2019)\citenamefont {Wang},
  \citenamefont {He}, \citenamefont {Yin}, \citenamefont {Lu}, \citenamefont
  {Cui}, \citenamefont {Chen}, \citenamefont {Zhou}, \citenamefont {Guo},\ and\
  \citenamefont {Han}}]{PhysRevX.9.021046}%
  \BibitemOpen
  \bibfield  {author} {\bibinfo {author} {\bibfnamefont {S.}~\bibnamefont
  {Wang}}, \bibinfo {author} {\bibfnamefont {D.-Y.}\ \bibnamefont {He}},
  \bibinfo {author} {\bibfnamefont {Z.-Q.}\ \bibnamefont {Yin}}, \bibinfo
  {author} {\bibfnamefont {F.-Y.}\ \bibnamefont {Lu}}, \bibinfo {author}
  {\bibfnamefont {C.-H.}\ \bibnamefont {Cui}}, \bibinfo {author} {\bibfnamefont
  {W.}~\bibnamefont {Chen}}, \bibinfo {author} {\bibfnamefont {Z.}~\bibnamefont
  {Zhou}}, \bibinfo {author} {\bibfnamefont {G.-C.}\ \bibnamefont {Guo}},\ and\
  \bibinfo {author} {\bibfnamefont {Z.-F.}\ \bibnamefont {Han}},\ }\bibfield
  {title} {\bibinfo {title} {Beating the fundamental rate-distance limit in a
  proof-of-principle quantum key distribution system},\ }\href
  {https://doi.org/10.1103/PhysRevX.9.021046} {\bibfield  {journal} {\bibinfo
  {journal} {Phys. Rev. X}\ }\textbf {\bibinfo {volume} {9}},\ \bibinfo {pages}
  {021046} (\bibinfo {year} {2019})}\BibitemShut {NoStop}%
\bibitem [{\citenamefont {Chen}\ \emph {et~al.}(2020)\citenamefont {Chen},
  \citenamefont {Zhang}, \citenamefont {Liu}, \citenamefont {Jiang},
  \citenamefont {Zhang}, \citenamefont {Hu}, \citenamefont {Guan},
  \citenamefont {Yu}, \citenamefont {Xu}, \citenamefont {Lin}, \citenamefont
  {Li}, \citenamefont {Chen}, \citenamefont {Li}, \citenamefont {You},
  \citenamefont {Wang}, \citenamefont {Wang}, \citenamefont {Zhang},\ and\
  \citenamefont {Pan}}]{Chen2020TF509km}%
  \BibitemOpen
  \bibfield  {author} {\bibinfo {author} {\bibfnamefont {J.-P.}\ \bibnamefont
  {Chen}}, \bibinfo {author} {\bibfnamefont {C.}~\bibnamefont {Zhang}},
  \bibinfo {author} {\bibfnamefont {Y.}~\bibnamefont {Liu}}, \bibinfo {author}
  {\bibfnamefont {C.}~\bibnamefont {Jiang}}, \bibinfo {author} {\bibfnamefont
  {W.}~\bibnamefont {Zhang}}, \bibinfo {author} {\bibfnamefont {X.-L.}\
  \bibnamefont {Hu}}, \bibinfo {author} {\bibfnamefont {J.-Y.}\ \bibnamefont
  {Guan}}, \bibinfo {author} {\bibfnamefont {Z.-W.}\ \bibnamefont {Yu}},
  \bibinfo {author} {\bibfnamefont {H.}~\bibnamefont {Xu}}, \bibinfo {author}
  {\bibfnamefont {J.}~\bibnamefont {Lin}}, \bibinfo {author} {\bibfnamefont
  {M.-J.}\ \bibnamefont {Li}}, \bibinfo {author} {\bibfnamefont
  {H.}~\bibnamefont {Chen}}, \bibinfo {author} {\bibfnamefont {H.}~\bibnamefont
  {Li}}, \bibinfo {author} {\bibfnamefont {L.}~\bibnamefont {You}}, \bibinfo
  {author} {\bibfnamefont {Z.}~\bibnamefont {Wang}}, \bibinfo {author}
  {\bibfnamefont {X.-B.}\ \bibnamefont {Wang}}, \bibinfo {author}
  {\bibfnamefont {Q.}~\bibnamefont {Zhang}},\ and\ \bibinfo {author}
  {\bibfnamefont {J.-W.}\ \bibnamefont {Pan}},\ }\bibfield  {title} {\bibinfo
  {title} {Sending-or-not-sending with independent lasers: Secure twin-field
  quantum key distribution over 509 km},\ }\href
  {https://doi.org/10.1103/PhysRevLett.124.070501} {\bibfield  {journal}
  {\bibinfo  {journal} {Phys. Rev. Lett.}\ }\textbf {\bibinfo {volume} {124}},\
  \bibinfo {pages} {070501} (\bibinfo {year} {2020})}\BibitemShut {NoStop}%
\bibitem [{\citenamefont {Wang}\ \emph {et~al.}(2022)\citenamefont {Wang},
  \citenamefont {Yin}, \citenamefont {He}, \citenamefont {Chen}, \citenamefont
  {Wang}, \citenamefont {Ye}, \citenamefont {Zhou}, \citenamefont {Fan-Yuan},
  \citenamefont {Wang}, \citenamefont {Zhu}, \citenamefont {Morozov},
  \citenamefont {Divochiy}, \citenamefont {Zhou}, \citenamefont {Guo},\ and\
  \citenamefont {Han}}]{wang2022_830km}%
  \BibitemOpen
  \bibfield  {author} {\bibinfo {author} {\bibfnamefont {S.}~\bibnamefont
  {Wang}}, \bibinfo {author} {\bibfnamefont {Z.-Q.}\ \bibnamefont {Yin}},
  \bibinfo {author} {\bibfnamefont {D.-Y.}\ \bibnamefont {He}}, \bibinfo
  {author} {\bibfnamefont {W.}~\bibnamefont {Chen}}, \bibinfo {author}
  {\bibfnamefont {R.-Q.}\ \bibnamefont {Wang}}, \bibinfo {author}
  {\bibfnamefont {P.}~\bibnamefont {Ye}}, \bibinfo {author} {\bibfnamefont
  {Y.}~\bibnamefont {Zhou}}, \bibinfo {author} {\bibfnamefont {G.-J.}\
  \bibnamefont {Fan-Yuan}}, \bibinfo {author} {\bibfnamefont {F.-X.}\
  \bibnamefont {Wang}}, \bibinfo {author} {\bibfnamefont {Y.-G.}\ \bibnamefont
  {Zhu}}, \bibinfo {author} {\bibfnamefont {P.~V.}\ \bibnamefont {Morozov}},
  \bibinfo {author} {\bibfnamefont {A.~V.}\ \bibnamefont {Divochiy}}, \bibinfo
  {author} {\bibfnamefont {Z.}~\bibnamefont {Zhou}}, \bibinfo {author}
  {\bibfnamefont {G.-C.}\ \bibnamefont {Guo}},\ and\ \bibinfo {author}
  {\bibfnamefont {Z.-F.}\ \bibnamefont {Han}},\ }\bibfield  {title} {\bibinfo
  {title} {Twin-field quantum key distribution over 830-km fibre},\ }\href
  {https://doi.org/10.1038/s41566-021-00928-2} {\bibfield  {journal} {\bibinfo
  {journal} {Nature Photonics}\ }\textbf {\bibinfo {volume} {16}},\ \bibinfo
  {pages} {154} (\bibinfo {year} {2022})}\BibitemShut {NoStop}%
\bibitem [{\citenamefont {Liu}\ \emph {et~al.}(2021{\natexlab{a}})\citenamefont
  {Liu}, \citenamefont {Jiang}, \citenamefont {Zhu}, \citenamefont {Zou},
  \citenamefont {Yu}, \citenamefont {Hu}, \citenamefont {Xu}, \citenamefont
  {Ma}, \citenamefont {Han}, \citenamefont {Chen}, \citenamefont {Dai},
  \citenamefont {Tang}, \citenamefont {Zhang}, \citenamefont {Li},
  \citenamefont {You}, \citenamefont {Wang}, \citenamefont {Hua}, \citenamefont
  {Hu}, \citenamefont {Zhang}, \citenamefont {Zhou}, \citenamefont {Zhang},
  \citenamefont {Wang}, \citenamefont {Chen},\ and\ \citenamefont
  {Pan}}]{LiuHui2021}%
  \BibitemOpen
  \bibfield  {author} {\bibinfo {author} {\bibfnamefont {H.}~\bibnamefont
  {Liu}}, \bibinfo {author} {\bibfnamefont {C.}~\bibnamefont {Jiang}}, \bibinfo
  {author} {\bibfnamefont {H.-T.}\ \bibnamefont {Zhu}}, \bibinfo {author}
  {\bibfnamefont {M.}~\bibnamefont {Zou}}, \bibinfo {author} {\bibfnamefont
  {Z.-W.}\ \bibnamefont {Yu}}, \bibinfo {author} {\bibfnamefont {X.-L.}\
  \bibnamefont {Hu}}, \bibinfo {author} {\bibfnamefont {H.}~\bibnamefont {Xu}},
  \bibinfo {author} {\bibfnamefont {S.}~\bibnamefont {Ma}}, \bibinfo {author}
  {\bibfnamefont {Z.}~\bibnamefont {Han}}, \bibinfo {author} {\bibfnamefont
  {J.-P.}\ \bibnamefont {Chen}}, \bibinfo {author} {\bibfnamefont
  {Y.}~\bibnamefont {Dai}}, \bibinfo {author} {\bibfnamefont {S.-B.}\
  \bibnamefont {Tang}}, \bibinfo {author} {\bibfnamefont {W.}~\bibnamefont
  {Zhang}}, \bibinfo {author} {\bibfnamefont {H.}~\bibnamefont {Li}}, \bibinfo
  {author} {\bibfnamefont {L.}~\bibnamefont {You}}, \bibinfo {author}
  {\bibfnamefont {Z.}~\bibnamefont {Wang}}, \bibinfo {author} {\bibfnamefont
  {Y.}~\bibnamefont {Hua}}, \bibinfo {author} {\bibfnamefont {H.}~\bibnamefont
  {Hu}}, \bibinfo {author} {\bibfnamefont {H.}~\bibnamefont {Zhang}}, \bibinfo
  {author} {\bibfnamefont {F.}~\bibnamefont {Zhou}}, \bibinfo {author}
  {\bibfnamefont {Q.}~\bibnamefont {Zhang}}, \bibinfo {author} {\bibfnamefont
  {X.-B.}\ \bibnamefont {Wang}}, \bibinfo {author} {\bibfnamefont {T.-Y.}\
  \bibnamefont {Chen}},\ and\ \bibinfo {author} {\bibfnamefont {J.-W.}\
  \bibnamefont {Pan}},\ }\bibfield  {title} {\bibinfo {title} {Field test of
  twin-field quantum key distribution through sending-or-not-sending over 428
  km},\ }\href {https://doi.org/10.1103/PhysRevLett.126.250502} {\bibfield
  {journal} {\bibinfo  {journal} {Phys. Rev. Lett.}\ }\textbf {\bibinfo
  {volume} {126}},\ \bibinfo {pages} {250502} (\bibinfo {year}
  {2021}{\natexlab{a}})}\BibitemShut {NoStop}%
\bibitem [{\citenamefont {Chen}\ \emph {et~al.}(2021)\citenamefont {Chen},
  \citenamefont {Zhang}, \citenamefont {Liu}, \citenamefont {Jiang},
  \citenamefont {Zhang}, \citenamefont {Han}, \citenamefont {Ma}, \citenamefont
  {Hu}, \citenamefont {Li}, \citenamefont {Liu}, \citenamefont {Zhou},
  \citenamefont {Jiang}, \citenamefont {Chen}, \citenamefont {Li},
  \citenamefont {You}, \citenamefont {Wang}, \citenamefont {Wang},
  \citenamefont {Zhang},\ and\ \citenamefont {Pan}}]{Chen2021_511km}%
  \BibitemOpen
  \bibfield  {author} {\bibinfo {author} {\bibfnamefont {J.-P.}\ \bibnamefont
  {Chen}}, \bibinfo {author} {\bibfnamefont {C.}~\bibnamefont {Zhang}},
  \bibinfo {author} {\bibfnamefont {Y.}~\bibnamefont {Liu}}, \bibinfo {author}
  {\bibfnamefont {C.}~\bibnamefont {Jiang}}, \bibinfo {author} {\bibfnamefont
  {W.-J.}\ \bibnamefont {Zhang}}, \bibinfo {author} {\bibfnamefont {Z.-Y.}\
  \bibnamefont {Han}}, \bibinfo {author} {\bibfnamefont {S.-Z.}\ \bibnamefont
  {Ma}}, \bibinfo {author} {\bibfnamefont {X.-L.}\ \bibnamefont {Hu}}, \bibinfo
  {author} {\bibfnamefont {Y.-H.}\ \bibnamefont {Li}}, \bibinfo {author}
  {\bibfnamefont {H.}~\bibnamefont {Liu}}, \bibinfo {author} {\bibfnamefont
  {F.}~\bibnamefont {Zhou}}, \bibinfo {author} {\bibfnamefont {H.-F.}\
  \bibnamefont {Jiang}}, \bibinfo {author} {\bibfnamefont {T.-Y.}\ \bibnamefont
  {Chen}}, \bibinfo {author} {\bibfnamefont {H.}~\bibnamefont {Li}}, \bibinfo
  {author} {\bibfnamefont {L.-X.}\ \bibnamefont {You}}, \bibinfo {author}
  {\bibfnamefont {Z.}~\bibnamefont {Wang}}, \bibinfo {author} {\bibfnamefont
  {X.-B.}\ \bibnamefont {Wang}}, \bibinfo {author} {\bibfnamefont
  {Q.}~\bibnamefont {Zhang}},\ and\ \bibinfo {author} {\bibfnamefont {J.-W.}\
  \bibnamefont {Pan}},\ }\bibfield  {title} {\bibinfo {title} {Twin-field
  quantum key distribution over a 511{\thinspace}km optical fibre linking two
  distant metropolitan areas},\ }\href
  {https://doi.org/10.1038/s41566-021-00828-5} {\bibfield  {journal} {\bibinfo
  {journal} {Nature Photonics}\ }\textbf {\bibinfo {volume} {15}},\ \bibinfo
  {pages} {570} (\bibinfo {year} {2021})}\BibitemShut {NoStop}%
\bibitem [{\citenamefont {Zhou}\ \emph {et~al.}(2024)\citenamefont {Zhou},
  \citenamefont {Lin}, \citenamefont {Ge}, \citenamefont {Fan}, \citenamefont
  {Yuan}, \citenamefont {Dong}, \citenamefont {Liu}, \citenamefont {Ma},
  \citenamefont {Chen}, \citenamefont {Jiang}, \citenamefont {Wang},
  \citenamefont {You}, \citenamefont {Zhang},\ and\ \citenamefont
  {Pan}}]{Zhou2024_546km}%
  \BibitemOpen
  \bibfield  {author} {\bibinfo {author} {\bibfnamefont {L.}~\bibnamefont
  {Zhou}}, \bibinfo {author} {\bibfnamefont {J.}~\bibnamefont {Lin}}, \bibinfo
  {author} {\bibfnamefont {C.}~\bibnamefont {Ge}}, \bibinfo {author}
  {\bibfnamefont {Y.}~\bibnamefont {Fan}}, \bibinfo {author} {\bibfnamefont
  {Z.}~\bibnamefont {Yuan}}, \bibinfo {author} {\bibfnamefont {H.}~\bibnamefont
  {Dong}}, \bibinfo {author} {\bibfnamefont {Y.}~\bibnamefont {Liu}}, \bibinfo
  {author} {\bibfnamefont {D.}~\bibnamefont {Ma}}, \bibinfo {author}
  {\bibfnamefont {J.-P.}\ \bibnamefont {Chen}}, \bibinfo {author}
  {\bibfnamefont {C.}~\bibnamefont {Jiang}}, \bibinfo {author} {\bibfnamefont
  {X.-B.}\ \bibnamefont {Wang}}, \bibinfo {author} {\bibfnamefont {L.-X.}\
  \bibnamefont {You}}, \bibinfo {author} {\bibfnamefont {Q.}~\bibnamefont
  {Zhang}},\ and\ \bibinfo {author} {\bibfnamefont {J.-W.}\ \bibnamefont
  {Pan}},\ }\bibfield  {title} {\bibinfo {title}
  {Independent-optical-frequency-comb-powered 546-km field test of twin-field
  quantum key distribution},\ }\href
  {https://doi.org/10.1103/PhysRevApplied.22.064057} {\bibfield  {journal}
  {\bibinfo  {journal} {Phys. Rev. Appl.}\ }\textbf {\bibinfo {volume} {22}},\
  \bibinfo {pages} {064057} (\bibinfo {year} {2024})}\BibitemShut {NoStop}%
\bibitem [{\citenamefont {Pittaluga}\ \emph {et~al.}(2025)\citenamefont
  {Pittaluga}, \citenamefont {Lo}, \citenamefont {Brzosko}, \citenamefont
  {Woodward}, \citenamefont {Scalcon}, \citenamefont {Winnel}, \citenamefont
  {Roger}, \citenamefont {Dynes}, \citenamefont {Owen}, \citenamefont
  {Ju{\'a}rez}, \citenamefont {Rydlichowski}, \citenamefont {Vicinanza},
  \citenamefont {Roberts},\ and\ \citenamefont {Shields}}]{Pittaluga2025}%
  \BibitemOpen
  \bibfield  {author} {\bibinfo {author} {\bibfnamefont {M.}~\bibnamefont
  {Pittaluga}}, \bibinfo {author} {\bibfnamefont {Y.~S.}\ \bibnamefont {Lo}},
  \bibinfo {author} {\bibfnamefont {A.}~\bibnamefont {Brzosko}}, \bibinfo
  {author} {\bibfnamefont {R.~I.}\ \bibnamefont {Woodward}}, \bibinfo {author}
  {\bibfnamefont {D.}~\bibnamefont {Scalcon}}, \bibinfo {author} {\bibfnamefont
  {M.~S.}\ \bibnamefont {Winnel}}, \bibinfo {author} {\bibfnamefont
  {T.}~\bibnamefont {Roger}}, \bibinfo {author} {\bibfnamefont {J.~F.}\
  \bibnamefont {Dynes}}, \bibinfo {author} {\bibfnamefont {K.~A.}\ \bibnamefont
  {Owen}}, \bibinfo {author} {\bibfnamefont {S.}~\bibnamefont {Ju{\'a}rez}},
  \bibinfo {author} {\bibfnamefont {P.}~\bibnamefont {Rydlichowski}}, \bibinfo
  {author} {\bibfnamefont {D.}~\bibnamefont {Vicinanza}}, \bibinfo {author}
  {\bibfnamefont {G.}~\bibnamefont {Roberts}},\ and\ \bibinfo {author}
  {\bibfnamefont {A.~J.}\ \bibnamefont {Shields}},\ }\bibfield  {title}
  {\bibinfo {title} {Long-distance coherent quantum communications in deployed
  telecom networks},\ }\href {https://doi.org/10.1038/s41586-025-08801-w}
  {\bibfield  {journal} {\bibinfo  {journal} {Nature}\ }\textbf {\bibinfo
  {volume} {640}},\ \bibinfo {pages} {911} (\bibinfo {year}
  {2025})}\BibitemShut {NoStop}%
\bibitem [{\citenamefont {Wang}\ \emph
  {et~al.}(2018{\natexlab{a}})\citenamefont {Wang}, \citenamefont {Yu},\ and\
  \citenamefont {Hu}}]{Wang_SNS2018}%
  \BibitemOpen
  \bibfield  {author} {\bibinfo {author} {\bibfnamefont {X.-B.}\ \bibnamefont
  {Wang}}, \bibinfo {author} {\bibfnamefont {Z.-W.}\ \bibnamefont {Yu}},\ and\
  \bibinfo {author} {\bibfnamefont {X.-L.}\ \bibnamefont {Hu}},\ }\bibfield
  {title} {\bibinfo {title} {Twin-field quantum key distribution with large
  misalignment error},\ }\href {https://doi.org/10.1103/PhysRevA.98.062323}
  {\bibfield  {journal} {\bibinfo  {journal} {Phys. Rev. A}\ }\textbf {\bibinfo
  {volume} {98}},\ \bibinfo {pages} {062323} (\bibinfo {year}
  {2018}{\natexlab{a}})}\BibitemShut {NoStop}%
\bibitem [{\citenamefont {Hu}\ \emph {et~al.}(2018)\citenamefont {Hu},
  \citenamefont {Da~Ros}, \citenamefont {Pu}, \citenamefont {Ye}, \citenamefont
  {Ingerslev}, \citenamefont {Porto~da Silva}, \citenamefont {Nooruzzaman},
  \citenamefont {Amma}, \citenamefont {Sasaki}, \citenamefont {Mizuno},
  \citenamefont {Miyamoto}, \citenamefont {Ottaviano}, \citenamefont
  {Semenova}, \citenamefont {Guan}, \citenamefont {Zibar}, \citenamefont
  {Galili}, \citenamefont {Yvind}, \citenamefont {Morioka},\ and\ \citenamefont
  {Oxenl{\o}we}}]{Hu2018}%
  \BibitemOpen
  \bibfield  {author} {\bibinfo {author} {\bibfnamefont {H.}~\bibnamefont
  {Hu}}, \bibinfo {author} {\bibfnamefont {F.}~\bibnamefont {Da~Ros}}, \bibinfo
  {author} {\bibfnamefont {M.}~\bibnamefont {Pu}}, \bibinfo {author}
  {\bibfnamefont {F.}~\bibnamefont {Ye}}, \bibinfo {author} {\bibfnamefont
  {K.}~\bibnamefont {Ingerslev}}, \bibinfo {author} {\bibfnamefont
  {E.}~\bibnamefont {Porto~da Silva}}, \bibinfo {author} {\bibfnamefont
  {M.}~\bibnamefont {Nooruzzaman}}, \bibinfo {author} {\bibfnamefont
  {Y.}~\bibnamefont {Amma}}, \bibinfo {author} {\bibfnamefont {Y.}~\bibnamefont
  {Sasaki}}, \bibinfo {author} {\bibfnamefont {T.}~\bibnamefont {Mizuno}},
  \bibinfo {author} {\bibfnamefont {Y.}~\bibnamefont {Miyamoto}}, \bibinfo
  {author} {\bibfnamefont {L.}~\bibnamefont {Ottaviano}}, \bibinfo {author}
  {\bibfnamefont {E.}~\bibnamefont {Semenova}}, \bibinfo {author}
  {\bibfnamefont {P.}~\bibnamefont {Guan}}, \bibinfo {author} {\bibfnamefont
  {D.}~\bibnamefont {Zibar}}, \bibinfo {author} {\bibfnamefont
  {M.}~\bibnamefont {Galili}}, \bibinfo {author} {\bibfnamefont
  {K.}~\bibnamefont {Yvind}}, \bibinfo {author} {\bibfnamefont
  {T.}~\bibnamefont {Morioka}},\ and\ \bibinfo {author} {\bibfnamefont {L.~K.}\
  \bibnamefont {Oxenl{\o}we}},\ }\bibfield  {title} {\bibinfo {title}
  {Single-source chip-based frequency comb enabling extreme parallel data
  transmission},\ }\href {https://doi.org/10.1038/s41566-018-0205-5} {\bibfield
   {journal} {\bibinfo  {journal} {Nature Photonics}\ }\textbf {\bibinfo
  {volume} {12}},\ \bibinfo {pages} {469} (\bibinfo {year} {2018})}\BibitemShut
  {NoStop}%
\bibitem [{\citenamefont {Kemal}\ \emph {et~al.}(2019)\citenamefont {Kemal},
  \citenamefont {Marin-Palomo}, \citenamefont {Panapakkam}, \citenamefont
  {Trocha}, \citenamefont {Wolf}, \citenamefont {Merghem}, \citenamefont
  {Lelarge}, \citenamefont {Ramdane}, \citenamefont {Randel}, \citenamefont
  {Freude},\ and\ \citenamefont {Koos}}]{Kemal:19}%
  \BibitemOpen
  \bibfield  {author} {\bibinfo {author} {\bibfnamefont {J.~N.}\ \bibnamefont
  {Kemal}}, \bibinfo {author} {\bibfnamefont {P.}~\bibnamefont {Marin-Palomo}},
  \bibinfo {author} {\bibfnamefont {V.}~\bibnamefont {Panapakkam}}, \bibinfo
  {author} {\bibfnamefont {P.}~\bibnamefont {Trocha}}, \bibinfo {author}
  {\bibfnamefont {S.}~\bibnamefont {Wolf}}, \bibinfo {author} {\bibfnamefont
  {K.}~\bibnamefont {Merghem}}, \bibinfo {author} {\bibfnamefont
  {F.}~\bibnamefont {Lelarge}}, \bibinfo {author} {\bibfnamefont
  {A.}~\bibnamefont {Ramdane}}, \bibinfo {author} {\bibfnamefont
  {S.}~\bibnamefont {Randel}}, \bibinfo {author} {\bibfnamefont
  {W.}~\bibnamefont {Freude}},\ and\ \bibinfo {author} {\bibfnamefont
  {C.}~\bibnamefont {Koos}},\ }\bibfield  {title} {\bibinfo {title} {Coherent
  wdm transmission using quantum-dash mode-locked laser diodes as
  multi-wavelength source and local oscillator},\ }\href
  {https://doi.org/10.1364/OE.27.031164} {\bibfield  {journal} {\bibinfo
  {journal} {Opt. Express}\ }\textbf {\bibinfo {volume} {27}},\ \bibinfo
  {pages} {31164} (\bibinfo {year} {2019})}\BibitemShut {NoStop}%
\bibitem [{\citenamefont {Liu}\ \emph {et~al.}(2023{\natexlab{a}})\citenamefont
  {Liu}, \citenamefont {Zhang}, \citenamefont {Jiang}, \citenamefont {Chen},
  \citenamefont {Zhang}, \citenamefont {Pan}, \citenamefont {Ma}, \citenamefont
  {Dong}, \citenamefont {Xiong}, \citenamefont {Zhang}, \citenamefont {Li},
  \citenamefont {Wang}, \citenamefont {Wu}, \citenamefont {Chen}, \citenamefont
  {You}, \citenamefont {Wang}, \citenamefont {Zhang},\ and\ \citenamefont
  {Pan}}]{Liu2023TF1000km}%
  \BibitemOpen
  \bibfield  {author} {\bibinfo {author} {\bibfnamefont {Y.}~\bibnamefont
  {Liu}}, \bibinfo {author} {\bibfnamefont {W.-J.}\ \bibnamefont {Zhang}},
  \bibinfo {author} {\bibfnamefont {C.}~\bibnamefont {Jiang}}, \bibinfo
  {author} {\bibfnamefont {J.-P.}\ \bibnamefont {Chen}}, \bibinfo {author}
  {\bibfnamefont {C.}~\bibnamefont {Zhang}}, \bibinfo {author} {\bibfnamefont
  {W.-X.}\ \bibnamefont {Pan}}, \bibinfo {author} {\bibfnamefont
  {D.}~\bibnamefont {Ma}}, \bibinfo {author} {\bibfnamefont {H.}~\bibnamefont
  {Dong}}, \bibinfo {author} {\bibfnamefont {J.-M.}\ \bibnamefont {Xiong}},
  \bibinfo {author} {\bibfnamefont {C.-J.}\ \bibnamefont {Zhang}}, \bibinfo
  {author} {\bibfnamefont {H.}~\bibnamefont {Li}}, \bibinfo {author}
  {\bibfnamefont {R.-C.}\ \bibnamefont {Wang}}, \bibinfo {author}
  {\bibfnamefont {J.}~\bibnamefont {Wu}}, \bibinfo {author} {\bibfnamefont
  {T.-Y.}\ \bibnamefont {Chen}}, \bibinfo {author} {\bibfnamefont
  {L.}~\bibnamefont {You}}, \bibinfo {author} {\bibfnamefont {X.-B.}\
  \bibnamefont {Wang}}, \bibinfo {author} {\bibfnamefont {Q.}~\bibnamefont
  {Zhang}},\ and\ \bibinfo {author} {\bibfnamefont {J.-W.}\ \bibnamefont
  {Pan}},\ }\bibfield  {title} {\bibinfo {title} {Experimental twin-field
  quantum key distribution over 1000 km fiber distance},\ }\href
  {https://doi.org/10.1103/PhysRevLett.130.210801} {\bibfield  {journal}
  {\bibinfo  {journal} {Phys. Rev. Lett.}\ }\textbf {\bibinfo {volume} {130}},\
  \bibinfo {pages} {210801} (\bibinfo {year} {2023}{\natexlab{a}})}\BibitemShut
  {NoStop}%
\bibitem [{\citenamefont {Yi}\ \emph {et~al.}(2015)\citenamefont {Yi},
  \citenamefont {Yang}, \citenamefont {Yang}, \citenamefont {Suh},\ and\
  \citenamefont {Vahala}}]{Yi:15}%
  \BibitemOpen
  \bibfield  {author} {\bibinfo {author} {\bibfnamefont {X.}~\bibnamefont
  {Yi}}, \bibinfo {author} {\bibfnamefont {Q.-F.}\ \bibnamefont {Yang}},
  \bibinfo {author} {\bibfnamefont {K.~Y.}\ \bibnamefont {Yang}}, \bibinfo
  {author} {\bibfnamefont {M.-G.}\ \bibnamefont {Suh}},\ and\ \bibinfo {author}
  {\bibfnamefont {K.}~\bibnamefont {Vahala}},\ }\bibfield  {title} {\bibinfo
  {title} {Soliton frequency comb at microwave rates in a high-q silica
  microresonator},\ }\href {https://doi.org/10.1364/OPTICA.2.001078} {\bibfield
   {journal} {\bibinfo  {journal} {Optica}\ }\textbf {\bibinfo {volume} {2}},\
  \bibinfo {pages} {1078} (\bibinfo {year} {2015})}\BibitemShut {NoStop}%
\bibitem [{\citenamefont {Brasch}\ \emph {et~al.}(2016)\citenamefont {Brasch},
  \citenamefont {Geiselmann}, \citenamefont {Herr}, \citenamefont {Lihachev},
  \citenamefont {Pfeiffer}, \citenamefont {Gorodetsky},\ and\ \citenamefont
  {Kippenberg}}]{Brasch:15}%
  \BibitemOpen
  \bibfield  {author} {\bibinfo {author} {\bibfnamefont {V.}~\bibnamefont
  {Brasch}}, \bibinfo {author} {\bibfnamefont {M.}~\bibnamefont {Geiselmann}},
  \bibinfo {author} {\bibfnamefont {T.}~\bibnamefont {Herr}}, \bibinfo {author}
  {\bibfnamefont {G.}~\bibnamefont {Lihachev}}, \bibinfo {author}
  {\bibfnamefont {M.~H.~P.}\ \bibnamefont {Pfeiffer}}, \bibinfo {author}
  {\bibfnamefont {M.~L.}\ \bibnamefont {Gorodetsky}},\ and\ \bibinfo {author}
  {\bibfnamefont {T.~J.}\ \bibnamefont {Kippenberg}},\ }\bibfield  {title}
  {\bibinfo {title} {Photonic chip{\textendash}based optical frequency comb
  using soliton cherenkov radiation},\ }\href
  {https://doi.org/10.1126/science.aad4811} {\bibfield  {journal} {\bibinfo
  {journal} {Science}\ }\textbf {\bibinfo {volume} {351}},\ \bibinfo {pages}
  {357} (\bibinfo {year} {2016})}\BibitemShut {NoStop}%
\bibitem [{\citenamefont {Joshi}\ \emph {et~al.}(2016)\citenamefont {Joshi},
  \citenamefont {Jang}, \citenamefont {Luke}, \citenamefont {Ji}, \citenamefont
  {Miller}, \citenamefont {Klenner}, \citenamefont {Okawachi}, \citenamefont
  {Lipson},\ and\ \citenamefont {Gaeta}}]{Joshi:16}%
  \BibitemOpen
  \bibfield  {author} {\bibinfo {author} {\bibfnamefont {C.}~\bibnamefont
  {Joshi}}, \bibinfo {author} {\bibfnamefont {J.~K.}\ \bibnamefont {Jang}},
  \bibinfo {author} {\bibfnamefont {K.}~\bibnamefont {Luke}}, \bibinfo {author}
  {\bibfnamefont {X.}~\bibnamefont {Ji}}, \bibinfo {author} {\bibfnamefont
  {S.~A.}\ \bibnamefont {Miller}}, \bibinfo {author} {\bibfnamefont
  {A.}~\bibnamefont {Klenner}}, \bibinfo {author} {\bibfnamefont
  {Y.}~\bibnamefont {Okawachi}}, \bibinfo {author} {\bibfnamefont
  {M.}~\bibnamefont {Lipson}},\ and\ \bibinfo {author} {\bibfnamefont {A.~L.}\
  \bibnamefont {Gaeta}},\ }\bibfield  {title} {\bibinfo {title} {Thermally
  controlled comb generation and soliton modelocking in microresonators},\
  }\href {https://doi.org/10.1364/OL.41.002565} {\bibfield  {journal} {\bibinfo
   {journal} {Opt. Lett.}\ }\textbf {\bibinfo {volume} {41}},\ \bibinfo {pages}
  {2565} (\bibinfo {year} {2016})}\BibitemShut {NoStop}%
\bibitem [{\citenamefont {Bao}\ \emph {et~al.}(2019)\citenamefont {Bao},
  \citenamefont {Cooper}, \citenamefont {Rowley}, \citenamefont {Di~Lauro},
  \citenamefont {Totero~Gongora}, \citenamefont {Chu}, \citenamefont {Little},
  \citenamefont {Oppo}, \citenamefont {Morandotti}, \citenamefont {Moss},
  \citenamefont {Wetzel}, \citenamefont {Peccianti},\ and\ \citenamefont
  {Pasquazi}}]{Bao:19}%
  \BibitemOpen
  \bibfield  {author} {\bibinfo {author} {\bibfnamefont {H.}~\bibnamefont
  {Bao}}, \bibinfo {author} {\bibfnamefont {A.}~\bibnamefont {Cooper}},
  \bibinfo {author} {\bibfnamefont {M.}~\bibnamefont {Rowley}}, \bibinfo
  {author} {\bibfnamefont {L.}~\bibnamefont {Di~Lauro}}, \bibinfo {author}
  {\bibfnamefont {J.~S.}\ \bibnamefont {Totero~Gongora}}, \bibinfo {author}
  {\bibfnamefont {S.~T.}\ \bibnamefont {Chu}}, \bibinfo {author} {\bibfnamefont
  {B.~E.}\ \bibnamefont {Little}}, \bibinfo {author} {\bibfnamefont {G.-L.}\
  \bibnamefont {Oppo}}, \bibinfo {author} {\bibfnamefont {R.}~\bibnamefont
  {Morandotti}}, \bibinfo {author} {\bibfnamefont {D.~J.}\ \bibnamefont
  {Moss}}, \bibinfo {author} {\bibfnamefont {B.}~\bibnamefont {Wetzel}},
  \bibinfo {author} {\bibfnamefont {M.}~\bibnamefont {Peccianti}},\ and\
  \bibinfo {author} {\bibfnamefont {A.}~\bibnamefont {Pasquazi}},\ }\bibfield
  {title} {\bibinfo {title} {Laser cavity-soliton microcombs},\ }\href
  {https://doi.org/10.1038/s41566-019-0379-5} {\bibfield  {journal} {\bibinfo
  {journal} {Nature Photonics}\ }\textbf {\bibinfo {volume} {13}},\ \bibinfo
  {pages} {384} (\bibinfo {year} {2019})}\BibitemShut {NoStop}%
\bibitem [{\citenamefont {Liu}\ \emph {et~al.}(2021{\natexlab{b}})\citenamefont
  {Liu}, \citenamefont {Gong}, \citenamefont {Bruch}, \citenamefont {Surya},
  \citenamefont {Lu},\ and\ \citenamefont {Tang}}]{LiuX:21}%
  \BibitemOpen
  \bibfield  {author} {\bibinfo {author} {\bibfnamefont {X.}~\bibnamefont
  {Liu}}, \bibinfo {author} {\bibfnamefont {Z.}~\bibnamefont {Gong}}, \bibinfo
  {author} {\bibfnamefont {A.~W.}\ \bibnamefont {Bruch}}, \bibinfo {author}
  {\bibfnamefont {J.~B.}\ \bibnamefont {Surya}}, \bibinfo {author}
  {\bibfnamefont {J.}~\bibnamefont {Lu}},\ and\ \bibinfo {author}
  {\bibfnamefont {H.~X.}\ \bibnamefont {Tang}},\ }\bibfield  {title} {\bibinfo
  {title} {Aluminum nitride nanophotonics for beyond-octave soliton microcomb
  generation and self-referencing},\ }\href
  {https://doi.org/10.1038/s41467-021-25751-9} {\bibfield  {journal} {\bibinfo
  {journal} {Nature Communications}\ }\textbf {\bibinfo {volume} {12}},\
  \bibinfo {pages} {5428} (\bibinfo {year} {2021}{\natexlab{b}})}\BibitemShut
  {NoStop}%
\bibitem [{\citenamefont {He}\ \emph {et~al.}(2019)\citenamefont {He},
  \citenamefont {Yang}, \citenamefont {Ling}, \citenamefont {Luo},
  \citenamefont {Liang}, \citenamefont {Li}, \citenamefont {Shen},
  \citenamefont {Wang}, \citenamefont {Vahala},\ and\ \citenamefont
  {Lin}}]{He:19}%
  \BibitemOpen
  \bibfield  {author} {\bibinfo {author} {\bibfnamefont {Y.}~\bibnamefont
  {He}}, \bibinfo {author} {\bibfnamefont {Q.-F.}\ \bibnamefont {Yang}},
  \bibinfo {author} {\bibfnamefont {J.}~\bibnamefont {Ling}}, \bibinfo {author}
  {\bibfnamefont {R.}~\bibnamefont {Luo}}, \bibinfo {author} {\bibfnamefont
  {H.}~\bibnamefont {Liang}}, \bibinfo {author} {\bibfnamefont
  {M.}~\bibnamefont {Li}}, \bibinfo {author} {\bibfnamefont {B.}~\bibnamefont
  {Shen}}, \bibinfo {author} {\bibfnamefont {H.}~\bibnamefont {Wang}}, \bibinfo
  {author} {\bibfnamefont {K.}~\bibnamefont {Vahala}},\ and\ \bibinfo {author}
  {\bibfnamefont {Q.}~\bibnamefont {Lin}},\ }\bibfield  {title} {\bibinfo
  {title} {Self-starting bi-chromatic linbo3 soliton microcomb},\ }\href
  {https://doi.org/10.1364/OPTICA.6.001138} {\bibfield  {journal} {\bibinfo
  {journal} {Optica}\ }\textbf {\bibinfo {volume} {6}},\ \bibinfo {pages}
  {1138} (\bibinfo {year} {2019})}\BibitemShut {NoStop}%
\bibitem [{\citenamefont {Gong}\ \emph {et~al.}(2020)\citenamefont {Gong},
  \citenamefont {Liu}, \citenamefont {Xu},\ and\ \citenamefont
  {Tang}}]{Gong:20}%
  \BibitemOpen
  \bibfield  {author} {\bibinfo {author} {\bibfnamefont {Z.}~\bibnamefont
  {Gong}}, \bibinfo {author} {\bibfnamefont {X.}~\bibnamefont {Liu}}, \bibinfo
  {author} {\bibfnamefont {Y.}~\bibnamefont {Xu}},\ and\ \bibinfo {author}
  {\bibfnamefont {H.~X.}\ \bibnamefont {Tang}},\ }\bibfield  {title} {\bibinfo
  {title} {Near-octave lithium niobate soliton microcomb},\ }\href
  {https://doi.org/10.1364/OPTICA.400994} {\bibfield  {journal} {\bibinfo
  {journal} {Optica}\ }\textbf {\bibinfo {volume} {7}},\ \bibinfo {pages}
  {1275} (\bibinfo {year} {2020})}\BibitemShut {NoStop}%
\bibitem [{\citenamefont {Guidry}\ \emph {et~al.}(2022)\citenamefont {Guidry},
  \citenamefont {Lukin}, \citenamefont {Yang}, \citenamefont {Trivedi},\ and\
  \citenamefont {Vu{\v c}kovi{\'c}}}]{Guidry:21}%
  \BibitemOpen
  \bibfield  {author} {\bibinfo {author} {\bibfnamefont {M.~A.}\ \bibnamefont
  {Guidry}}, \bibinfo {author} {\bibfnamefont {D.~M.}\ \bibnamefont {Lukin}},
  \bibinfo {author} {\bibfnamefont {K.~Y.}\ \bibnamefont {Yang}}, \bibinfo
  {author} {\bibfnamefont {R.}~\bibnamefont {Trivedi}},\ and\ \bibinfo {author}
  {\bibfnamefont {J.}~\bibnamefont {Vu{\v c}kovi{\'c}}},\ }\bibfield  {title}
  {\bibinfo {title} {Quantum optics of soliton microcombs},\ }\href
  {https://doi.org/10.1038/s41566-021-00901-z} {\bibfield  {journal} {\bibinfo
  {journal} {Nature Photonics}\ }\textbf {\bibinfo {volume} {16}},\ \bibinfo
  {pages} {52} (\bibinfo {year} {2022})}\BibitemShut {NoStop}%
\bibitem [{\citenamefont {Liu}\ \emph {et~al.}(2021{\natexlab{c}})\citenamefont
  {Liu}, \citenamefont {Huang}, \citenamefont {Wang}, \citenamefont {He},
  \citenamefont {Raja}, \citenamefont {Liu}, \citenamefont {Engelsen},\ and\
  \citenamefont {Kippenberg}}]{Liu:21}%
  \BibitemOpen
  \bibfield  {author} {\bibinfo {author} {\bibfnamefont {J.}~\bibnamefont
  {Liu}}, \bibinfo {author} {\bibfnamefont {G.}~\bibnamefont {Huang}}, \bibinfo
  {author} {\bibfnamefont {R.~N.}\ \bibnamefont {Wang}}, \bibinfo {author}
  {\bibfnamefont {J.}~\bibnamefont {He}}, \bibinfo {author} {\bibfnamefont
  {A.~S.}\ \bibnamefont {Raja}}, \bibinfo {author} {\bibfnamefont
  {T.}~\bibnamefont {Liu}}, \bibinfo {author} {\bibfnamefont {N.~J.}\
  \bibnamefont {Engelsen}},\ and\ \bibinfo {author} {\bibfnamefont {T.~J.}\
  \bibnamefont {Kippenberg}},\ }\bibfield  {title} {\bibinfo {title}
  {High-yield, wafer-scale fabrication of ultralow-loss, dispersion-engineered
  silicon nitride photonic circuits},\ }\href
  {https://doi.org/10.1038/s41467-021-21973-z} {\bibfield  {journal} {\bibinfo
  {journal} {Nature Communications}\ }\textbf {\bibinfo {volume} {12}},\
  \bibinfo {pages} {2236} (\bibinfo {year} {2021}{\natexlab{c}})}\BibitemShut
  {NoStop}%
\bibitem [{\citenamefont {Ye}\ \emph {et~al.}(2023)\citenamefont {Ye},
  \citenamefont {Jia}, \citenamefont {Huang}, \citenamefont {Shen},
  \citenamefont {Long}, \citenamefont {Shi}, \citenamefont {Luo}, \citenamefont
  {Gao}, \citenamefont {Sun}, \citenamefont {Guo}, \citenamefont {He},\ and\
  \citenamefont {Liu}}]{Ye:23}%
  \BibitemOpen
  \bibfield  {author} {\bibinfo {author} {\bibfnamefont {Z.}~\bibnamefont
  {Ye}}, \bibinfo {author} {\bibfnamefont {H.}~\bibnamefont {Jia}}, \bibinfo
  {author} {\bibfnamefont {Z.}~\bibnamefont {Huang}}, \bibinfo {author}
  {\bibfnamefont {C.}~\bibnamefont {Shen}}, \bibinfo {author} {\bibfnamefont
  {J.}~\bibnamefont {Long}}, \bibinfo {author} {\bibfnamefont {B.}~\bibnamefont
  {Shi}}, \bibinfo {author} {\bibfnamefont {Y.-H.}\ \bibnamefont {Luo}},
  \bibinfo {author} {\bibfnamefont {L.}~\bibnamefont {Gao}}, \bibinfo {author}
  {\bibfnamefont {W.}~\bibnamefont {Sun}}, \bibinfo {author} {\bibfnamefont
  {H.}~\bibnamefont {Guo}}, \bibinfo {author} {\bibfnamefont {J.}~\bibnamefont
  {He}},\ and\ \bibinfo {author} {\bibfnamefont {J.}~\bibnamefont {Liu}},\
  }\bibfield  {title} {\bibinfo {title} {Foundry manufacturing of
  tight-confinement, dispersion-engineered, ultralow-loss silicon nitride
  photonic integrated circuits},\ }\href {https://doi.org/10.1364/PRJ.486379}
  {\bibfield  {journal} {\bibinfo  {journal} {Photon. Res.}\ }\textbf {\bibinfo
  {volume} {11}},\ \bibinfo {pages} {558} (\bibinfo {year} {2023})}\BibitemShut
  {NoStop}%
\bibitem [{\citenamefont {Girardi}\ \emph {et~al.}(2025)\citenamefont
  {Girardi}, \citenamefont {\'{O}skar B.~Helgason}, \citenamefont
  {L\'{o}pez-Ortega}, \citenamefont {Rebolledo-Salgado},\ and\ \citenamefont
  {Torres-Company}}]{Girardi:25}%
  \BibitemOpen
  \bibfield  {author} {\bibinfo {author} {\bibfnamefont {M.}~\bibnamefont
  {Girardi}}, \bibinfo {author} {\bibnamefont {\'{O}skar B.~Helgason}},
  \bibinfo {author} {\bibfnamefont {C.~H.}\ \bibnamefont {L\'{o}pez-Ortega}},
  \bibinfo {author} {\bibfnamefont {I.}~\bibnamefont {Rebolledo-Salgado}},\
  and\ \bibinfo {author} {\bibfnamefont {V.}~\bibnamefont {Torres-Company}},\
  }\bibfield  {title} {\bibinfo {title} {Superefficient microcombs at the wafer
  level},\ }\href {https://doi.org/10.1364/OE.563489} {\bibfield  {journal}
  {\bibinfo  {journal} {Opt. Express}\ }\textbf {\bibinfo {volume} {33}},\
  \bibinfo {pages} {27451} (\bibinfo {year} {2025})}\BibitemShut {NoStop}%
\bibitem [{\citenamefont {Marin-Palomo}\ \emph {et~al.}(2017)\citenamefont
  {Marin-Palomo}, \citenamefont {Kemal}, \citenamefont {Karpov}, \citenamefont
  {Kordts}, \citenamefont {Pfeifle}, \citenamefont {Pfeiffer}, \citenamefont
  {Trocha}, \citenamefont {Wolf}, \citenamefont {Brasch}, \citenamefont
  {Anderson}, \citenamefont {Rosenberger}, \citenamefont {Vijayan},
  \citenamefont {Freude}, \citenamefont {Kippenberg},\ and\ \citenamefont
  {Koos}}]{Marin-Palomo:17}%
  \BibitemOpen
  \bibfield  {author} {\bibinfo {author} {\bibfnamefont {P.}~\bibnamefont
  {Marin-Palomo}}, \bibinfo {author} {\bibfnamefont {J.~N.}\ \bibnamefont
  {Kemal}}, \bibinfo {author} {\bibfnamefont {M.}~\bibnamefont {Karpov}},
  \bibinfo {author} {\bibfnamefont {A.}~\bibnamefont {Kordts}}, \bibinfo
  {author} {\bibfnamefont {J.}~\bibnamefont {Pfeifle}}, \bibinfo {author}
  {\bibfnamefont {M.~H.~P.}\ \bibnamefont {Pfeiffer}}, \bibinfo {author}
  {\bibfnamefont {P.}~\bibnamefont {Trocha}}, \bibinfo {author} {\bibfnamefont
  {S.}~\bibnamefont {Wolf}}, \bibinfo {author} {\bibfnamefont {V.}~\bibnamefont
  {Brasch}}, \bibinfo {author} {\bibfnamefont {M.~H.}\ \bibnamefont
  {Anderson}}, \bibinfo {author} {\bibfnamefont {R.}~\bibnamefont
  {Rosenberger}}, \bibinfo {author} {\bibfnamefont {K.}~\bibnamefont
  {Vijayan}}, \bibinfo {author} {\bibfnamefont {W.}~\bibnamefont {Freude}},
  \bibinfo {author} {\bibfnamefont {T.~J.}\ \bibnamefont {Kippenberg}},\ and\
  \bibinfo {author} {\bibfnamefont {C.}~\bibnamefont {Koos}},\ }\bibfield
  {title} {\bibinfo {title} {Microresonator-based solitons for massively
  parallel coherent optical communications},\ }\href
  {https://doi.org/10.1038/nature22387} {\bibfield  {journal} {\bibinfo
  {journal} {Nature}\ }\textbf {\bibinfo {volume} {546}},\ \bibinfo {pages}
  {274} (\bibinfo {year} {2017})}\BibitemShut {NoStop}%
\bibitem [{\citenamefont {Corcoran}\ \emph {et~al.}(2020)\citenamefont
  {Corcoran}, \citenamefont {Tan}, \citenamefont {Xu}, \citenamefont {Boes},
  \citenamefont {Wu}, \citenamefont {Nguyen}, \citenamefont {Chu},
  \citenamefont {Little}, \citenamefont {Morandotti}, \citenamefont
  {Mitchell},\ and\ \citenamefont {Moss}}]{Corcoran:20}%
  \BibitemOpen
  \bibfield  {author} {\bibinfo {author} {\bibfnamefont {B.}~\bibnamefont
  {Corcoran}}, \bibinfo {author} {\bibfnamefont {M.}~\bibnamefont {Tan}},
  \bibinfo {author} {\bibfnamefont {X.}~\bibnamefont {Xu}}, \bibinfo {author}
  {\bibfnamefont {A.}~\bibnamefont {Boes}}, \bibinfo {author} {\bibfnamefont
  {J.}~\bibnamefont {Wu}}, \bibinfo {author} {\bibfnamefont {T.~G.}\
  \bibnamefont {Nguyen}}, \bibinfo {author} {\bibfnamefont {S.~T.}\
  \bibnamefont {Chu}}, \bibinfo {author} {\bibfnamefont {B.~E.}\ \bibnamefont
  {Little}}, \bibinfo {author} {\bibfnamefont {R.}~\bibnamefont {Morandotti}},
  \bibinfo {author} {\bibfnamefont {A.}~\bibnamefont {Mitchell}},\ and\
  \bibinfo {author} {\bibfnamefont {D.~J.}\ \bibnamefont {Moss}},\ }\bibfield
  {title} {\bibinfo {title} {Ultra-dense optical data transmission over
  standard fibre with a single chip source},\ }\href
  {https://doi.org/10.1038/s41467-020-16265-x} {\bibfield  {journal} {\bibinfo
  {journal} {Nature Communications}\ }\textbf {\bibinfo {volume} {11}},\
  \bibinfo {pages} {2568} (\bibinfo {year} {2020})}\BibitemShut {NoStop}%
\bibitem [{\citenamefont {Mazur}\ \emph {et~al.}(2021)\citenamefont {Mazur},
  \citenamefont {Suh}, \citenamefont {F{\"u}l{\"o}p}, \citenamefont
  {Schr{\"o}der}, \citenamefont {Torres-Company}, \citenamefont {Karlsson},
  \citenamefont {Vahala},\ and\ \citenamefont {Andrekson}}]{Mazur:21}%
  \BibitemOpen
  \bibfield  {author} {\bibinfo {author} {\bibfnamefont {M.}~\bibnamefont
  {Mazur}}, \bibinfo {author} {\bibfnamefont {M.-G.}\ \bibnamefont {Suh}},
  \bibinfo {author} {\bibfnamefont {A.}~\bibnamefont {F{\"u}l{\"o}p}}, \bibinfo
  {author} {\bibfnamefont {J.}~\bibnamefont {Schr{\"o}der}}, \bibinfo {author}
  {\bibfnamefont {V.}~\bibnamefont {Torres-Company}}, \bibinfo {author}
  {\bibfnamefont {M.}~\bibnamefont {Karlsson}}, \bibinfo {author}
  {\bibfnamefont {K.}~\bibnamefont {Vahala}},\ and\ \bibinfo {author}
  {\bibfnamefont {P.}~\bibnamefont {Andrekson}},\ }\bibfield  {title} {\bibinfo
  {title} {High spectral efficiency coherent superchannel transmission with
  soliton microcombs},\ }\href {https://doi.org/10.1109/JLT.2021.3073567}
  {\bibfield  {journal} {\bibinfo  {journal} {Journal of Lightwave Technology}\
  }\textbf {\bibinfo {volume} {39}},\ \bibinfo {pages} {4367} (\bibinfo {year}
  {2021})}\BibitemShut {NoStop}%
\bibitem [{\citenamefont {Wang}\ \emph {et~al.}(2020)\citenamefont {Wang},
  \citenamefont {Wang}, \citenamefont {Niu}, \citenamefont {Wang},
  \citenamefont {Zou}, \citenamefont {Dong}, \citenamefont {Little},
  \citenamefont {Chu}, \citenamefont {Liu}, \citenamefont {Hao}, \citenamefont
  {Liu}, \citenamefont {Wang}, \citenamefont {Yin}, \citenamefont {He},
  \citenamefont {Zhang}, \citenamefont {Zhao}, \citenamefont {Han},
  \citenamefont {Guo},\ and\ \citenamefont {Chen}}]{DKS_BB84_2019}%
  \BibitemOpen
  \bibfield  {author} {\bibinfo {author} {\bibfnamefont {F.-X.}\ \bibnamefont
  {Wang}}, \bibinfo {author} {\bibfnamefont {W.}~\bibnamefont {Wang}}, \bibinfo
  {author} {\bibfnamefont {R.}~\bibnamefont {Niu}}, \bibinfo {author}
  {\bibfnamefont {X.}~\bibnamefont {Wang}}, \bibinfo {author} {\bibfnamefont
  {C.-L.}\ \bibnamefont {Zou}}, \bibinfo {author} {\bibfnamefont {C.-H.}\
  \bibnamefont {Dong}}, \bibinfo {author} {\bibfnamefont {B.~E.}\ \bibnamefont
  {Little}}, \bibinfo {author} {\bibfnamefont {S.~T.}\ \bibnamefont {Chu}},
  \bibinfo {author} {\bibfnamefont {H.}~\bibnamefont {Liu}}, \bibinfo {author}
  {\bibfnamefont {P.}~\bibnamefont {Hao}}, \bibinfo {author} {\bibfnamefont
  {S.}~\bibnamefont {Liu}}, \bibinfo {author} {\bibfnamefont {S.}~\bibnamefont
  {Wang}}, \bibinfo {author} {\bibfnamefont {Z.-Q.}\ \bibnamefont {Yin}},
  \bibinfo {author} {\bibfnamefont {D.-Y.}\ \bibnamefont {He}}, \bibinfo
  {author} {\bibfnamefont {W.}~\bibnamefont {Zhang}}, \bibinfo {author}
  {\bibfnamefont {W.}~\bibnamefont {Zhao}}, \bibinfo {author} {\bibfnamefont
  {Z.-F.}\ \bibnamefont {Han}}, \bibinfo {author} {\bibfnamefont {G.-C.}\
  \bibnamefont {Guo}},\ and\ \bibinfo {author} {\bibfnamefont {W.}~\bibnamefont
  {Chen}},\ }\bibfield  {title} {\bibinfo {title} {Quantum key distribution
  with on-chip dissipative kerr soliton},\ }\href
  {https://doi.org/https://doi.org/10.1002/lpor.201900190} {\bibfield
  {journal} {\bibinfo  {journal} {Laser \& Photonics Reviews}\ }\textbf
  {\bibinfo {volume} {14}},\ \bibinfo {pages} {1900190} (\bibinfo {year}
  {2020})}\BibitemShut {NoStop}%
\bibitem [{\citenamefont {Huang}\ \emph {et~al.}(2025)\citenamefont {Huang},
  \citenamefont {Wang}, \citenamefont {Wang}, \citenamefont {Wang},
  \citenamefont {Zou}, \citenamefont {Tang}, \citenamefont {Little},
  \citenamefont {Zhao}, \citenamefont {Han}, \citenamefont {Yang},
  \citenamefont {Wang}, \citenamefont {Chen},\ and\ \citenamefont
  {Zhang}}]{doi:10.1126/sciadv.adq8982}%
  \BibitemOpen
  \bibfield  {author} {\bibinfo {author} {\bibfnamefont {L.}~\bibnamefont
  {Huang}}, \bibinfo {author} {\bibfnamefont {W.}~\bibnamefont {Wang}},
  \bibinfo {author} {\bibfnamefont {F.}~\bibnamefont {Wang}}, \bibinfo {author}
  {\bibfnamefont {Y.}~\bibnamefont {Wang}}, \bibinfo {author} {\bibfnamefont
  {C.}~\bibnamefont {Zou}}, \bibinfo {author} {\bibfnamefont {L.}~\bibnamefont
  {Tang}}, \bibinfo {author} {\bibfnamefont {B.~E.}\ \bibnamefont {Little}},
  \bibinfo {author} {\bibfnamefont {W.}~\bibnamefont {Zhao}}, \bibinfo {author}
  {\bibfnamefont {Z.}~\bibnamefont {Han}}, \bibinfo {author} {\bibfnamefont
  {J.}~\bibnamefont {Yang}}, \bibinfo {author} {\bibfnamefont {G.}~\bibnamefont
  {Wang}}, \bibinfo {author} {\bibfnamefont {W.}~\bibnamefont {Chen}},\ and\
  \bibinfo {author} {\bibfnamefont {W.}~\bibnamefont {Zhang}},\ }\bibfield
  {title} {\bibinfo {title} {Massively parallel hong-ou-mandel interference
  based on independent soliton microcombs},\ }\href
  {https://doi.org/10.1126/sciadv.adq8982} {\bibfield  {journal} {\bibinfo
  {journal} {Science Advances}\ }\textbf {\bibinfo {volume} {11}},\ \bibinfo
  {pages} {eadq8982} (\bibinfo {year} {2025})},\ \Eprint
  {https://arxiv.org/abs/https://www.science.org/doi/pdf/10.1126/sciadv.adq8982}
  {https://www.science.org/doi/pdf/10.1126/sciadv.adq8982} \BibitemShut
  {NoStop}%
\bibitem [{\citenamefont {Yan}\ \emph {et~al.}(2025)\citenamefont {Yan},
  \citenamefont {Zheng}, \citenamefont {Wen}, \citenamefont {Lu}, \citenamefont
  {Du}, \citenamefont {Lu}, \citenamefont {Zhu},\ and\ \citenamefont
  {Ma}}]{yan2025measurement}%
  \BibitemOpen
  \bibfield  {author} {\bibinfo {author} {\bibfnamefont {W.}~\bibnamefont
  {Yan}}, \bibinfo {author} {\bibfnamefont {X.}~\bibnamefont {Zheng}}, \bibinfo
  {author} {\bibfnamefont {W.}~\bibnamefont {Wen}}, \bibinfo {author}
  {\bibfnamefont {L.}~\bibnamefont {Lu}}, \bibinfo {author} {\bibfnamefont
  {Y.}~\bibnamefont {Du}}, \bibinfo {author} {\bibfnamefont {Y.-Q.}\
  \bibnamefont {Lu}}, \bibinfo {author} {\bibfnamefont {S.}~\bibnamefont
  {Zhu}},\ and\ \bibinfo {author} {\bibfnamefont {X.-S.}\ \bibnamefont {Ma}},\
  }\bibfield  {title} {\bibinfo {title} {A measurement-device-independent
  quantum key distribution network using optical frequency comb},\ }\href@noop
  {} {\bibfield  {journal} {\bibinfo  {journal} {npj Quantum Information}\
  }\textbf {\bibinfo {volume} {11}},\ \bibinfo {pages} {97} (\bibinfo {year}
  {2025})}\BibitemShut {NoStop}%
\bibitem [{\citenamefont {Zheng}\ \emph {et~al.}(2026)\citenamefont {Zheng},
  \citenamefont {Wang}, \citenamefont {Jia}, \citenamefont {Huang},
  \citenamefont {Yuan}, \citenamefont {Zhai}, \citenamefont {Dai},
  \citenamefont {Shi}, \citenamefont {Zhang}, \citenamefont {Zhang},
  \citenamefont {Zhuang}, \citenamefont {Liu}, \citenamefont {Mao},
  \citenamefont {Dai}, \citenamefont {Fu}, \citenamefont {Jiao}, \citenamefont
  {Shi}, \citenamefont {Dai}, \citenamefont {Wang}, \citenamefont {Li},
  \citenamefont {Gong}, \citenamefont {Yuan}, \citenamefont {Chang},\ and\
  \citenamefont {Wang}}]{microcomb_TFQKDnet2026}%
  \BibitemOpen
  \bibfield  {author} {\bibinfo {author} {\bibfnamefont {Y.}~\bibnamefont
  {Zheng}}, \bibinfo {author} {\bibfnamefont {H.}~\bibnamefont {Wang}},
  \bibinfo {author} {\bibfnamefont {X.}~\bibnamefont {Jia}}, \bibinfo {author}
  {\bibfnamefont {J.}~\bibnamefont {Huang}}, \bibinfo {author} {\bibfnamefont
  {H.}~\bibnamefont {Yuan}}, \bibinfo {author} {\bibfnamefont {C.}~\bibnamefont
  {Zhai}}, \bibinfo {author} {\bibfnamefont {J.}~\bibnamefont {Dai}}, \bibinfo
  {author} {\bibfnamefont {J.}~\bibnamefont {Shi}}, \bibinfo {author}
  {\bibfnamefont {L.}~\bibnamefont {Zhang}}, \bibinfo {author} {\bibfnamefont
  {X.}~\bibnamefont {Zhang}}, \bibinfo {author} {\bibfnamefont
  {M.}~\bibnamefont {Zhuang}}, \bibinfo {author} {\bibfnamefont
  {J.}~\bibnamefont {Liu}}, \bibinfo {author} {\bibfnamefont {J.}~\bibnamefont
  {Mao}}, \bibinfo {author} {\bibfnamefont {T.}~\bibnamefont {Dai}}, \bibinfo
  {author} {\bibfnamefont {Z.}~\bibnamefont {Fu}}, \bibinfo {author}
  {\bibfnamefont {Y.}~\bibnamefont {Jiao}}, \bibinfo {author} {\bibfnamefont
  {Y.}~\bibnamefont {Shi}}, \bibinfo {author} {\bibfnamefont {D.}~\bibnamefont
  {Dai}}, \bibinfo {author} {\bibfnamefont {X.}~\bibnamefont {Wang}}, \bibinfo
  {author} {\bibfnamefont {Y.}~\bibnamefont {Li}}, \bibinfo {author}
  {\bibfnamefont {Q.}~\bibnamefont {Gong}}, \bibinfo {author} {\bibfnamefont
  {Z.}~\bibnamefont {Yuan}}, \bibinfo {author} {\bibfnamefont {L.}~\bibnamefont
  {Chang}},\ and\ \bibinfo {author} {\bibfnamefont {J.}~\bibnamefont {Wang}},\
  }\bibfield  {title} {\bibinfo {title} {Large-scale quantum communication
  networks with integrated photonics},\ }\bibfield  {journal} {\bibinfo
  {journal} {Nature}\ }\href {https://doi.org/10.1038/s41586-026-10152-z}
  {10.1038/s41586-026-10152-z} (\bibinfo {year} {2026})\BibitemShut {NoStop}%
\bibitem [{\citenamefont {Gisin}\ \emph {et~al.}(2006)\citenamefont {Gisin},
  \citenamefont {Fasel}, \citenamefont {Kraus}, \citenamefont {Zbinden},\ and\
  \citenamefont {Ribordy}}]{gisin2006trojan}%
  \BibitemOpen
  \bibfield  {author} {\bibinfo {author} {\bibfnamefont {N.}~\bibnamefont
  {Gisin}}, \bibinfo {author} {\bibfnamefont {S.}~\bibnamefont {Fasel}},
  \bibinfo {author} {\bibfnamefont {B.}~\bibnamefont {Kraus}}, \bibinfo
  {author} {\bibfnamefont {H.}~\bibnamefont {Zbinden}},\ and\ \bibinfo {author}
  {\bibfnamefont {G.}~\bibnamefont {Ribordy}},\ }\bibfield  {title} {\bibinfo
  {title} {Trojan-horse attacks on quantum-key-distribution systems},\
  }\href@noop {} {\bibfield  {journal} {\bibinfo  {journal} {Physical Review
  A—Atomic, Molecular, and Optical Physics}\ }\textbf {\bibinfo {volume}
  {73}},\ \bibinfo {pages} {022320} (\bibinfo {year} {2006})}\BibitemShut
  {NoStop}%
\bibitem [{\citenamefont {Jain}\ \emph {et~al.}(2014)\citenamefont {Jain},
  \citenamefont {Anisimova}, \citenamefont {Khan}, \citenamefont {Makarov},
  \citenamefont {Marquardt},\ and\ \citenamefont {Leuchs}}]{jain2014trojan}%
  \BibitemOpen
  \bibfield  {author} {\bibinfo {author} {\bibfnamefont {N.}~\bibnamefont
  {Jain}}, \bibinfo {author} {\bibfnamefont {E.}~\bibnamefont {Anisimova}},
  \bibinfo {author} {\bibfnamefont {I.}~\bibnamefont {Khan}}, \bibinfo {author}
  {\bibfnamefont {V.}~\bibnamefont {Makarov}}, \bibinfo {author} {\bibfnamefont
  {C.}~\bibnamefont {Marquardt}},\ and\ \bibinfo {author} {\bibfnamefont
  {G.}~\bibnamefont {Leuchs}},\ }\bibfield  {title} {\bibinfo {title}
  {Trojan-horse attacks threaten the security of practical quantum
  cryptography},\ }\href@noop {} {\bibfield  {journal} {\bibinfo  {journal}
  {New Journal of Physics}\ }\textbf {\bibinfo {volume} {16}},\ \bibinfo
  {pages} {123030} (\bibinfo {year} {2014})}\BibitemShut {NoStop}%
\bibitem [{\citenamefont {Wiesemann}\ \emph {et~al.}(2025)\citenamefont
  {Wiesemann}, \citenamefont {Gr{\"u}nenfelder}, \citenamefont
  {Bl{\'a}zquez~Co{\'\i}do}, \citenamefont {Walenta},\ and\ \citenamefont
  {Rusca}}]{wiesemann2025evaluation}%
  \BibitemOpen
  \bibfield  {author} {\bibinfo {author} {\bibfnamefont {J.}~\bibnamefont
  {Wiesemann}}, \bibinfo {author} {\bibfnamefont {F.}~\bibnamefont
  {Gr{\"u}nenfelder}}, \bibinfo {author} {\bibfnamefont {A.}~\bibnamefont
  {Bl{\'a}zquez~Co{\'\i}do}}, \bibinfo {author} {\bibfnamefont
  {N.}~\bibnamefont {Walenta}},\ and\ \bibinfo {author} {\bibfnamefont
  {D.}~\bibnamefont {Rusca}},\ }\bibfield  {title} {\bibinfo {title}
  {Evaluation of quantum key distribution systems against injection-locking
  attacks},\ }\href@noop {} {\bibfield  {journal} {\bibinfo  {journal} {APL
  photonics}\ }\textbf {\bibinfo {volume} {10}} (\bibinfo {year}
  {2025})}\BibitemShut {NoStop}%
\bibitem [{\citenamefont {Ju{\'a}rez}\ \emph {et~al.}(2026)\citenamefont
  {Ju{\'a}rez}, \citenamefont {Marcomini}, \citenamefont {Petrov},
  \citenamefont {Woodward}, \citenamefont {Dowling}, \citenamefont {Stevenson},
  \citenamefont {Curty},\ and\ \citenamefont {Rusca}}]{juarez2026reference}%
  \BibitemOpen
  \bibfield  {author} {\bibinfo {author} {\bibfnamefont {S.}~\bibnamefont
  {Ju{\'a}rez}}, \bibinfo {author} {\bibfnamefont {A.}~\bibnamefont
  {Marcomini}}, \bibinfo {author} {\bibfnamefont {M.}~\bibnamefont {Petrov}},
  \bibinfo {author} {\bibfnamefont {R.~I.}\ \bibnamefont {Woodward}}, \bibinfo
  {author} {\bibfnamefont {T.~J.}\ \bibnamefont {Dowling}}, \bibinfo {author}
  {\bibfnamefont {R.~M.}\ \bibnamefont {Stevenson}}, \bibinfo {author}
  {\bibfnamefont {M.}~\bibnamefont {Curty}},\ and\ \bibinfo {author}
  {\bibfnamefont {D.}~\bibnamefont {Rusca}},\ }\bibfield  {title} {\bibinfo
  {title} {Reference-beam attacks against twin-field quantum key distribution
  using optical injection locking},\ }\href@noop {} {\bibfield  {journal}
  {\bibinfo  {journal} {Physical Review A}\ }\textbf {\bibinfo {volume}
  {113}},\ \bibinfo {pages} {032613} (\bibinfo {year} {2026})}\BibitemShut
  {NoStop}%
\bibitem [{\citenamefont {Guo}\ \emph {et~al.}(2017)\citenamefont {Guo},
  \citenamefont {Karpov}, \citenamefont {Lucas}, \citenamefont {Kordts},
  \citenamefont {Pfeiffer}, \citenamefont {Brasch}, \citenamefont {Lihachev},
  \citenamefont {Lobanov}, \citenamefont {Gorodetsky},\ and\ \citenamefont
  {Kippenberg}}]{Guo2017}%
  \BibitemOpen
  \bibfield  {author} {\bibinfo {author} {\bibfnamefont {H.}~\bibnamefont
  {Guo}}, \bibinfo {author} {\bibfnamefont {M.}~\bibnamefont {Karpov}},
  \bibinfo {author} {\bibfnamefont {E.}~\bibnamefont {Lucas}}, \bibinfo
  {author} {\bibfnamefont {A.}~\bibnamefont {Kordts}}, \bibinfo {author}
  {\bibfnamefont {M.~H.~P.}\ \bibnamefont {Pfeiffer}}, \bibinfo {author}
  {\bibfnamefont {V.}~\bibnamefont {Brasch}}, \bibinfo {author} {\bibfnamefont
  {G.}~\bibnamefont {Lihachev}}, \bibinfo {author} {\bibfnamefont {V.~E.}\
  \bibnamefont {Lobanov}}, \bibinfo {author} {\bibfnamefont {M.~L.}\
  \bibnamefont {Gorodetsky}},\ and\ \bibinfo {author} {\bibfnamefont {T.~J.}\
  \bibnamefont {Kippenberg}},\ }\bibfield  {title} {\bibinfo {title} {Universal
  dynamics and deterministic switching of dissipative kerr solitons in optical
  microresonators},\ }\href {https://doi.org/10.1038/nphys3893} {\bibfield
  {journal} {\bibinfo  {journal} {Nature Physics}\ }\textbf {\bibinfo {volume}
  {13}},\ \bibinfo {pages} {94} (\bibinfo {year} {2017})}\BibitemShut {NoStop}%
\bibitem [{\citenamefont {Zheng}\ \emph
  {et~al.}(2023{\natexlab{a}})\citenamefont {Zheng}, \citenamefont {Sun},
  \citenamefont {Ding}, \citenamefont {Wen}, \citenamefont {Chen},
  \citenamefont {Shi}, \citenamefont {Luo}, \citenamefont {Long}, \citenamefont
  {Shen}, \citenamefont {Meng}, \citenamefont {Guo},\ and\ \citenamefont
  {Liu}}]{Zheng2023}%
  \BibitemOpen
  \bibfield  {author} {\bibinfo {author} {\bibfnamefont {H.}~\bibnamefont
  {Zheng}}, \bibinfo {author} {\bibfnamefont {W.}~\bibnamefont {Sun}}, \bibinfo
  {author} {\bibfnamefont {X.}~\bibnamefont {Ding}}, \bibinfo {author}
  {\bibfnamefont {H.}~\bibnamefont {Wen}}, \bibinfo {author} {\bibfnamefont
  {R.}~\bibnamefont {Chen}}, \bibinfo {author} {\bibfnamefont {B.}~\bibnamefont
  {Shi}}, \bibinfo {author} {\bibfnamefont {Y.-H.}\ \bibnamefont {Luo}},
  \bibinfo {author} {\bibfnamefont {J.}~\bibnamefont {Long}}, \bibinfo {author}
  {\bibfnamefont {C.}~\bibnamefont {Shen}}, \bibinfo {author} {\bibfnamefont
  {S.}~\bibnamefont {Meng}}, \bibinfo {author} {\bibfnamefont {H.}~\bibnamefont
  {Guo}},\ and\ \bibinfo {author} {\bibfnamefont {J.}~\bibnamefont {Liu}},\
  }\bibfield  {title} {\bibinfo {title} {Programmable access to microresonator
  solitons with modulational sideband heating},\ }\href
  {https://doi.org/10.1063/5.0173243} {\bibfield  {journal} {\bibinfo
  {journal} {APL Photonics}\ }\textbf {\bibinfo {volume} {8}},\ \bibinfo
  {pages} {126110} (\bibinfo {year} {2023}{\natexlab{a}})}\BibitemShut
  {NoStop}%
\bibitem [{\citenamefont {Lei}\ \emph {et~al.}(2022)\citenamefont {Lei},
  \citenamefont {Ye}, \citenamefont {Helgason}, \citenamefont {F{\"u}l{\"o}p},
  \citenamefont {Girardi},\ and\ \citenamefont {Torres-Company}}]{Lei2022}%
  \BibitemOpen
  \bibfield  {author} {\bibinfo {author} {\bibfnamefont {F.}~\bibnamefont
  {Lei}}, \bibinfo {author} {\bibfnamefont {Z.}~\bibnamefont {Ye}}, \bibinfo
  {author} {\bibfnamefont {{\'O}.~B.}\ \bibnamefont {Helgason}}, \bibinfo
  {author} {\bibfnamefont {A.}~\bibnamefont {F{\"u}l{\"o}p}}, \bibinfo {author}
  {\bibfnamefont {M.}~\bibnamefont {Girardi}},\ and\ \bibinfo {author}
  {\bibfnamefont {V.}~\bibnamefont {Torres-Company}},\ }\bibfield  {title}
  {\bibinfo {title} {Optical linewidth of soliton microcombs},\ }\href
  {https://doi.org/10.1038/s41467-022-30726-5} {\bibfield  {journal} {\bibinfo
  {journal} {Nature Communications}\ }\textbf {\bibinfo {volume} {13}},\
  \bibinfo {pages} {3161} (\bibinfo {year} {2022})}\BibitemShut {NoStop}%
\bibitem [{\citenamefont {Jiang}\ \emph {et~al.}(2020)\citenamefont {Jiang},
  \citenamefont {Hu}, \citenamefont {Xu}, \citenamefont {Yu},\ and\
  \citenamefont {Wang}}]{jiang_2020}%
  \BibitemOpen
  \bibfield  {author} {\bibinfo {author} {\bibfnamefont {C.}~\bibnamefont
  {Jiang}}, \bibinfo {author} {\bibfnamefont {X.-L.}\ \bibnamefont {Hu}},
  \bibinfo {author} {\bibfnamefont {H.}~\bibnamefont {Xu}}, \bibinfo {author}
  {\bibfnamefont {Z.-W.}\ \bibnamefont {Yu}},\ and\ \bibinfo {author}
  {\bibfnamefont {X.-B.}\ \bibnamefont {Wang}},\ }\bibfield  {title} {\bibinfo
  {title} {Zigzag approach to higher key rate of sending-or-not-sending twin
  field quantum key distribution with finite-key effects},\ }\href
  {https://doi.org/10.1088/1367-2630/ab81b7} {\bibfield  {journal} {\bibinfo
  {journal} {New Journal of Physics}\ }\textbf {\bibinfo {volume} {22}},\
  \bibinfo {pages} {053048} (\bibinfo {year} {2020})}\BibitemShut {NoStop}%
\bibitem [{\citenamefont {Jiang}\ \emph {et~al.}(2021)\citenamefont {Jiang},
  \citenamefont {Hu}, \citenamefont {Yu},\ and\ \citenamefont
  {Wang}}]{jiang_2021}%
  \BibitemOpen
  \bibfield  {author} {\bibinfo {author} {\bibfnamefont {C.}~\bibnamefont
  {Jiang}}, \bibinfo {author} {\bibfnamefont {X.-L.}\ \bibnamefont {Hu}},
  \bibinfo {author} {\bibfnamefont {Z.-W.}\ \bibnamefont {Yu}},\ and\ \bibinfo
  {author} {\bibfnamefont {X.-B.}\ \bibnamefont {Wang}},\ }\bibfield  {title}
  {\bibinfo {title} {Composable security for practical quantum key distribution
  with two way classical communication},\ }\href
  {https://doi.org/10.1088/1367-2630/ac0285} {\bibfield  {journal} {\bibinfo
  {journal} {New Journal of Physics}\ }\textbf {\bibinfo {volume} {23}},\
  \bibinfo {pages} {063038} (\bibinfo {year} {2021})}\BibitemShut {NoStop}%
\bibitem [{\citenamefont {Hu}\ \emph {et~al.}(2022)\citenamefont {Hu},
  \citenamefont {Jiang}, \citenamefont {Yu},\ and\ \citenamefont
  {Wang}}]{hu_2022}%
  \BibitemOpen
  \bibfield  {author} {\bibinfo {author} {\bibfnamefont {X.-L.}\ \bibnamefont
  {Hu}}, \bibinfo {author} {\bibfnamefont {C.}~\bibnamefont {Jiang}}, \bibinfo
  {author} {\bibfnamefont {Z.-W.}\ \bibnamefont {Yu}},\ and\ \bibinfo {author}
  {\bibfnamefont {X.-B.}\ \bibnamefont {Wang}},\ }\bibfield  {title} {\bibinfo
  {title} {Universal approach to sending-or-not-sending twin field quantum key
  distribution},\ }\href {https://doi.org/10.1088/2058-9565/ac8e90} {\bibfield
  {journal} {\bibinfo  {journal} {Quantum Science and Technology}\ }\textbf
  {\bibinfo {volume} {7}},\ \bibinfo {pages} {045031} (\bibinfo {year}
  {2022})}\BibitemShut {NoStop}%
\bibitem [{\citenamefont {Liu}\ \emph {et~al.}(2023{\natexlab{b}})\citenamefont
  {Liu}, \citenamefont {Ma}, \citenamefont {Ding}, \citenamefont {Zhang},
  \citenamefont {Zhou},\ and\ \citenamefont {Wang}}]{PhysRevA.108.022605}%
  \BibitemOpen
  \bibfield  {author} {\bibinfo {author} {\bibfnamefont {J.-Y.}\ \bibnamefont
  {Liu}}, \bibinfo {author} {\bibfnamefont {X.}~\bibnamefont {Ma}}, \bibinfo
  {author} {\bibfnamefont {H.-J.}\ \bibnamefont {Ding}}, \bibinfo {author}
  {\bibfnamefont {C.-H.}\ \bibnamefont {Zhang}}, \bibinfo {author}
  {\bibfnamefont {X.-Y.}\ \bibnamefont {Zhou}},\ and\ \bibinfo {author}
  {\bibfnamefont {Q.}~\bibnamefont {Wang}},\ }\bibfield  {title} {\bibinfo
  {title} {Experimental demonstration of five-intensity
  measurement-device-independent quantum key distribution over 442 km},\ }\href
  {https://doi.org/10.1103/PhysRevA.108.022605} {\bibfield  {journal} {\bibinfo
   {journal} {Phys. Rev. A}\ }\textbf {\bibinfo {volume} {108}},\ \bibinfo
  {pages} {022605} (\bibinfo {year} {2023}{\natexlab{b}})}\BibitemShut
  {NoStop}%
\bibitem [{\citenamefont {Shao}\ \emph {et~al.}(2025)\citenamefont {Shao},
  \citenamefont {Zhou}, \citenamefont {Lin}, \citenamefont {Minder},
  \citenamefont {Ge}, \citenamefont {Xie}, \citenamefont {Shen}, \citenamefont
  {Yan}, \citenamefont {Yin},\ and\ \citenamefont {Yuan}}]{PhysRevX.15.021066}%
  \BibitemOpen
  \bibfield  {author} {\bibinfo {author} {\bibfnamefont {S.-F.}\ \bibnamefont
  {Shao}}, \bibinfo {author} {\bibfnamefont {L.}~\bibnamefont {Zhou}}, \bibinfo
  {author} {\bibfnamefont {J.}~\bibnamefont {Lin}}, \bibinfo {author}
  {\bibfnamefont {M.}~\bibnamefont {Minder}}, \bibinfo {author} {\bibfnamefont
  {C.}~\bibnamefont {Ge}}, \bibinfo {author} {\bibfnamefont {Y.-M.}\
  \bibnamefont {Xie}}, \bibinfo {author} {\bibfnamefont {A.}~\bibnamefont
  {Shen}}, \bibinfo {author} {\bibfnamefont {Z.}~\bibnamefont {Yan}}, \bibinfo
  {author} {\bibfnamefont {H.-L.}\ \bibnamefont {Yin}},\ and\ \bibinfo {author}
  {\bibfnamefont {Z.}~\bibnamefont {Yuan}},\ }\bibfield  {title} {\bibinfo
  {title} {High-rate measurement-device-independent quantum communication
  without optical reference light},\ }\href
  {https://doi.org/10.1103/PhysRevX.15.021066} {\bibfield  {journal} {\bibinfo
  {journal} {Phys. Rev. X}\ }\textbf {\bibinfo {volume} {15}},\ \bibinfo
  {pages} {021066} (\bibinfo {year} {2025})}\BibitemShut {NoStop}%
\bibitem [{\citenamefont {Pittaluga}\ \emph {et~al.}(2021)\citenamefont
  {Pittaluga}, \citenamefont {Minder}, \citenamefont {Lucamarini},
  \citenamefont {Sanzaro}, \citenamefont {Woodward}, \citenamefont {Li},
  \citenamefont {Yuan},\ and\ \citenamefont {Shields}}]{Pittaluga2021}%
  \BibitemOpen
  \bibfield  {author} {\bibinfo {author} {\bibfnamefont {M.}~\bibnamefont
  {Pittaluga}}, \bibinfo {author} {\bibfnamefont {M.}~\bibnamefont {Minder}},
  \bibinfo {author} {\bibfnamefont {M.}~\bibnamefont {Lucamarini}}, \bibinfo
  {author} {\bibfnamefont {M.}~\bibnamefont {Sanzaro}}, \bibinfo {author}
  {\bibfnamefont {R.~I.}\ \bibnamefont {Woodward}}, \bibinfo {author}
  {\bibfnamefont {M.-J.}\ \bibnamefont {Li}}, \bibinfo {author} {\bibfnamefont
  {Z.}~\bibnamefont {Yuan}},\ and\ \bibinfo {author} {\bibfnamefont {A.~J.}\
  \bibnamefont {Shields}},\ }\bibfield  {title} {\bibinfo {title} {600-km
  repeater-like quantum communications with dual-band stabilization},\ }\href
  {https://doi.org/10.1038/s41566-021-00811-0} {\bibfield  {journal} {\bibinfo
  {journal} {Nature Photonics}\ }\textbf {\bibinfo {volume} {15}},\ \bibinfo
  {pages} {530} (\bibinfo {year} {2021})}\BibitemShut {NoStop}%
\bibitem [{\citenamefont {Liu}\ \emph {et~al.}(2023{\natexlab{c}})\citenamefont
  {Liu}, \citenamefont {Zhang}, \citenamefont {Jiang}, \citenamefont {Chen},
  \citenamefont {Ma}, \citenamefont {Zhang}, \citenamefont {Pan}, \citenamefont
  {Dong}, \citenamefont {Xiong}, \citenamefont {Zhang}, \citenamefont {Li},
  \citenamefont {Wang}, \citenamefont {Lu}, \citenamefont {Wu}, \citenamefont
  {Chen}, \citenamefont {You}, \citenamefont {Wang}, \citenamefont {Zhang},\
  and\ \citenamefont {Pan}}]{Liu2023_1000km_finite-key_analysis}%
  \BibitemOpen
  \bibfield  {author} {\bibinfo {author} {\bibfnamefont {Y.}~\bibnamefont
  {Liu}}, \bibinfo {author} {\bibfnamefont {W.-J.}\ \bibnamefont {Zhang}},
  \bibinfo {author} {\bibfnamefont {C.}~\bibnamefont {Jiang}}, \bibinfo
  {author} {\bibfnamefont {J.-P.}\ \bibnamefont {Chen}}, \bibinfo {author}
  {\bibfnamefont {D.}~\bibnamefont {Ma}}, \bibinfo {author} {\bibfnamefont
  {C.}~\bibnamefont {Zhang}}, \bibinfo {author} {\bibfnamefont {W.-X.}\
  \bibnamefont {Pan}}, \bibinfo {author} {\bibfnamefont {H.}~\bibnamefont
  {Dong}}, \bibinfo {author} {\bibfnamefont {J.-M.}\ \bibnamefont {Xiong}},
  \bibinfo {author} {\bibfnamefont {C.-J.}\ \bibnamefont {Zhang}}, \bibinfo
  {author} {\bibfnamefont {H.}~\bibnamefont {Li}}, \bibinfo {author}
  {\bibfnamefont {R.-C.}\ \bibnamefont {Wang}}, \bibinfo {author}
  {\bibfnamefont {C.-Y.}\ \bibnamefont {Lu}}, \bibinfo {author} {\bibfnamefont
  {J.}~\bibnamefont {Wu}}, \bibinfo {author} {\bibfnamefont {T.-Y.}\
  \bibnamefont {Chen}}, \bibinfo {author} {\bibfnamefont {L.}~\bibnamefont
  {You}}, \bibinfo {author} {\bibfnamefont {X.-B.}\ \bibnamefont {Wang}},
  \bibinfo {author} {\bibfnamefont {Q.}~\bibnamefont {Zhang}},\ and\ \bibinfo
  {author} {\bibfnamefont {J.-W.}\ \bibnamefont {Pan}},\ }\bibfield  {title}
  {\bibinfo {title} {1002 km twin-field quantum key distribution with
  finite-key analysis},\ }\href {https://doi.org/10.1007/s44214-023-00039-9}
  {\bibfield  {journal} {\bibinfo  {journal} {Quantum Frontiers}\ }\textbf
  {\bibinfo {volume} {2}},\ \bibinfo {pages} {16} (\bibinfo {year}
  {2023}{\natexlab{c}})}\BibitemShut {NoStop}%
\bibitem [{\citenamefont {Moille}\ \emph {et~al.}(2025)\citenamefont {Moille},
  \citenamefont {Shandilya}, \citenamefont {Stone}, \citenamefont {Menyuk},\
  and\ \citenamefont {Srinivasan}}]{moille2025all}%
  \BibitemOpen
  \bibfield  {author} {\bibinfo {author} {\bibfnamefont {G.}~\bibnamefont
  {Moille}}, \bibinfo {author} {\bibfnamefont {P.}~\bibnamefont {Shandilya}},
  \bibinfo {author} {\bibfnamefont {J.}~\bibnamefont {Stone}}, \bibinfo
  {author} {\bibfnamefont {C.}~\bibnamefont {Menyuk}},\ and\ \bibinfo {author}
  {\bibfnamefont {K.}~\bibnamefont {Srinivasan}},\ }\bibfield  {title}
  {\bibinfo {title} {All-optical noise quenching of an integrated frequency
  comb},\ }\href@noop {} {\bibfield  {journal} {\bibinfo  {journal} {Optica}\
  }\textbf {\bibinfo {volume} {12}},\ \bibinfo {pages} {1020} (\bibinfo {year}
  {2025})}\BibitemShut {NoStop}%
\bibitem [{\citenamefont {Lei}\ \emph {et~al.}(2024)\citenamefont {Lei},
  \citenamefont {Sun}, \citenamefont {Helgason}, \citenamefont {Ye},
  \citenamefont {Gao}, \citenamefont {Karlsson}, \citenamefont {Andrekson},\
  and\ \citenamefont {Torres-Company}}]{lei2024self}%
  \BibitemOpen
  \bibfield  {author} {\bibinfo {author} {\bibfnamefont {F.}~\bibnamefont
  {Lei}}, \bibinfo {author} {\bibfnamefont {Y.}~\bibnamefont {Sun}}, \bibinfo
  {author} {\bibfnamefont {{\'O}.~B.}\ \bibnamefont {Helgason}}, \bibinfo
  {author} {\bibfnamefont {Z.}~\bibnamefont {Ye}}, \bibinfo {author}
  {\bibfnamefont {Y.}~\bibnamefont {Gao}}, \bibinfo {author} {\bibfnamefont
  {M.}~\bibnamefont {Karlsson}}, \bibinfo {author} {\bibfnamefont {P.~A.}\
  \bibnamefont {Andrekson}},\ and\ \bibinfo {author} {\bibfnamefont
  {V.}~\bibnamefont {Torres-Company}},\ }\bibfield  {title} {\bibinfo {title}
  {Self-injection-locked optical parametric oscillator based on microcombs},\
  }\href@noop {} {\bibfield  {journal} {\bibinfo  {journal} {Optica}\ }\textbf
  {\bibinfo {volume} {11}},\ \bibinfo {pages} {420} (\bibinfo {year}
  {2024})}\BibitemShut {NoStop}%
\bibitem [{\citenamefont {Helgason}\ \emph {et~al.}(2023)\citenamefont
  {Helgason}, \citenamefont {Girardi}, \citenamefont {Ye}, \citenamefont {Lei},
  \citenamefont {Schr{\"o}der},\ and\ \citenamefont
  {{Torres-Company}}}]{HelgasonO:2023}%
  \BibitemOpen
  \bibfield  {author} {\bibinfo {author} {\bibfnamefont {{\'O}.~B.}\
  \bibnamefont {Helgason}}, \bibinfo {author} {\bibfnamefont {M.}~\bibnamefont
  {Girardi}}, \bibinfo {author} {\bibfnamefont {Z.}~\bibnamefont {Ye}},
  \bibinfo {author} {\bibfnamefont {F.}~\bibnamefont {Lei}}, \bibinfo {author}
  {\bibfnamefont {J.}~\bibnamefont {Schr{\"o}der}},\ and\ \bibinfo {author}
  {\bibfnamefont {V.}~\bibnamefont {{Torres-Company}}},\ }\bibfield  {title}
  {\bibinfo {title} {Surpassing the nonlinear conversion efficiency of soliton
  microcombs},\ }\href {https://doi.org/10.1038/s41566-023-01280-3} {\bibfield
  {journal} {\bibinfo  {journal} {Nature Photonics}\ }\textbf {\bibinfo
  {volume} {17}},\ \bibinfo {pages} {992} (\bibinfo {year} {2023})}\BibitemShut
  {NoStop}%
\bibitem [{\citenamefont {Kondratiev}\ \emph {et~al.}(2023)\citenamefont
  {Kondratiev}, \citenamefont {Lobanov}, \citenamefont {Shitikov},
  \citenamefont {Galiev}, \citenamefont {Chermoshentsev}, \citenamefont
  {Dmitriev}, \citenamefont {Danilin}, \citenamefont {Lonshakov}, \citenamefont
  {Min'kov}, \citenamefont {Sokol}, \citenamefont {Cordette}, \citenamefont
  {Luo}, \citenamefont {Liang}, \citenamefont {Liu},\ and\ \citenamefont
  {Bilenko}}]{Kondratiev:23}%
  \BibitemOpen
  \bibfield  {author} {\bibinfo {author} {\bibfnamefont {N.~M.}\ \bibnamefont
  {Kondratiev}}, \bibinfo {author} {\bibfnamefont {V.~E.}\ \bibnamefont
  {Lobanov}}, \bibinfo {author} {\bibfnamefont {A.~E.}\ \bibnamefont
  {Shitikov}}, \bibinfo {author} {\bibfnamefont {R.~R.}\ \bibnamefont
  {Galiev}}, \bibinfo {author} {\bibfnamefont {D.~A.}\ \bibnamefont
  {Chermoshentsev}}, \bibinfo {author} {\bibfnamefont {N.~Y.}\ \bibnamefont
  {Dmitriev}}, \bibinfo {author} {\bibfnamefont {A.~N.}\ \bibnamefont
  {Danilin}}, \bibinfo {author} {\bibfnamefont {E.~A.}\ \bibnamefont
  {Lonshakov}}, \bibinfo {author} {\bibfnamefont {K.~N.}\ \bibnamefont
  {Min'kov}}, \bibinfo {author} {\bibfnamefont {D.~M.}\ \bibnamefont {Sokol}},
  \bibinfo {author} {\bibfnamefont {S.~J.}\ \bibnamefont {Cordette}}, \bibinfo
  {author} {\bibfnamefont {Y.-H.}\ \bibnamefont {Luo}}, \bibinfo {author}
  {\bibfnamefont {W.}~\bibnamefont {Liang}}, \bibinfo {author} {\bibfnamefont
  {J.}~\bibnamefont {Liu}},\ and\ \bibinfo {author} {\bibfnamefont {I.~A.}\
  \bibnamefont {Bilenko}},\ }\bibfield  {title} {\bibinfo {title} {Recent
  advances in laser self-injection locking to high-q microresonators},\ }\href
  {https://doi.org/10.1007/s11467-022-1245-3} {\bibfield  {journal} {\bibinfo
  {journal} {Frontiers of Physics}\ }\textbf {\bibinfo {volume} {18}},\
  \bibinfo {pages} {21305} (\bibinfo {year} {2023})}\BibitemShut {NoStop}%
\bibitem [{\citenamefont {Stern}\ \emph {et~al.}(2018)\citenamefont {Stern},
  \citenamefont {Ji}, \citenamefont {Okawachi}, \citenamefont {Gaeta},\ and\
  \citenamefont {Lipson}}]{SternB:2018}%
  \BibitemOpen
  \bibfield  {author} {\bibinfo {author} {\bibfnamefont {B.}~\bibnamefont
  {Stern}}, \bibinfo {author} {\bibfnamefont {X.}~\bibnamefont {Ji}}, \bibinfo
  {author} {\bibfnamefont {Y.}~\bibnamefont {Okawachi}}, \bibinfo {author}
  {\bibfnamefont {A.~L.}\ \bibnamefont {Gaeta}},\ and\ \bibinfo {author}
  {\bibfnamefont {M.}~\bibnamefont {Lipson}},\ }\bibfield  {title} {\bibinfo
  {title} {Battery-operated integrated frequency comb generator},\ }\href
  {https://doi.org/10.1038/s41586-018-0598-9} {\bibfield  {journal} {\bibinfo
  {journal} {Nature}\ }\textbf {\bibinfo {volume} {562}},\ \bibinfo {pages}
  {401} (\bibinfo {year} {2018})}\BibitemShut {NoStop}%
\bibitem [{\citenamefont {Shen}\ \emph {et~al.}(2020)\citenamefont {Shen},
  \citenamefont {Chang}, \citenamefont {Liu}, \citenamefont {Wang},
  \citenamefont {Yang}, \citenamefont {Xiang}, \citenamefont {Wang},
  \citenamefont {He}, \citenamefont {Liu}, \citenamefont {Xie}, \citenamefont
  {Guo}, \citenamefont {Kinghorn}, \citenamefont {Wu}, \citenamefont {Ji},
  \citenamefont {Kippenberg}, \citenamefont {Vahala},\ and\ \citenamefont
  {Bowers}}]{Shen:20}%
  \BibitemOpen
  \bibfield  {author} {\bibinfo {author} {\bibfnamefont {B.}~\bibnamefont
  {Shen}}, \bibinfo {author} {\bibfnamefont {L.}~\bibnamefont {Chang}},
  \bibinfo {author} {\bibfnamefont {J.}~\bibnamefont {Liu}}, \bibinfo {author}
  {\bibfnamefont {H.}~\bibnamefont {Wang}}, \bibinfo {author} {\bibfnamefont
  {Q.-F.}\ \bibnamefont {Yang}}, \bibinfo {author} {\bibfnamefont
  {C.}~\bibnamefont {Xiang}}, \bibinfo {author} {\bibfnamefont {R.~N.}\
  \bibnamefont {Wang}}, \bibinfo {author} {\bibfnamefont {J.}~\bibnamefont
  {He}}, \bibinfo {author} {\bibfnamefont {T.}~\bibnamefont {Liu}}, \bibinfo
  {author} {\bibfnamefont {W.}~\bibnamefont {Xie}}, \bibinfo {author}
  {\bibfnamefont {J.}~\bibnamefont {Guo}}, \bibinfo {author} {\bibfnamefont
  {D.}~\bibnamefont {Kinghorn}}, \bibinfo {author} {\bibfnamefont
  {L.}~\bibnamefont {Wu}}, \bibinfo {author} {\bibfnamefont {Q.-X.}\
  \bibnamefont {Ji}}, \bibinfo {author} {\bibfnamefont {T.~J.}\ \bibnamefont
  {Kippenberg}}, \bibinfo {author} {\bibfnamefont {K.}~\bibnamefont {Vahala}},\
  and\ \bibinfo {author} {\bibfnamefont {J.~E.}\ \bibnamefont {Bowers}},\
  }\bibfield  {title} {\bibinfo {title} {Integrated turnkey soliton
  microcombs},\ }\href {https://doi.org/10.1038/s41586-020-2358-x} {\bibfield
  {journal} {\bibinfo  {journal} {Nature}\ }\textbf {\bibinfo {volume} {582}},\
  \bibinfo {pages} {365} (\bibinfo {year} {2020})}\BibitemShut {NoStop}%
\bibitem [{\citenamefont {Xiang}\ \emph {et~al.}(2021)\citenamefont {Xiang},
  \citenamefont {Liu}, \citenamefont {Guo}, \citenamefont {Chang},
  \citenamefont {Wang}, \citenamefont {Weng}, \citenamefont {Peters},
  \citenamefont {Xie}, \citenamefont {Zhang}, \citenamefont {Riemensberger},
  \citenamefont {Selvidge}, \citenamefont {Kippenberg},\ and\ \citenamefont
  {Bowers}}]{Xiang:21}%
  \BibitemOpen
  \bibfield  {author} {\bibinfo {author} {\bibfnamefont {C.}~\bibnamefont
  {Xiang}}, \bibinfo {author} {\bibfnamefont {J.}~\bibnamefont {Liu}}, \bibinfo
  {author} {\bibfnamefont {J.}~\bibnamefont {Guo}}, \bibinfo {author}
  {\bibfnamefont {L.}~\bibnamefont {Chang}}, \bibinfo {author} {\bibfnamefont
  {R.~N.}\ \bibnamefont {Wang}}, \bibinfo {author} {\bibfnamefont
  {W.}~\bibnamefont {Weng}}, \bibinfo {author} {\bibfnamefont {J.}~\bibnamefont
  {Peters}}, \bibinfo {author} {\bibfnamefont {W.}~\bibnamefont {Xie}},
  \bibinfo {author} {\bibfnamefont {Z.}~\bibnamefont {Zhang}}, \bibinfo
  {author} {\bibfnamefont {J.}~\bibnamefont {Riemensberger}}, \bibinfo {author}
  {\bibfnamefont {J.}~\bibnamefont {Selvidge}}, \bibinfo {author}
  {\bibfnamefont {T.~J.}\ \bibnamefont {Kippenberg}},\ and\ \bibinfo {author}
  {\bibfnamefont {J.~E.}\ \bibnamefont {Bowers}},\ }\bibfield  {title}
  {\bibinfo {title} {Laser soliton microcombs heterogeneously integrated on
  silicon},\ }\href {https://doi.org/10.1126/science.abh2076} {\bibfield
  {journal} {\bibinfo  {journal} {Science}\ }\textbf {\bibinfo {volume}
  {373}},\ \bibinfo {pages} {99} (\bibinfo {year} {2021})}\BibitemShut
  {NoStop}%
\bibitem [{\citenamefont {Sun}\ \emph {et~al.}(2025{\natexlab{a}})\citenamefont
  {Sun}, \citenamefont {Chen}, \citenamefont {Li}, \citenamefont {Shen},
  \citenamefont {Yu}, \citenamefont {Li}, \citenamefont {Long}, \citenamefont
  {Zheng}, \citenamefont {Wang}, \citenamefont {Long}, \citenamefont {Chen},
  \citenamefont {Zhang}, \citenamefont {Shi}, \citenamefont {Gao},
  \citenamefont {Luo}, \citenamefont {Chen},\ and\ \citenamefont
  {Liu}}]{Sun:25}%
  \BibitemOpen
  \bibfield  {author} {\bibinfo {author} {\bibfnamefont {W.}~\bibnamefont
  {Sun}}, \bibinfo {author} {\bibfnamefont {Z.}~\bibnamefont {Chen}}, \bibinfo
  {author} {\bibfnamefont {L.}~\bibnamefont {Li}}, \bibinfo {author}
  {\bibfnamefont {C.}~\bibnamefont {Shen}}, \bibinfo {author} {\bibfnamefont
  {K.}~\bibnamefont {Yu}}, \bibinfo {author} {\bibfnamefont {S.}~\bibnamefont
  {Li}}, \bibinfo {author} {\bibfnamefont {J.}~\bibnamefont {Long}}, \bibinfo
  {author} {\bibfnamefont {H.}~\bibnamefont {Zheng}}, \bibinfo {author}
  {\bibfnamefont {L.}~\bibnamefont {Wang}}, \bibinfo {author} {\bibfnamefont
  {T.}~\bibnamefont {Long}}, \bibinfo {author} {\bibfnamefont {Q.}~\bibnamefont
  {Chen}}, \bibinfo {author} {\bibfnamefont {Z.}~\bibnamefont {Zhang}},
  \bibinfo {author} {\bibfnamefont {B.}~\bibnamefont {Shi}}, \bibinfo {author}
  {\bibfnamefont {L.}~\bibnamefont {Gao}}, \bibinfo {author} {\bibfnamefont
  {Y.-H.}\ \bibnamefont {Luo}}, \bibinfo {author} {\bibfnamefont
  {B.}~\bibnamefont {Chen}},\ and\ \bibinfo {author} {\bibfnamefont
  {J.}~\bibnamefont {Liu}},\ }\bibfield  {title} {\bibinfo {title} {A
  chip-integrated comb-based microwave oscillator},\ }\href
  {https://doi.org/10.1038/s41377-025-01795-0} {\bibfield  {journal} {\bibinfo
  {journal} {Light: Science \& Applications}\ }\textbf {\bibinfo {volume}
  {14}},\ \bibinfo {pages} {179} (\bibinfo {year}
  {2025}{\natexlab{a}})}\BibitemShut {NoStop}%
\bibitem [{\citenamefont {Churaev}\ \emph {et~al.}(2023)\citenamefont
  {Churaev}, \citenamefont {Wang}, \citenamefont {Riedhauser}, \citenamefont
  {Snigirev}, \citenamefont {Bl{\'e}sin}, \citenamefont {M{\"o}hl},
  \citenamefont {Anderson}, \citenamefont {Siddharth}, \citenamefont {Popoff},
  \citenamefont {Drechsler}, \citenamefont {Caimi}, \citenamefont {H{\"o}nl},
  \citenamefont {Riemensberger}, \citenamefont {Liu}, \citenamefont {Seidler},\
  and\ \citenamefont {Kippenberg}}]{ChuraevM:2023}%
  \BibitemOpen
  \bibfield  {author} {\bibinfo {author} {\bibfnamefont {M.}~\bibnamefont
  {Churaev}}, \bibinfo {author} {\bibfnamefont {R.~N.}\ \bibnamefont {Wang}},
  \bibinfo {author} {\bibfnamefont {A.}~\bibnamefont {Riedhauser}}, \bibinfo
  {author} {\bibfnamefont {V.}~\bibnamefont {Snigirev}}, \bibinfo {author}
  {\bibfnamefont {T.}~\bibnamefont {Bl{\'e}sin}}, \bibinfo {author}
  {\bibfnamefont {C.}~\bibnamefont {M{\"o}hl}}, \bibinfo {author}
  {\bibfnamefont {M.~H.}\ \bibnamefont {Anderson}}, \bibinfo {author}
  {\bibfnamefont {A.}~\bibnamefont {Siddharth}}, \bibinfo {author}
  {\bibfnamefont {Y.}~\bibnamefont {Popoff}}, \bibinfo {author} {\bibfnamefont
  {U.}~\bibnamefont {Drechsler}}, \bibinfo {author} {\bibfnamefont
  {D.}~\bibnamefont {Caimi}}, \bibinfo {author} {\bibfnamefont
  {S.}~\bibnamefont {H{\"o}nl}}, \bibinfo {author} {\bibfnamefont
  {J.}~\bibnamefont {Riemensberger}}, \bibinfo {author} {\bibfnamefont
  {J.}~\bibnamefont {Liu}}, \bibinfo {author} {\bibfnamefont {P.}~\bibnamefont
  {Seidler}},\ and\ \bibinfo {author} {\bibfnamefont {T.~J.}\ \bibnamefont
  {Kippenberg}},\ }\bibfield  {title} {\bibinfo {title} {A heterogeneously
  integrated lithium niobate-on-silicon nitride photonic platform},\ }\href
  {https://doi.org/10.1038/s41467-023-39047-7} {\bibfield  {journal} {\bibinfo
  {journal} {Nature Communications}\ }\textbf {\bibinfo {volume} {14}},\
  \bibinfo {pages} {3499} (\bibinfo {year} {2023})}\BibitemShut {NoStop}%
\bibitem [{\citenamefont {Niels}\ \emph {et~al.}(2026)\citenamefont {Niels},
  \citenamefont {Vanackere}, \citenamefont {Vissers}, \citenamefont {Zhai},
  \citenamefont {Nenezic}, \citenamefont {Declercq}, \citenamefont
  {Bruynsteen}, \citenamefont {Niu}, \citenamefont {Moerman}, \citenamefont
  {Caytan}, \citenamefont {Singh}, \citenamefont {Lemey}, \citenamefont {Yin},
  \citenamefont {Janssen}, \citenamefont {Verheyen}, \citenamefont {Singh},
  \citenamefont {Bode}, \citenamefont {Davi}, \citenamefont {Ferraro},
  \citenamefont {Absil}, \citenamefont {Balakrishnan}, \citenamefont
  {Van~Campenhout}, \citenamefont {Roelkens}, \citenamefont {Kuyken},\ and\
  \citenamefont {Billet}}]{NielsM:2026}%
  \BibitemOpen
  \bibfield  {author} {\bibinfo {author} {\bibfnamefont {M.}~\bibnamefont
  {Niels}}, \bibinfo {author} {\bibfnamefont {T.}~\bibnamefont {Vanackere}},
  \bibinfo {author} {\bibfnamefont {E.}~\bibnamefont {Vissers}}, \bibinfo
  {author} {\bibfnamefont {T.}~\bibnamefont {Zhai}}, \bibinfo {author}
  {\bibfnamefont {P.}~\bibnamefont {Nenezic}}, \bibinfo {author} {\bibfnamefont
  {J.}~\bibnamefont {Declercq}}, \bibinfo {author} {\bibfnamefont
  {C.}~\bibnamefont {Bruynsteen}}, \bibinfo {author} {\bibfnamefont
  {S.}~\bibnamefont {Niu}}, \bibinfo {author} {\bibfnamefont {A.}~\bibnamefont
  {Moerman}}, \bibinfo {author} {\bibfnamefont {O.}~\bibnamefont {Caytan}},
  \bibinfo {author} {\bibfnamefont {N.}~\bibnamefont {Singh}}, \bibinfo
  {author} {\bibfnamefont {S.}~\bibnamefont {Lemey}}, \bibinfo {author}
  {\bibfnamefont {X.}~\bibnamefont {Yin}}, \bibinfo {author} {\bibfnamefont
  {S.}~\bibnamefont {Janssen}}, \bibinfo {author} {\bibfnamefont
  {P.}~\bibnamefont {Verheyen}}, \bibinfo {author} {\bibfnamefont
  {N.}~\bibnamefont {Singh}}, \bibinfo {author} {\bibfnamefont
  {D.}~\bibnamefont {Bode}}, \bibinfo {author} {\bibfnamefont {M.}~\bibnamefont
  {Davi}}, \bibinfo {author} {\bibfnamefont {F.}~\bibnamefont {Ferraro}},
  \bibinfo {author} {\bibfnamefont {P.}~\bibnamefont {Absil}}, \bibinfo
  {author} {\bibfnamefont {S.}~\bibnamefont {Balakrishnan}}, \bibinfo {author}
  {\bibfnamefont {J.}~\bibnamefont {Van~Campenhout}}, \bibinfo {author}
  {\bibfnamefont {G.}~\bibnamefont {Roelkens}}, \bibinfo {author}
  {\bibfnamefont {B.}~\bibnamefont {Kuyken}},\ and\ \bibinfo {author}
  {\bibfnamefont {M.}~\bibnamefont {Billet}},\ }\bibfield  {title} {\bibinfo
  {title} {A high-speed heterogeneous lithium tantalate silicon photonics
  platform},\ }\href {https://doi.org/10.1038/s41566-025-01832-9} {\bibfield
  {journal} {\bibinfo  {journal} {Nature Photonics}\ }\textbf {\bibinfo
  {volume} {20}},\ \bibinfo {pages} {225} (\bibinfo {year} {2026})}\BibitemShut
  {NoStop}%
\bibitem [{\citenamefont {Wang}\ \emph
  {et~al.}(2018{\natexlab{b}})\citenamefont {Wang}, \citenamefont {Yu},\ and\
  \citenamefont {Hu}}]{physreva.98.062323}%
  \BibitemOpen
  \bibfield  {author} {\bibinfo {author} {\bibfnamefont {X.-B.}\ \bibnamefont
  {Wang}}, \bibinfo {author} {\bibfnamefont {Z.-W.}\ \bibnamefont {Yu}},\ and\
  \bibinfo {author} {\bibfnamefont {X.-L.}\ \bibnamefont {Hu}},\ }\bibfield
  {title} {\bibinfo {title} {Twin-field quantum key distribution with large
  misalignment error},\ }\href {https://doi.org/10.1103/PhysRevA.98.062323}
  {\bibfield  {journal} {\bibinfo  {journal} {Phys. Rev. A}\ }\textbf {\bibinfo
  {volume} {98}},\ \bibinfo {pages} {062323} (\bibinfo {year}
  {2018}{\natexlab{b}})}\BibitemShut {NoStop}%
\bibitem [{\citenamefont {Vitanov}\ \emph {et~al.}(2013)\citenamefont
  {Vitanov}, \citenamefont {Dupuis}, \citenamefont {Tomamichel},\ and\
  \citenamefont {Renner}}]{vitanov2013chain}%
  \BibitemOpen
  \bibfield  {author} {\bibinfo {author} {\bibfnamefont {A.}~\bibnamefont
  {Vitanov}}, \bibinfo {author} {\bibfnamefont {F.}~\bibnamefont {Dupuis}},
  \bibinfo {author} {\bibfnamefont {M.}~\bibnamefont {Tomamichel}},\ and\
  \bibinfo {author} {\bibfnamefont {R.}~\bibnamefont {Renner}},\ }\bibfield
  {title} {\bibinfo {title} {Chain rules for smooth min- and max-entropies},\
  }\href {https://doi.org/10.1109/TIT.2013.2238656} {\bibfield  {journal}
  {\bibinfo  {journal} {IEEE Transactions on Information Theory}\ }\textbf
  {\bibinfo {volume} {59}},\ \bibinfo {pages} {2603} (\bibinfo {year}
  {2013})}\BibitemShut {NoStop}%
\bibitem [{\citenamefont {Sun}\ \emph {et~al.}(2025{\natexlab{b}})\citenamefont
  {Sun}, \citenamefont {Chen}, \citenamefont {Li}, \citenamefont {Shen},
  \citenamefont {Yu}, \citenamefont {Li}, \citenamefont {Long}, \citenamefont
  {Zheng}, \citenamefont {Wang}, \citenamefont {Long}, \citenamefont {Chen},
  \citenamefont {Zhang}, \citenamefont {Shi}, \citenamefont {Gao},
  \citenamefont {Luo}, \citenamefont {Chen},\ and\ \citenamefont
  {Liu}}]{Sun2025}%
  \BibitemOpen
  \bibfield  {author} {\bibinfo {author} {\bibfnamefont {W.}~\bibnamefont
  {Sun}}, \bibinfo {author} {\bibfnamefont {Z.}~\bibnamefont {Chen}}, \bibinfo
  {author} {\bibfnamefont {L.}~\bibnamefont {Li}}, \bibinfo {author}
  {\bibfnamefont {C.}~\bibnamefont {Shen}}, \bibinfo {author} {\bibfnamefont
  {K.}~\bibnamefont {Yu}}, \bibinfo {author} {\bibfnamefont {S.}~\bibnamefont
  {Li}}, \bibinfo {author} {\bibfnamefont {J.}~\bibnamefont {Long}}, \bibinfo
  {author} {\bibfnamefont {H.}~\bibnamefont {Zheng}}, \bibinfo {author}
  {\bibfnamefont {L.}~\bibnamefont {Wang}}, \bibinfo {author} {\bibfnamefont
  {T.}~\bibnamefont {Long}}, \bibinfo {author} {\bibfnamefont {Q.}~\bibnamefont
  {Chen}}, \bibinfo {author} {\bibfnamefont {Z.}~\bibnamefont {Zhang}},
  \bibinfo {author} {\bibfnamefont {B.}~\bibnamefont {Shi}}, \bibinfo {author}
  {\bibfnamefont {L.}~\bibnamefont {Gao}}, \bibinfo {author} {\bibfnamefont
  {Y.-H.}\ \bibnamefont {Luo}}, \bibinfo {author} {\bibfnamefont
  {B.}~\bibnamefont {Chen}},\ and\ \bibinfo {author} {\bibfnamefont
  {J.}~\bibnamefont {Liu}},\ }\bibfield  {title} {\bibinfo {title} {A
  chip-integrated comb-based microwave oscillator},\ }\href
  {https://doi.org/10.1038/s41377-025-01795-0} {\bibfield  {journal} {\bibinfo
  {journal} {Light: Science {\&} Applications}\ }\textbf {\bibinfo {volume}
  {14}},\ \bibinfo {pages} {179} (\bibinfo {year}
  {2025}{\natexlab{b}})}\BibitemShut {NoStop}%
\bibitem [{\citenamefont {Luo}\ \emph {et~al.}(2024)\citenamefont {Luo},
  \citenamefont {Shi}, \citenamefont {Sun}, \citenamefont {Chen}, \citenamefont
  {Huang}, \citenamefont {Wang}, \citenamefont {Long}, \citenamefont {Shen},
  \citenamefont {Ye}, \citenamefont {Guo},\ and\ \citenamefont
  {Liu}}]{Luo2024}%
  \BibitemOpen
  \bibfield  {author} {\bibinfo {author} {\bibfnamefont {Y.-H.}\ \bibnamefont
  {Luo}}, \bibinfo {author} {\bibfnamefont {B.}~\bibnamefont {Shi}}, \bibinfo
  {author} {\bibfnamefont {W.}~\bibnamefont {Sun}}, \bibinfo {author}
  {\bibfnamefont {R.}~\bibnamefont {Chen}}, \bibinfo {author} {\bibfnamefont
  {S.}~\bibnamefont {Huang}}, \bibinfo {author} {\bibfnamefont
  {Z.}~\bibnamefont {Wang}}, \bibinfo {author} {\bibfnamefont {J.}~\bibnamefont
  {Long}}, \bibinfo {author} {\bibfnamefont {C.}~\bibnamefont {Shen}}, \bibinfo
  {author} {\bibfnamefont {Z.}~\bibnamefont {Ye}}, \bibinfo {author}
  {\bibfnamefont {H.}~\bibnamefont {Guo}},\ and\ \bibinfo {author}
  {\bibfnamefont {J.}~\bibnamefont {Liu}},\ }\bibfield  {title} {\bibinfo
  {title} {A wideband, high-resolution vector spectrum analyzer for integrated
  photonics},\ }\href {https://doi.org/10.1038/s41377-024-01435-z} {\bibfield
  {journal} {\bibinfo  {journal} {Light: Science {\&} Applications}\ }\textbf
  {\bibinfo {volume} {13}},\ \bibinfo {pages} {83} (\bibinfo {year}
  {2024})}\BibitemShut {NoStop}%
\bibitem [{\citenamefont {Shi}\ \emph {et~al.}(2025)\citenamefont {Shi},
  \citenamefont {Zheng}, \citenamefont {Hu}, \citenamefont {Zhao},
  \citenamefont {Shang}, \citenamefont {Zhong}, \citenamefont {Chen},
  \citenamefont {Luo}, \citenamefont {Long}, \citenamefont {Sun}, \citenamefont
  {Ma}, \citenamefont {Xie}, \citenamefont {Gao}, \citenamefont {Shen},
  \citenamefont {Wang}, \citenamefont {Liang}, \citenamefont {Zhang},\ and\
  \citenamefont {Liu}}]{Shi2025}%
  \BibitemOpen
  \bibfield  {author} {\bibinfo {author} {\bibfnamefont {B.}~\bibnamefont
  {Shi}}, \bibinfo {author} {\bibfnamefont {M.-Y.}\ \bibnamefont {Zheng}},
  \bibinfo {author} {\bibfnamefont {Y.}~\bibnamefont {Hu}}, \bibinfo {author}
  {\bibfnamefont {Y.}~\bibnamefont {Zhao}}, \bibinfo {author} {\bibfnamefont
  {Z.}~\bibnamefont {Shang}}, \bibinfo {author} {\bibfnamefont
  {Z.}~\bibnamefont {Zhong}}, \bibinfo {author} {\bibfnamefont
  {Z.}~\bibnamefont {Chen}}, \bibinfo {author} {\bibfnamefont {Y.-H.}\
  \bibnamefont {Luo}}, \bibinfo {author} {\bibfnamefont {J.}~\bibnamefont
  {Long}}, \bibinfo {author} {\bibfnamefont {W.}~\bibnamefont {Sun}}, \bibinfo
  {author} {\bibfnamefont {W.}~\bibnamefont {Ma}}, \bibinfo {author}
  {\bibfnamefont {X.-P.}\ \bibnamefont {Xie}}, \bibinfo {author} {\bibfnamefont
  {L.}~\bibnamefont {Gao}}, \bibinfo {author} {\bibfnamefont {C.}~\bibnamefont
  {Shen}}, \bibinfo {author} {\bibfnamefont {A.}~\bibnamefont {Wang}}, \bibinfo
  {author} {\bibfnamefont {W.}~\bibnamefont {Liang}}, \bibinfo {author}
  {\bibfnamefont {Q.}~\bibnamefont {Zhang}},\ and\ \bibinfo {author}
  {\bibfnamefont {J.}~\bibnamefont {Liu}},\ }\bibfield  {title} {\bibinfo
  {title} {A hyperfine-transition-referenced vector spectrum analyzer for
  visible-light integrated photonics},\ }\href
  {https://doi.org/10.1038/s41467-025-61970-0} {\bibfield  {journal} {\bibinfo
  {journal} {Nature Communications}\ }\textbf {\bibinfo {volume} {16}},\
  \bibinfo {pages} {7025} (\bibinfo {year} {2025})}\BibitemShut {NoStop}%
\bibitem [{\citenamefont {Shi}\ \emph {et~al.}(2026)\citenamefont {Shi},
  \citenamefont {Zhang}, \citenamefont {Zheng}, \citenamefont {Hu},
  \citenamefont {Zhong}, \citenamefont {Shang}, \citenamefont {Ma},
  \citenamefont {Xie}, \citenamefont {Bai}, \citenamefont {Luo}, \citenamefont
  {Wang}, \citenamefont {Guo}, \citenamefont {Zhang},\ and\ \citenamefont
  {Liu}}]{shi2026}%
  \BibitemOpen
  \bibfield  {author} {\bibinfo {author} {\bibfnamefont {B.}~\bibnamefont
  {Shi}}, \bibinfo {author} {\bibfnamefont {C.}~\bibnamefont {Zhang}}, \bibinfo
  {author} {\bibfnamefont {M.-Y.}\ \bibnamefont {Zheng}}, \bibinfo {author}
  {\bibfnamefont {Y.}~\bibnamefont {Hu}}, \bibinfo {author} {\bibfnamefont
  {Z.}~\bibnamefont {Zhong}}, \bibinfo {author} {\bibfnamefont
  {Z.}~\bibnamefont {Shang}}, \bibinfo {author} {\bibfnamefont
  {W.}~\bibnamefont {Ma}}, \bibinfo {author} {\bibfnamefont {X.-P.}\
  \bibnamefont {Xie}}, \bibinfo {author} {\bibfnamefont {X.}~\bibnamefont
  {Bai}}, \bibinfo {author} {\bibfnamefont {Y.-H.}\ \bibnamefont {Luo}},
  \bibinfo {author} {\bibfnamefont {A.}~\bibnamefont {Wang}}, \bibinfo {author}
  {\bibfnamefont {H.}~\bibnamefont {Guo}}, \bibinfo {author} {\bibfnamefont
  {Q.}~\bibnamefont {Zhang}},\ and\ \bibinfo {author} {\bibfnamefont
  {J.}~\bibnamefont {Liu}},\ }\href {https://arxiv.org/abs/2602.00958}
  {\bibinfo {title} {Metrology-grade mid-infrared spectroscopy for
  multi-dimensional perception}} (\bibinfo {year} {2026}),\ \Eprint
  {https://arxiv.org/abs/2602.00958} {arXiv:2602.00958 [physics.optics]}
  \BibitemShut {NoStop}%
\bibitem [{\citenamefont {Zheng}\ \emph
  {et~al.}(2023{\natexlab{b}})\citenamefont {Zheng}, \citenamefont {Sun},
  \citenamefont {Ding}, \citenamefont {Wen}, \citenamefont {Chen},
  \citenamefont {Shi}, \citenamefont {Luo}, \citenamefont {Long}, \citenamefont
  {Shen}, \citenamefont {Meng}, \citenamefont {Guo},\ and\ \citenamefont
  {Liu}}]{Zheng23}%
  \BibitemOpen
  \bibfield  {author} {\bibinfo {author} {\bibfnamefont {H.}~\bibnamefont
  {Zheng}}, \bibinfo {author} {\bibfnamefont {W.}~\bibnamefont {Sun}}, \bibinfo
  {author} {\bibfnamefont {X.}~\bibnamefont {Ding}}, \bibinfo {author}
  {\bibfnamefont {H.}~\bibnamefont {Wen}}, \bibinfo {author} {\bibfnamefont
  {R.}~\bibnamefont {Chen}}, \bibinfo {author} {\bibfnamefont {B.}~\bibnamefont
  {Shi}}, \bibinfo {author} {\bibfnamefont {Y.-H.}\ \bibnamefont {Luo}},
  \bibinfo {author} {\bibfnamefont {J.}~\bibnamefont {Long}}, \bibinfo {author}
  {\bibfnamefont {C.}~\bibnamefont {Shen}}, \bibinfo {author} {\bibfnamefont
  {S.}~\bibnamefont {Meng}}, \bibinfo {author} {\bibfnamefont {H.}~\bibnamefont
  {Guo}},\ and\ \bibinfo {author} {\bibfnamefont {J.}~\bibnamefont {Liu}},\
  }\bibfield  {title} {\bibinfo {title} {Programmable access to microresonator
  solitons with modulational sideband heating},\ }\href
  {https://doi.org/10.1063/5.0173243} {\bibfield  {journal} {\bibinfo
  {journal} {APL Photonics}\ }\textbf {\bibinfo {volume} {8}},\ \bibinfo
  {pages} {126110} (\bibinfo {year} {2023}{\natexlab{b}})}\BibitemShut
  {NoStop}%
\bibitem [{\citenamefont {Arbabi}\ and\ \citenamefont
  {Goddard}(2013)}]{Arbabi:13}%
  \BibitemOpen
  \bibfield  {author} {\bibinfo {author} {\bibfnamefont {A.}~\bibnamefont
  {Arbabi}}\ and\ \bibinfo {author} {\bibfnamefont {L.~L.}\ \bibnamefont
  {Goddard}},\ }\bibfield  {title} {\bibinfo {title} {Measurements of the
  refractive indices and thermo-optic coefficients of si3n4 and siox using
  microring resonances},\ }\href {https://doi.org/10.1364/OL.38.003878}
  {\bibfield  {journal} {\bibinfo  {journal} {Opt. Lett.}\ }\textbf {\bibinfo
  {volume} {38}},\ \bibinfo {pages} {3878} (\bibinfo {year}
  {2013})}\BibitemShut {NoStop}%
\end{thebibliography}%
\end{document}